\documentclass[preprint,letter]{revtex4}

\usepackage{amssymb}
\usepackage{amsfonts}
\usepackage{amsmath}
\usepackage{mathrsfs}
\usepackage{eufrak}
\usepackage{fancyhdr}
\usepackage{verbatim}
\usepackage[dvips]{graphicx}

\DeclareMathAlphabet{\mathpzc}{OT1}{pzc}{m}{it}
\newcommand{\Zgr}{\mathbb Z}
\newcommand{\cep}{\mathcal{C}_\epsilon}
\newcommand{\dep}{\mathcal{D}_\epsilon}

\newcommand{\oneover}[1]{\frac{1}{#1}}
\newcommand{\lagra}{\mathcal{L}}

\newcommand{\intd}{\mathrm{d}}
\newcommand{\intD}{\mathcal{D}}
\newcommand{\ano}{\mathcal{A}_{(6)}}
\newcommand{\sano}{\mathcal{A}_{(6)\star}}
\newcommand{\kb}{\frac{\k}{b^2}}

\newcommand{\ka}{\frac{k}{a^2}}

\newcommand{\sT}{T_\star}
\newcommand{\sH}{H_\star}
\newcommand{\sF}{F_\star}

\newcommand{\sB}{B_\star}

\newcommand{\sR}{R_\star}
\newcommand{\sbb}{b_\star}
\newcommand{\srho}{\rho_\star}
\newcommand{\ssig}{\s_\star}
\newcommand{\ssigp}{\s_{p\star}}
\newcommand{\ssigpp}{\s_{\p\star}}
\newcommand{\schi}{\chi_\star}
\newcommand{\rhofo}{\varrho}
\newcommand{\chifo}{\mathpzc{x}}
\newcommand{\rhofz}{\varrho_0}
\newcommand{\chifz}{\mathpzc{x}_0}
\newcommand{\cHH}{\check H}
\newcommand{\cTT}{\check T}
\newcommand{\cAA}{\check A}
\newcommand{\Afo}{\mathpzc{A}}
\newcommand{\cAAfo}{\check{\mathpzc{A}}}

\newcommand{\ck}{\check\k}

\newcommand{\cqq}{\check q}

\newcommand{\crho}{\check\rho}
\newcommand{\cchi}{\check\chi}
\newcommand{\crhofo}{\check\varrho}
\newcommand{\cchifo}{\check{\mathpzc{x}}}
\newcommand{\csH}{\check H_\star}

\newcommand{\csrho}{\check\rho_\star}
\newcommand{\cschi}{\check\chi_\star}

\newcommand{\stano}{\mathcal{\tilde A}_{(6)\star}}
\newcommand{\sY}{{Y}_{\star}}
\newcommand{\stY}{{\tilde Y}_{\star}}
\newcommand{\tn}{\tilde\nu}

\newcommand{\tcreq}{\tilde c_{\rho\scriptscriptstyle{(eq)}}}
\newcommand{\tcrst}{\tilde c_{\rho\scriptscriptstyle{(st)}}}
\newcommand{\tcrx}{\tilde c_{\rho,\scriptscriptstyle\x}}
\newcommand{\tcrd}{\tilde c_{\rho,\scriptscriptstyle d}}
\newcommand{\tcrsix}{\tilde c_{\rho,\scriptscriptstyle{d=6}}}
\newcommand{\tcv}{\tilde c_V}
\newcommand{\tcveq}{\tilde c_{V\scriptscriptstyle{(eq)}}}
\newcommand{\tcvst}{\tilde c_{V\scriptscriptstyle{(st)}}}
\newcommand{\tcvx}{\tilde c_{V,\scriptscriptstyle\x}}
\newcommand{\tcvd}{\tilde c_{V,\scriptscriptstyle d}}
\newcommand{\tcvsix}{\tilde c_{V,\scriptscriptstyle{d=6}}}

\newcommand{\Mpl}{M_{Pl}^4}
\newcommand{\mpl}{M_{Pl}^2}
\newcommand{\MPl}{M_{Pl}^6}
\newcommand{\sgn}{\mathrm{sgn}}

\newcommand{\diag}{\mathrm{diag}}

\newcommand{\ex}{\mathrm{e}}

\newcommand{\ogr}{\mathcal{O}}

\newcommand{\cd}{\cdot}
\newcommand{\pa}{\partial}

\newcommand{\bea}{\begin{eqnarray}}
\newcommand{\ena}{\end{eqnarray}}
\newcommand {\non}{\nonumber}

\newcommand{\refeq}[1]{(\ref{#1})}

\renewcommand{\a}{\alpha}
\renewcommand{\b}{\beta}
\renewcommand{\d}{\delta}
\renewcommand{\th}{\theta}

\newcommand{\g}{\gamma}

\newcommand{\e}{\epsilon}
\newcommand{\z}{\zeta}

\renewcommand{\k}{\kappa}

\renewcommand{\l}{\lambda}
\renewcommand{\L}{\Lambda}
\newcommand{\m}{\mu}

\newcommand{\n}{\nu}

\newcommand{\x}{\xi}

\newcommand{\p}{\pi}

\newcommand{\s}{\sigma}

\renewcommand{\S}{\Sigma}
\renewcommand{\t}{\tau}

\renewcommand{\o}{\omega}
\renewcommand{\O}{\Omega}

\begin{document}

\date{\today}
\preprint{CPHT-RR008.0207\\Bicocca-FT-07-4}

\title{7D Randall--Sundrum cosmology, brane--bulk energy exchange and holography}
\author{Liuba Mazzanti}
\affiliation{Dip. di Fisica G. Occhialini, Milano--Bicocca University, 20126 Milano, ITALY,\\INFN, Milano--Bicocca,\\ and CPhT, \'Ecole Polytechnique, 91128 Palaiseau, FRANCE.}
\email{liuba.mazzanti@cpht.polytechnique.fr}

\begin{abstract}
We discuss the cosmological implications and the holographic dual theory of the 7D Randall--Sundrum (RS) gravitational set--up. Adding generic matter in the bulk on the 7D gravity side, we study the
cosmological evolution inferred by the non vanishing value of the brane--bulk energy exchange parameter. This analysis is achieved in detail for specific assumptions on the internal space evolution,
including analytical considerations and numerical results.  The dual theory is then constructed, making use of the holographic renormalization procedure. The resulting renormalized 6D CFT is anomalous and
coupled to 6D gravity plus higher order corrections. The critical point analysis on the brane is performed. Finally, we sketch a comparison between the two dual descriptions. We moreover generalize the
AdS/CFT dual theory to the non conformal and interacting case, relating the energy exchange parameter of the bulk gravity description to the new interactions between hidden and visible sectors.
\end{abstract}


\maketitle

\section{Introduction}\label{introduction} 
Many aspects of brane models have been recently developed both from the point of view of Standard Model building (see for example \cite{Kiritsis:2003mc,Cascales:2003wn} and references therein) and of
cosmology (\cite{Langlois:2002bb}--\cite{Kiritsis:2005bm} and references therein). In particular, Randall-Sundrum \cite{Randall:1999ee} cosmology \cite{Binetruy:1999ut} is among the most interesting insights
related to brane--world cosmology (other are mirage cosmology \cite{Kehagias:1999vr}, brane inflation \cite{Dvali:1998pa}, brane induced gravity \cite{Gabadadze:2006tf}--\cite{Maartens:2003tw},
brane/anti-brane inflation \cite{Burgess:2001fx}--\cite{Buchel:2006em}, cosmologies from higher derivative corrections \cite{Nojiri:2000gv}--\cite{Sami:2005zc}, particular examples with varying speed of
light \cite{Alexander:1999cb} and cosmological evolution induced by the rolling tachion \cite{Gibbons:2002md}, inflation in flux compactification scenarios \cite{Kachru:2003sx}--\cite{Conlon:2005jm} and
recent brane--world models \cite{Setare:2006pj}--\cite{Maartens:2006qf}). Branes can be used to give origin to four dimensional gauge theories (living on a 3--brane) and hence localizing matter in four
dimensions embedded in the ten dimensional space--time in which string theory lives (or eleven dimensional for M theory). On the gravity side, since gravitons propagate in all dimensions admitted by string
theory, the validity of Newton's law seems to constrain the number of non compact dimensions to be four.

Randall and Sundrum \cite{Randall:1999ee} proposed an alternative way of localizing gravity in four dimensions without compactifying the extra dimensions. This was achieved in RSII model by assuming a
warped extra direction, instead of a compact one (compact extra dimensions have been used \cite{Arkani-Hamed:1998rs} as an attempt of giving an explanation to the hierarchy problem, to which RSI
\cite{Randall:1999ee} represents an alternative way out). The set--up of RSII is five dimensional gravity in a bulk space--time cut by a 3--brane. There is, in addition, a $\mathbb Z_2$ reflection of the
extra dimension tranverse to the brane, having as fixed point the location of the brane. The result is a bound state graviton mode localized on the brane and a tower of KK modes, without mass gap, that give
negligible corrections to the effective 4D gravity description. The metric solving the equations of motion for the five dimensional action is a slice of $AdS_5$ copied and reflected w.r.t. the $\mathbb Z_2$
symmetry.  It can be viewed as coming from the Type IIB string theory background for a stack of $N$ D3--branes, in the low energy effective field theory approximation, which is indeed dual to the
$AdS_5\times S^5$ near horizon geometry for a 3--brane supergravity solution. In RS analysis only gravity on $AdS_5$ is considered, since the $S^5$ is factored out from the anti de Sitter space, giving KK
modes. The truncation of $AdS_5$ with the 3--brane cuts out its boundary. 

In a recent work \cite{Bao:2005ni}, a different string background related to RS set--up has been considered. The analogous analysis to the 5D RS model has been made by Bao and Lykken \cite{Bao:2005ni} in a
seven dimensional anti de Sitter background rather than in the five dimensional original model. The background may come from the near horizon geometry of M5--branes in the eleven dimensional M theory, which
gives $AdS_7\times S^4$. As for the five dimensional model, the sphere is factored out and only the physics of gravity in $AdS_7$ plus KK modes is considered. A further step performed in \cite{Bao:2005ni}
is to reduce $AdS_7\rightarrow AdS_5\times \S^2$, where $\S^2$ is a two dimensional internal space (namely a two--sphere or a torus). In \cite{Bao:2005ni} the RS spectrum of KK modes gets modified and
supplemental KK and winding modes appear, because of the $\S^2$ compactification. In \cite{Avramis:2004cn} the 7D supergravity orbifold compactification on $S^1/\mathbb Z_2$ is considered in the context of
anomaly cancellation on the boundary of the background, showing that the matter contents of the theory cannot be completely generic.

On the cosmological side, RS models can give new descriptions of the cosmological evolution of our universe. A realistic model should be able to explain the existence of dark energy and the nature of dark
matter, early time inflation and eventually the exit from this phase, as well as the late time acceleration coming from the present observations --- additional issues are related to the cosmological constant,
temperature anisotropies, etc. (see \cite{Padmanabhan:2006kz}--\cite{Trodden:2004st} for recent reviews on the observed cosmology). RS cosmologies have been studied and found to exhibit some of these
features. Generally, brane--world cosmological models should take account of the energy exchange between brane and bulk that naturally arises because of the non factorized extra direction.  The implications
of this energy exchange has been analyzed in \cite{Kiritsis:2002zf}. The authors propose some scenaria describing the cosmological evolution of a universe with two accelerating phases, as we expect from
experimental data. Moerover, most of the fixed points were shown to be stable.  Earlier attempts include \cite{Kiritsis:2001bc}. Subsequent papers \cite{Hebecker:2001nv}--\cite{Cai:2005qm} have been written
on the subject, also finding new solutions, some of which exact.

The aim of this paper is to investigate the 7D RS cosmology with brane--bulk energy exchange and to explore the model from the holographic point of view, making an explicit comparison between the two
descriptions. Our starting point is gravity in the 7D bulk cut by a five--brane and with the usual RS $\mathbb Z_2$ reflection plus a generic matter term on the brane. In order to study the cosmological
evolution of the brane--world, the ansatz for the metric is time dependent. Besides, the direction tranverse to the 5--brane is the warped direction characterizing RS models. Unlike in \cite{Bao:2005ni}, we
have different warp factors for the 3D extended space and for the two dimensional compact internal space. It is worth noticing that the gravitational coupling constant of the 4D space--time is dynamically
related to the 7D Newton constant, since the compact space volume is generally time dependent. Indeed, the 4D energy density and pressures are also dynamical functions of the density and the pressures defined
on the brane. We analyze the generic Friedmann and (non) conservation equations, also including the energy exchange terms, in order to get the expressions for the Hubble parameter of the 4D space--time as a
function of the density and to describe realistic cosmologies. It's interesting to study both analytically and numerically the system of Einstein equations. The analogous critical point analysis in the 5D
bulk was performed in \cite{Kiritsis:2002zf}. Some explicit solutions, derived with simplifying assumptions on the parameter of the internal space and on its geometry are also illustrated in our work. 

Further investigations on the 5D RS cosmology with brane--bulk energy exchange have been made from the holographic point of view \cite{Kiritsis:2005bm}, studying the dual theory in one lower dimension. The
gauge/gravity duality \cite{Maldacena:1997re,Witten:1998qj} (see \cite{Aharony:1999ti} for a complete review) has undergone great improvements over the last ten years and provides a new approach to the
analysis of brane--world models. As it is explained in \cite{Hawking:2000kj}--\cite{Perez-Victoria:2001pa}, the truncation of the $AdS_{d+1}$ space is equivalent to introducing in the dual picture a UV cutoff
for the $d$--dimensional gauge theory (conformal field theory) coupled to $d$--dimensional gravity. Earlier suggestions about this idea are present in \cite{Nojiri:2000eb}. The presence of brane--bulk
exchange corresponds to interactions between the gauge theory and the matter fields while the bulk ``self--interaction'' is shown to be related to the perturbation of the CFT (that becomes a strongly coupled
gauge theory). In \cite{Kiritsis:2005bm} explicit examples of cosmological evolutions in the holographic 5D/4D picture as well as comparison between the two dual theories have been discussed. Other
cosmological models have been analyzed in the context of the holographic correspondence \cite{Hawking:2000bb}--\cite{Gubser:1999vj}.

Exploiting the AdS/CFT results, we build the holographic theory corresponding to the 7D RS background. The 7D RS dual theory (see \cite{Kiritsis:2005bm} for the analogous 5D/4D derivation) is then a
renormalized 6D CFT (the theory corresponding to the M5 system is an anomalous \cite{Henningson:1998gx}--\cite{deHaro:2000xn} (0,2) SCFT, but any other six dimensional large--N CFT can be chosen) coupled to
6D gravity. See also \cite{Nojiri:1998dh} for other examples of holographic Weyl anomaly derivation. There are in addition higher order corrections to gravity and the six dimensional matter action.  Higher
derivative terms driven by conformal four dimensional anomaly \cite{Deser:1993yx} were proved to lead to an inflationary critical point in the 4D Starobinsky model \cite{Starobinsky:1980te} and to a
successive greaceful exit from the long primordial inflation. As illustrated in \cite{Vilenkin:1985md} higher derivative contributions to the Einstein equations cause the universe to enter a matter dominated
era where the scale factor oscillates after inflation and to proceed through thermalization to a radiation dominated era. In our 6D holographic cosmological model we specially look for de Sitter fixed point
solutions to the equations of motion describing late time acceleration or critical points suitable for early time inflation studying the associated stability matrix. A comparison with the 7D bulk analysis
results shows some peculiar features of the 7D/6D set--up. 

Interesting results for cosmologies with compactification emerge in the context of dynamical compactification \cite{Mohammedi:2002tv,Andrew:2006yt}. The compact space is treated with a different scale factor
(as in the approach that will be used in this paper). In particular, in the context of dynamical compactification, the scale factor for the internal space has an inverse power dependence on the scale factor
for the visible directions. The extra dimensions thus contract as the extended space expands. We also include some remarks on dynamical compactification applications in our set--up. Other attempts to reduce
to conventional cosmology and investigate issues such as the cosmological constant from models with arbitrary number of extra dimensions are given in \cite{Kamani:2006tv,Zhuk:2006kh}. In particular,
cosmologies in six dimesions are analyzed in \cite{Papantonopoulos:2006uj}.

The structure of the paper is as follows. Next section will describe the set--up of the seven dimensional RS model. Section \ref{bulk cosmology} will be focused on the cosmological evolution from the 7D
point of view, admitting brane--bulk energy exchange and particularly investigating the form of Einstein equations with some specific ansatz, while in section \ref{critical bulk} the critical point analysis
will be illustrated, including numerical phase space portrait and explicit solutions. In section \ref{dual} and the following we will derive the 6D holographic dual to the 7D RS and the associated equations
of motion. Section \ref{critical brane} summarizes the fixed points in the holographic description and their stability. Some examples of the correspondence between the brane and bulk points of view will be
given in section \ref{examples}. The generalization to non conformal and interacting theory, corresponding to non vanishing brane--bulk energy exchange and bulk self interaction in the 7D approach, will be
exposed in section \ref{general}. Finally, the last section will summarize the results, give some conclusions and further considerations.  The first appendix gives the form of the general anomaly for a 6D CFT
in a curved space and of the other trace terms in the 6D theory and appendix \ref{fixed points} is devoted to the critical point analysis in the six dimensional cosmology. 
\section{7D RS set--up}\label{setup} 
As we announced in the introduction, we will work in a seven dimensional bulk with a 5--brane located at the origin of the direction $z$ tranverse to the brane itself and with a $z\rightarrow-z$ $\Zgr_2$
identification. In analogy to the 5D RS model, the action in this seven dimensional set--up is given by the sum of the Einstein--Hilbert action with 7D cosmological constant plus a contribution localised on
the brane that represents the brane tension. Besides, we also put a matter term both in the bulk and on the brane. In formula we thus have 
\bea\label{7D action}
S&=&S_{EH}+S_{GH}+S_{m,B}+S_{tens}+S_m=\non\\&=&\int\intd^7 x\sqrt{-G}\left(M^5R-\L_7+\lagra_B^{mat}\right)+\int\intd^6 x\sqrt{-\g}\left(-V+\lagra_b^{mat}\right)+S_{GH}
\ena 
where $V$ is the brane tension and we will call the associated contribution to the action $S_{tens}$. $S_{EH}$ is the usual Einstein--Hilbert action with the seven dimensional bulk cosmological constant
$\L_7$ and $\lagra_B^{mat},\lagra_b^{mat}$ are, respectively, the bulk and brane matter lagrangians. $S_{GH}=\int\intd^6x\sqrt{-\g}K$, where $K$ denotes the trace of the extrinsic curvature on the boundary,
is the Gibbons--Hawking action added in order to cancel the boundary term arising computing the variation on the Einstein--Hilbert action and to get the usual Einstein equations. The action for the matter in
the bulk $S_{m,B}$ is an additional term with respect to the usual RS set--up, whereas the matter on the brane contribution will be refered to as $S_m$. The metric $\g_{\m\n}$ is the induced metric on the
brane. The brane tension is necessary in the RS models in order to compensate for the presence of the cosmological constant in the bulk.

The classical solution of the equations of motion for the theory above, neglecting all the matter terms and with a warped geometry of the kind $\intd s^2=\ex^{-W}\intd x^2+\intd z^2$ ($W=W(z)$ is the warp
factor and the 6D $x$--directions are flat), is the analogue of the solution described by Randall and Sundrum \cite{Randall:1999ee} for the 5D RSII model. The 7D solution gives as a result
$W(z)=2|z|\sqrt{-\frac{\L_7}{30M^5}}$, so that the space--time is a slice of $AdS_7$ with the $\Zgr_2$ typical reflection, where there should exist a relation between the brane tension and the bulk
cosmological constant $3V^2=-40M^5\L_7$. 

The aim of this section and of the next one is to generalize the RS ansatz to a time dependent background and to wrap the 5--brane over a two dimensional internal space, ending up with an effective 4D
cosmology.  Taking account of the seventh warped extra dimension and of the compactification over the other two extra dimensions, giving two different warp factors to the 3D space and the internal 2D
space, the time dependent ansatz for the metric is of the form
\bea\label{7D metric}
\intd s^2=-n^2(t,z)\intd t^2+a^2(t,z)\z_{ij}\intd x^i\intd x^j+b^2(t,z)\xi_{ab}\intd y^a\intd y^b+f^2(t,z)\intd z^2
\ena
with the maximally symmetric $\z_{ij}$ background in three spatial dimensions (with spatial curvature $k$) and $\xi_{ab}$ for the 2D internal space (with spatial curvature $\k$). We use capital indices
$A,B,\dots$ to run over the seven dimensions, $i,j,\dots$ for the three spatial dimensions of the 4D space--time, $a,b,\dots$ for the two internal dimensions. In our notations $z$ represents the seventh
warped extra direction, the $y$ coordinates belong to the 2D internal space, while the $\{x_\m\}=\{t,x_i,y_a\}$ run over the 6D space--time on the brane. Summarizing, the structure of the bulk is thus made of
a time coordinate, three extended maximally symmetric spatial dimensions (that gives, toghether with the time, the visible 4D space--time), two compact dimensions and a warped direction. The 3D and 2D spaces
have two different scale factors $a(t,z)$ and $b(t,z)$ respectively, while a gauge choice can be made for the values of the $n(t,z)$ and $f(t,z)$ factors on the brane, i.e. when $z=0$.

A less physically meaningful background, but better understood, would be to have a five dimensional maximally symmetric space with some 5D metric $\tilde\z_{ij}$ and just one scale factor $\tilde a(t,z)$,
without compactifying on any two dimensional internal space. The solution to the equations of motion in this case is much simpler. The results related to this background will be briefly mentioned along with
the more realistic analysis with brane wraping over the two dimensional internal space.
\section{Cosmological evolution in the bulk}\label{bulk cosmology}
In this section we will analyze some aspects of the cosmological evolution from the 7D bulk point of view. We will write the equations of motion for the bulk action and solve them by making assumptions to
simplify their form and get explicit results evaluated on the brane.

Given the set--up described in the previous section, we parametrize all the contributions to the stress--energy tensor as 
\bea\label{stress energy para bulk}
T^A_C|_{v,b}&=&\frac{\delta(z)}{f}\,\diag\left(-V,-V,-V,-V,-V,-V,0\right)\non\\
T^A_C|_{v,B}&=&\diag\left(-\L_7,-\L_7,-\L_7,-\L_7,-\L_7,-\L_7,-\L_7\right)\non\\
T^A_C|_{m,b}&=&\frac{\delta(z)}{f}\,\diag\left(-\rho,p,p,p,\p,\p,0\right)\non\\ T^A_C|_{m,B}&=&T^A_C 
\ena 
with the subindices $v$ and $m$ indicating the vacuum and matter stress--energy tensors, while $b$ and $B$ stand for the brane and bulk contributions respectively.  A difference between this (4+2+1)D
background and the simpler (6+1)D analysis without the 2D compactification cited at the end of the previous section, is having in \refeq{stress energy para bulk} two different pressures in the 3D space and in
the 2D compact dimensions for the matter on the brane, while for the (6+1)D background we would put $\p=p$. This generalization is due to the fact that we don't assume homogeneity for the matter fluid in the
whole (3+2)--dimensional space, but only in the 3D and 2D spaces separetely. 

Having calculated the Einstein tensor, we can put the explicit expression in the equation
\bea
G_{AC}=\oneover{2M^5}T_{AC}
\ena
evaluated on the brane (from now on all the functions are evaluated on the brane, i.e. at $z\rightarrow0$), in the specific background (\ref{7D metric}). As a consequence, for the $00$, $ij$ and $ab$
components we obtain the jump equations
\bea\label{jump eqs}
a'_+=-a'_-&=&-\frac{fa}{20M^5}\left(V+\rho+2p-2\p\right)\non\\
b'_+=-b'_-&=&-\frac{fb}{20M^5}\left(V+\rho-3p+3\p\right)\\
n'_+=-n'_-&=&\frac{fn}{20M^5}\left(-V+4\rho+3p+2\p\right)\non
\ena
These are the values of the warp factors in the limit $z\to0$, where the subscripts $+$ and $-$ dinstinguish the limit taken from below from the limit taken from above. The prime denotes the partial
derivative with respect to the $z$ coordinate, while the dot indicates the time derivative.  For the $07$ and $77$ components, substituing the expressions (\ref{jump eqs}) and choosing a gauge with $f(t,0)=1$
and $n(t,0)=1$, we get the (non) conservation equation
\bea\label{conservation 7D}
&&\dot\rho+3\frac{\dot a}{a}\left(\rho+p\right)+2\frac{\dot b}{b}\left(\rho+\p\right)=2T_{07}
\ena
and the Friedmann equation
\bea\label{Friedmann 7D}
&&3\frac{\ddot a}{a}+2\frac{\ddot b}{b}+3\frac{\dot a^2}{a^2}+\frac{\dot b^2}{b^2}+6\frac{\dot a}{a}\frac{\dot b}{b}+3\frac{k}{a^2}+\frac{\k}{b^2}=\non\\
&=&-\frac{5}{(20M^5)^2}\bigg[V\left(6p-\p-\rho\right)+\rho\left(6p-\p+\rho\right)+\rho^2+\non\\
&&\left.+\oneover{5}\left(p-\p\right)\left(V-19\rho-3p-7\p\right)\right]+15\l_{RS}-\oneover{2M^5}T^7_7
\ena
We have defined the constant
\bea
\l_{RS}=\frac{1}{30M^5}\left(\L_7+\frac{3}{40M^5}V^2\right)
\ena
which plays the role of an effective cosmological constant on the brane. These \refeq{conservation 7D}--\refeq{Friedmann 7D} are two equations in five variables $H,F,\rho,p,\p$. We will thus have to make an
ansatz for some of those variables. 

The pure RS system correspond to setting $T^0_7=T^7_7=0$, that means putting to zero the brane--bulk energy exchange and no cosmological term on the brane, i.e. $\l_{RS}=0$, to restore RS fine--tuning.

We can now write a simplified version of the differential equations (\ref{conservation 7D})--(\ref{Friedmann 7D}), using the usual ansatz for the equation of state of the matter fluid on the brane, i.e.
\bea\label{ansatz pressures 7D}
p=w\rho,\qquad \p=w_\p\rho
\ena
The set of equations, in terms of the Hubble parameters $H\equiv\frac{\dot a}{a},F\equiv\frac{\dot b}{b}$, is
\bea\label{Friedmann 7D ansatz}
3\dot H+2\dot
F+6H^2+6HF+3F^2+3\frac{k}{a^2}+\frac{\k}{b^2}
&=&-\oneover{M^{10}}\left(c_VV+c_\rho\rho\right)\rho+15\l_{RS}-\frac{T^7_7}{2M^5}\\
\dot\rho+\left[3(1+w)H+2(1+w_\p)F\right]\rho&=&2T_{07}\label{conservation 7D ansatz}
\ena
with
\bea\label{cV crho}
c_V=\frac{31w-6w_\p-5}{400},\quad c_\rho=\frac{11w+14w_\p+10-(w-w_\p)(3w-7w_\p)}{400}
\ena

Looking at the definition of the two coefficients $c_V,c_\rho$, we can note that equation (\ref{Friedmann 7D ansatz}) gets further simplified when the two pressures $p$ and $\p$ are equal. This can be seen
also from the previous equation (\ref{Friedmann 7D}), where the ``non--standard'' term on the r.h.s. (standard with respect to the homogeneous background analysis) is proportional to $(p-\p)$. We will first
examine some cosmological solutions assuming $p=\p$ and then we will drop this equal pressure condition to find an expression for $H$, in terms of the energy density, in the particular limits of static
compact extra dimensions and equal scale factors.
\subsection{Equal pressures in 3D and 2D compact space}\label{equal pressures}
We can first try to find an interesting solution by simplifying the computation assuming $\p=p$. In this case, the equation (\ref{Friedmann 7D}) written in terms of the Hubble parameters of the 3D space and
2D extra dimensions, defined as $H=\dot a/a$, $F=\dot b/b$ respectively, toghether with the (non) conservation equation (\ref{conservation 7D}), becomes
\bea
3\dot H+2\dot F+6H^2+6HF+3F^2+3\frac{k}{a^2}+\frac{\k}{b^2}&=&  \non\\
=-\oneover{80M^{10}}\left[V\left(5p-\rho\right)+\rho\left(5p+2\rho\right)\right]+15\l_{RS}-\frac{T^7_7}{2M^5}&&  \label{Friedmann 7D equal pressures before}\\\non\\
\dot\rho+\left(3H+2F\right)\left(\rho+p\right)&=&2T_{07}  \label{conservation 7D equal pressures}
\ena
We note that the system of equations written above still contains three variables $H(t),F(t),\rho(t)$ but only two equations. So we are able to just determine the value of the 3D Hubble parameter $H(t)$ as
a function of the 2D one $F(t)$. Moreover, given the complicated form of this set of equations, we will make some assumptions on the internal space, such as flat compact extra dimensions ($\k=0$) or static
extra dimensions ($F(t)\equiv0$) in the following subsections.

Manipulating \refeq{Friedmann 7D equal pressures before} the system takes the form
\bea\label{Friedmann 7D equal pressures}
5\frac{\intd}{\intd t}\left(3H+2F\right)^2+6\left(3H+2F\right)^3+6\left(3H+2F\right)\left(H-F\right)^2&=&\non\\
=\frac{1}{8M^{10}}\left[5V\dot\rho+6\left(3H+2F\right)V\rho+5\rho\dot\rho+3\left(3H+2F\right)\rho^2\right]+150\left(3H+2F\right)\l_{RS}+&&\non\\
+\frac{5}{8M^{10}}\left(V+\rho\right)2T^0_7-\frac{5}{M^5}\left(3H+2F\right)T^7_7-10\left(3H+2F\right)\left(3\frac{k}{a^2}+\frac{\k}{b^2}\right)&&
\ena
It is interesting to derive the first order ODEs from the second order one \refeq{Friedmann 7D equal pressures}, in order to find the expression for $H^2$ as a function of the localized matter energy density
$\rho$ and to perform the critical point analysis.
\subsubsection{Flat compact extra dimensions with equal pressures}\label{bulk flat}
In flat compact extra dimensions ($\k=0$) and flat 3D space ($k=0$), equation (\ref{Friedmann 7D equal pressures}) shows that in the limit in which the two Hubble parameters are almost equal we can deduce the
solution for $(3H+2F)$ in terms of the localized energy density $\rho$ and of a mirage density $\chi$ that we will define below through a differential equation. In fact, in this case the third term on the
l.h.s.  of (\ref{Friedmann 7D equal pressures}) is negligeable, leaving an analogous differential expression on both sides of the equality.

The solution for $\left(H-F\right)\ll\left(3H+2F\right)$ and $k=\k=0$ is given by
\bea\label{7D flat Hubble}
\left(3H+2F\right)^2=\frac{1}{16M^{10}}\rho^2+\frac{V}{8M^{10}}\left(\rho+\chi\right)+25\l_{RS}
\ena
The solution is written in terms of the mirage density $\chi$ and the localized energy density $\rho$. The mirage density must satisfy
\bea\label{7D flat chi}
\dot\chi+\frac{6}{5}\left(3H+2F\right)\chi=\left(\frac{\rho}{V}+1\right)2T^0_7-\frac{8M^5}{V}\left(3H+2F\right)T^7_7
\ena
and the equation for $\rho$ is the (non) conservation equation
\bea\label{7D flat rho}
\dot\rho+\left(3H+2F\right)\left(\rho+p\right)=2T_{07}
\ena

Here we get a linear and quadratic $\rho$ dependence for the Hubble parameter $H^2$, as well as a depedence on the mirage density $\chi$ and on the hidden sector Hubble parameter $F$. The quadratic and linear
terms in $\rho$ are analogous to those in the 5D analysis \cite{Kiritsis:2005bm}, implying that for $\rho\ll V$ the cosmological evolution looks four dimensional, while it moves away from the 4D behavior
for $\rho\gg V$. The term in $\chi$ also already appears in the 5D model, as well as the $\l_{RS}$ constant term. Besides, the mirage energy density dynamics are controlled by the bulk parameters $T^0_7$ and
$T^7_7$, that represent the brane--bulk energy exchange and bulk pressure, as in \cite{Kiritsis:2005bm}.  However, a new variable $F$, the internal dimension Hubble parameter, arises and remains undetermined
unless we make an ansatz for it. We can argue that the solution (\ref{7D flat Hubble})--(\ref{7D flat chi}) is written in terms of a ``total'' Hubble parameter $\oneover{5}(3H+2F)$, that carries the same
characteristics as the $H$ Hubble parameter in the 5D model, but also includes the dynamics of the evolution of the extra dimensions. For equal scale factors, $F=H$, this ``total'' Hubble parameter reduces to
$H$ alone, giving the exact analogue to the 5D RS cosmology. 
\subsubsection{Equal scale factors with equal pressures}\label{bulk equal pressures equal scale factors} 
A special case in which the $\left(H-F\right)\ll\left(3H+2F\right)$ limit is valid is the equal scale factor case $F=H$. The results can directly be obtained from the previous subsection, yielding 
\bea\label{H 7D equal equal pressures}
H^2&=&\frac{1}{400M^{10}}\rho^2+\frac{V}{200M^{10}}\left(\rho+\chi\right)+\l_{RS}-\oneover{10}\left(3\frac{k}{a^2}+\frac{\k}{a^2}\right)\\
\dot\chi+6H\chi&=&\left(\frac{\rho}{V}+1\right)2T^0_7-\frac{40M^5}{V}HT^7_7\\ \dot\rho+5H\left(\rho+p\right)&=&2T_{07} 
\ena 
We added the curvature contributions that can be computed exactly in this limit. This solution is particularly simple thanks to the simultaneous vanishing of the $(H-F)$ and $(p-\p)$ terms.  It shows the
quadratic depandence of $H^2$ on $\rho$ and the linear term in $\left(\rho+\chi\right)$. The mirage density reduces to free radiation in 6D space--time when we restrict to pure RS configuration, with no
energy exchange. In this same limit, the localized matter energy density obeys to standard conservation equation in 6D. 

We will now drop the equal pressure ansatz and derive the expression for the Hubble parameter of the 3D space making particular assumptions on the internal space scale factor. We suppose from now on to live
in a spatially flat universe ($k=0$), where nevertheless the extra dimensions may be curved ($\k\neq0$ generally). 
\subsection{Equal scale factors (generic pressures)}\label{bulk equal scale factors}
For generic pressures, we make use of the parametrization by means of $w,w_\p$ for the pressures of both the non compact and internal dimensions $p,\p$ that we introduced in (\ref{ansatz pressures 7D}),
coming from the generalization of the equation of state for a fluid with energy density $\rho$. 

We evaluate the Friedmann and (non) conservation differential equations (\ref{Friedmann 7D ansatz})--(\ref{conservation 7D ansatz}) assuming the scale factors of the 3D space and of the 2D internal space to
be equal and consequently assuming the Hubble parameters governing the cosmological evolution of the two spaces to be equal, $F=H$. The system (\ref{Friedmann 7D ansatz})--(\ref{conservation 7D ansatz}) takes
the form
\bea
\frac{5}{2}\left(\dot{H^2}+6H^3\right)&=&-\oneover{M^{10}}\left(c_VV+c_\rho\rho\right)H\rho-\frac{T^7_7}{2M^5}+15H\l_{RS}-\frac{\k}{a^2}H\label{Friedmann 7D equal}\\
\dot\rho+(3(1+w)+2(1+w_\p))H\rho&=&2T_{07}\label{conservation 7D equal}
\ena
with the coefficients $c_V,c_\rho$ still given by (\ref{cV crho}). With the help of (\ref{conservation 7D equal}), (\ref{Friedmann 7D equal}) can be brought in a form from which we can explicitly deduce $H$
as a function of $\rho$ and $\chi$
\bea\label{Friedmann 7D equal explicit}
\frac{5}{2}\left(\dot{H^2}+6H^3\right)&=&\frac{1}{M^{10}}\left[\tcveq V\left(\dot\rho+6H\rho\right)+\tcreq \left(\dot{\left(\rho^2\right)}+6H\rho^2\right)\right]+\non\\
&&+\frac{2T^0_7}{M^{10}}\left(\tcveq V+\tcreq \rho\right)-H\frac{T^7_7}{2M^5}+15H\l_{RS}-\frac{\k}{a^2}H
\ena
yelding
\bea
H^2&=&\frac{\tcreq }{5M^{10}}\rho^2+\frac{2\tcveq V}{5M^{10}}\left(\rho+\chi\right)+\l_{RS}-\oneover{10}\frac{\k}{a^2}\label{H 7D equal}\\
\dot\chi+6H\chi&=&2T^0_7\left(1+\frac{\tcreq }{\tcveq }\frac{\rho}{V}\right)-\frac{M^5}{2\tcveq V}HT^7_7  \label{chi 7D equal}\\
\dot\rho+(3(1+w)+2(1+w_\p))H\rho&=&2T_{07}\label{rho 7D equal}
\ena
The two constants $\tcveq ,\tcreq $ must satisfy
\bea\label{ctilde equal}
\tcveq =\frac{c_V}{3(1+w)+2(1+w_\p)-6},\qquad \tcreq =\frac{c_\rho}{3(1+w)+2(1+w_\p)-3}
\ena
in order to have the right coefficients in equation (\ref{Friedmann 7D equal explicit}). For some values of $w,w_\p$ the denominator of $\tcveq$ or $\tcreq$ may vanish. However we can fix $w_\p$ such that
$c_V$ (or $c_\rho$) becomes proportional to $3(1+w)+2(1+w_\p)-6$ (or $3(1+w)+2(1+w_\p)-3$ for $c_\rho$) and $\tcveq$ (or $\tcreq $) is finite. As an example consider $w_\p=w$ (equal pressure in the internal
space and 3D space, $\p=p$) and check that both $\tcveq$ and $\tcreq$ remains finite and equal to $1/80$. Clearly, when $\tcveq$ (or $\tcreq$) diverges we cannot write the Friedmann equation \refeq{Friedmann
7D equal} in the form
\refeq{H 7D equal}.  
\subsection{Static compact extra dimensions (generic pressures)}\label{bulk static extra}
We can follow the same procedure as in the equal scale factor limit for the case of static compact extra dimensions $F=0$. While in the previous subsection the two internal and observed spaces were evolving
according to the same dynamics, in this limit the extra dimensions do not evolve and remain static.

The two differential equations of motion for the gravity action in this case become
\bea
\frac{3}{2}\left(\dot{H^2}+4H^3\right)&=&-\oneover{M^{10}}\left(c_VV+c_\rho\rho\right)H\rho-\frac{T^7_7}{2M^5}+15H\l_{RS}-\frac{\k}{b_0^2}H\label{Friedmann 7D static}\\
\dot\rho+3(w+1)H\rho&=&2T_{07}\label{conservation 7D static}
\ena
where $c_V,c_\rho$ are as before \refeq{cV crho}. We introduce the new coefficients $\tcvst ,\tcrst $ defined by
\bea\label{ctilde static}
\tcvst =\frac{c_V}{3w-1},\qquad \tcrst =\frac{c_\rho}{3w+1}
\ena
After plugging (\ref{conservation 7D static}) into the Friedmann equation (\ref{Friedmann 7D static}) we come to the expressions for $H$ and $\chi$ 
\bea\label{H 7D static}
H^2&=&\frac{\tcrst }{3M^{10}}\rho^2+\frac{2\tcvst V}{3M^{10}}\left(\rho+\chi\right)+\frac{5\l_{RS}}{2}-\oneover{6}\frac{\k}{b_0^2}\\
\dot\chi+4H\chi&=&\left(1+\frac{\tcrst }{\tcvst }\frac{\rho}{V}\right)2T^0_7-\frac{M^5}{2\tcvst V}HT^7_7  \label{chi 7D static}\\
\dot\rho+3(w+1)H\rho&=&2T_{07}\label{rho 7D static}
\ena
In analogy to the equal scale factor case, these expressions are valid as long as we don't have $w=1/3$ ($w=-1/3$) with $c_V\ne0$ ($c_\rho\ne0$).
\subsection{Proportional Hubble parameters}\label{proportional hubble}
We are going to combine in the same description the two limits of $a(t)=b(t)$ and $F=0$, impliying, in the first case, an equal cosmological evolution for the internal space and the 3D visible spatial
dimensions and, in the second case, the absence of evolution for the compact space. 

Both the two systems of differential equations (\ref{H 7D equal})--(\ref{rho 7D equal}) and (\ref{H 7D static})--(\ref{rho 7D static}) obtained in the two different limits can be written in a unified
formulation that encloses the two just cited sets of equations, defining some appropriate constant parameters. We introduce an ``effective'' number of dimensions $d$ that takes the values $d=6$ in the equal
scale factor limit and $d=4$ in the static compact extra dimensions. If we look at the equations (\ref{chi 7D equal}) and (\ref{chi 7D static}), we see that $d$ appears as the number of dimensions for which
the energy density $\chi$ satisfies the free radiation conservation equation in the limit of pure RS ($T^7_7=T^0_7=0$). In fact, the Friedmann equation plus the two (non) conservation equations can be
rewritten as
\bea
H^2&=&\frac{\tcrd }{(d-1)M^{10}}\rho^2+\frac{2\tcvd V}{(d-1)M^{10}}\left(\rho+\chi\right)-\oneover{2(d-1)}\frac{\k}{b^2}+\frac{30}{d(d-1)}\l_{RS}  \label{bulk limits H}\\
\dot\chi+d H\chi&=&2T^0_7\left(1+\frac{\tcrd }{\tcvd }\frac{\rho}{V}\right)-\frac{M^5}{2\tcvd V}HT^7_7  \label{bulk limits chi}\\
\dot\rho+w_d H\rho&=&2T_{07}  \label{bulk limits rho}
\ena
We have in addition defined $w_d=3(1+w)+(d-4)(1+w_\p)$ and $\tcvd =c_V/(w_d-d),\tcrd =c_\rho/(w_d-d/2)$ where $c_V$ and $c_\rho$ are given in (\ref{cV crho}). We remind that in order to get an algebraic
equation for $H^2$ as a function of the energy densities $\rho$ and $\chi$ (\ref{bulk limits H}), we have to keep $\tcvd,\tcrd$ finite, i.e.  respectively $w_d\neq d,w_d\neq d/2$ unless $c_V=0,c_\rho=0$. For
example we cannot write $H$ in the form (\ref{bulk limits H}) if $w=1/3,w_\p=0$ in both the equal scale factor and the static compact extra dimension limit, since the linear term in $\rho$ has a diverging
coefficient. 

Moreover, we can futher generalize this analysis introducing a parameter $\x$ such that $F=\xi H$. This description contains all the above studied limits. The analogous of previous relations
(\ref{bulk limits H})--(\ref{bulk limits rho}) can be written as
\bea
H^2&=&\frac{\tcrx }{(3+2\xi)M^{10}}\rho^2+\frac{2\tcvx V}{(3+2\xi)M^{10}}\left(\rho+\chi\right)-\frac{1}{\x^2+3\x+6}\frac{\k}{a^{2\x}}+\frac{5}{\x^2+2\x+2}\l_{RS} \non\\ \label{bulk limits H prop}\\
\dot\chi+d_\xi H\chi&=&2T^0_7\left(1+\frac{\tcrx }{\tcvx }\frac{\rho}{V}\right)-\frac{M^5}{2\tcvx V}HT^7_7  \label{bulk limits chi prop}\\
\dot\rho+w_\x H\rho&=&2T_{07}  \label{bulk limits rho prop}
\ena
where now $d_\xi$ is a more complicated function of the proportionality constant $\xi$ between the two Hubble parameters $d_\xi\equiv6\frac{\xi^2+2\xi+2}{3+2\xi}$ and it reduces to $d=6,d=4$ in the two
previously examined limits of equal scale factors and static compact extra dimensions ($\xi=1,\xi=0$). The constant $w_\x$ reduces to $w_d$ for $\x=0,\x=1$ and is defined by $w_\x\equiv3(1+w)+2\x(1+w_\p)$.
The two coefficients $\tcvx,\tcrx$ are defined as $\tcvd,\tcrd$, with $w_d\to w_\x,d\to d_\x$. The result (\ref{bulk limits H prop}) is valid unless $\x=-3/2$. In that case the equation for $H$ becomes
algebraic --- though we still have the curvature term explicitely depending on the scale factor ---, thus
\bea\label{bulk limits H prop 3}
H^2&=&-\frac{\tcrx }{33M^{10}}\rho^2-\frac{\tcvx V}{33M^{10}}\rho-\frac{\k}{33a^{2\x}}+4\l_{RS}\\
\dot\rho+3(w-w_\p)H\rho&=&2T_{07}\non
\ena
No mirage density appears and the Hubble parameter is a quadratic polynomial in the localized energy density $\rho$ alone. Besides, if the pressures are equal $w_\p=w$ in the pure RS set--up $T^0_7=0$, the
energy density is constant in time and so is $H^2+\frac{\k a^3}{33}$, for $\x=-3/2$ \refeq{bulk limits H prop 3}. The set of equations \refeq{bulk limits H prop 3} also doesn't depend on $T^7_7$ at all. 

Again we have to restrict to $w_\x\ne d_\x,d_\x/2$ to keep $\tcvx,\tcrx$ finite (unless $c_V\propto w_\x-d_\x,c_\rho\propto w_\x-d_\x/2$). 

We remark that for the scale factors satisfying $b(t)=1/a(t)$, i.e. dynamical compactification with $\x=-1$, the equation for the mirage energy density $\chi$ in the pure RS set--up is still an effective 6D
free radiation conservation equation, as for $b(t)=a(t)$. In fact, the only solutions to $d_\x=6$ are $\x=\pm1$. To obtain a 4D free radiation equation for $\chi$ we have to require $\x=0$, since the second
solution to $d_\x=4$ is $\x=-3/2$, for wich we don't define a mirage density \refeq{bulk limits H prop 3}.

However, this is not the end of the story. Introducing the effective 4D densties $\rhofo=V_{(2)}\rho$ and $\chifo=V_{(2)}\chi$ (where $V_{(2)}=v\,b^2(t)=v\,a^{2\x}(t)$ is the volume of the 2D internal space),
we have to replace the l.h.s. of equations \refeq{bulk limits chi prop}--\refeq{bulk limits rho prop} repsectively by $\dot\chifo+(d_\x-2\x)H\chifo$ and $\dot\rhofo+(w_\x-2\x)H\rhofo$ (the r.h.s. are also
modified and we will explicitly write them at the end of subsection \ref{bulk critical points gen}). This tells us that the 4D mirage density is a free radiation energy density for pure RS in four dimensions
for $d_\x-2\x=4$, which has solutions $\x=0,\x=1$ --- i.e. static internal space or equal scale factors, justifying the study of these two limits.
\subsection{Comments}
We here summarize some considerations about the bulk evolution equations derived in the previous subsections and, in particular, about the explicit expressions we have found for the Hubble parameters in the
discussed limits.
\begin{itemize}
\item[-]
With the assumption of having the same pressure for the matter fluid in the two dimensional internal space and in the 3D visible space (i.e. $\p=p$), we found a form of the Friedmann equation that has the
advantage of keeping both the Hubble parameters not constrained by any particular ansatz. The Friedmann equation (\ref{Friedmann 7D equal pressures}) provides an expression for $(3H+2F)$ in terms of $\rho$
and $\chi$ (for spatially flat spaces). This solution, though, is satisfactory only in the limit of small $(H-F)$. When $(H-F)$ is not negligeable w.r.t. $(3H+2F)$, the mirage density equation may be written
introducing an extra term independent of the bulk parameters $T^0_7,T^7_7$. This prevents $\chi$ to obey to a free radiation equation in the pure RS set--up ($T^0_7=T^7_7=0$), as it should instead be in the
context of the AdS/CFT correspondence (we will discuss the comparison between the bulk and the dual brane analysis in section \ref{examples}).
\item[-]
In the simple limits of equal scale factors (\ref{H 7D equal}) and static compact extra dimensions (\ref{H 7D static}) we recovered an expression for $H^2$ containing a quadratic term in $\rho$ and a linear
term in $(\rho+\chi)$, where $\rho$ is the localized energy density and $\chi$ is an artificially introduced mirage density that accounts for the bulk dynamics. In fact it depends on the bulk parameters
$T^0_7,T^7_7$. When $T^0_7=T^7_7=0$, the mirage density obeys to 4D free radiation equation for the static compact extra dimension case and to 6D free radiation equation for the equal scale factor case. This
is in complete analogy to the 5D analysis \cite{Kiritsis:2002zf}, where the same dependence on $\rho$ and $\chi$ occurs and the mirage energy satisfies 4D free radiation for pure RS (i.e. $T^0_5=T^5_5=0$).
\item[-]
When $w_\p=w$ (or equivalently $\p=p$) in the equal scale factor limit (\ref{H 7D equal}), we find the results given by the equal pressures subsection in the case $F=H$ (\ref{H 7D equal equal pressures}). The
two limits of equal pressures and equal scale factors then commute and the results are consistent.
\item[-]
The description of section \ref{proportional hubble} encloses in a unifying way the results in the limits of static compact extra dimensions and equal scale factors. It moreover generalizes these results to
the case of evolutions governed by proportional Hubble parameters $F(t)=\x H(t)$. We will use the set of equations written in terms of the effective number of dimensions $d$ (that describes the two limits of
static internal space, with $d=4$, and equal scaling for the compactification space and the 3D space, with $d=6$) to study the corresponding cosmological evolution in the next section.  
\end{itemize}

We are now going to proceed to the analysis of the critical points for this seven dimensional universe in a 7D Randall--Sundrum set--up, including the energy exchange term. 
\section{Bulk critical point analysis with energy exchange}\label{critical bulk}
We have until now transformed the second order differential equation (\ref{Friedmann 7D}) plus the (non) conservation equation (\ref{conservation 7D}) in a set of three linear differential equations
(\ref{bulk limits H})--(\ref{bulk limits rho}) for combined equal scale factor and static compact extra dimension limits, or more generally (\ref{bulk limits H prop})--(\ref{bulk limits rho prop}) for
proportional Hubble parameters. We have introduced the mirage density $\chi$ defined by its differential equation. In this section we will use the first system of equations (\ref{bulk limits H})--(\ref{bulk
limits rho}), obtained to describe both the limit of equal scale factors and static compact extra dimensions, to find its fixed points and the corresponding stability. The critical point analysis will allow
us to study the cosmological evolution in the bulk description for $F=0$ or $F=H$. 

We make an assumption on the bulk components of the stress--energy tensor that appears in the differential equations for the energy densities $\rho$ and $\chi$. As in \cite{Kiritsis:2002zf}, we will take
the diagonal elements of the stress energy tensor to satisfy the relation
\bea\label{small diagonal components}
\bigg|\frac{T^{(\mathrm{diag})}_{m,B}}{T^{(\mathrm{diag})}_{v,B}}\bigg|\ll\bigg|\frac{T^{(\mathrm{diag})}_{m,b}}{T^{(\mathrm{diag})}_{v,b}}\bigg|
\ena 
This enforce the solution to the Friedmann equation to be reasonably independent of the bulk dynamics, since the $T^7_7$ term in (\ref{Friedmann 7D}) becomes negligeable with respect to the first term on the
r.h.s. of the same equation. Imposing such a relation, $T^7_7$ disappears from the sets of linear differential equations, while we remain left with the $T^0_7$ component. For the future bulk calculations we
will define $T\equiv2T^0_7$ to simplify the notation.

Before starting the critical point analysis we note that when $T=0$ the system of equations \refeq{bulk limits H}--\refeq{bulk limits rho} have only trivial critical points characterized by zero visible
Hubble parameter when the internal space is flat. There are two of these critical points. One is given by $\sH^2=-\k/2(d-1)b^2$, $\srho=\schi=0$ (which is valid only if we are compactifying on hyperbolic or
flat spaces) and the other is $\sH=0$, $\tcrd \srho^2+2\tcvd V\left(\srho+\schi\right)=M^{10}\k/b$. 

We will first restrict to small density
~\footnote{We are referring to the energy density localized on the 5--brane. The effective 4D density is $\rhofo=V_{(2)}\rho$, where $V_{(2)}$ is the volume of the internal compactification pace. Similar
relations are established for the mirage density and the pressures. We note that the volume of the 2D compact space varies in time, unless extra dimensions are static, since it is proportional to
$b^2(t)$: it contracts as the 4D visible space expands in the dynamical compactification approach \cite{Mohammedi:2002tv,Andrew:2006yt}}
 $\rho\ll V$ and flat internal space $\k=0$ (remind that the 3D space is already supposed to be flat, having put $k=0$) and then go through the generic density
analysis. Eventual de Sitter stable solutions (for the 4D visible space--time) could represent the present accelerated era, while inflationary phases at early times may be associated to primordial inflation.
\subsection{Small energy density and flat compact extra dimensions}
When the localized energy density is small and the internal space curvature vanishes, $\rho\ll V,\k=0$, the bulk Einstein equations (\ref{bulk limits H})--(\ref{bulk limits rho}) in terms of $H$, $\chi$ and
$\rho$ become
\bea
H^2&=&\frac{2\tcvd V}{(d-1)M^{10}}\left(\rho+\chi\right)  \label{bulk limits H small}\\
\dot\chi+d H\chi&=&T\\
\dot\rho+w_d H\rho&=&-T  \label{bulk limits rho small}
\ena
We note that in this approximation expression (\ref{bulk limits H small}) isn't valid for $w_d=d$, unless $c_V\propto(w_d-d)$ (this would for example determine a specific value for $w_\p$ as a function of
$w$). Nevertheless we will find in the rest of the section that $\tcvd$ always appears with the coefficient $(w_d-d)$ in the critical point analysis. We thus have to keep in mind that
$(w_d-d)\tcvd=c_V=(31w-6w_\p-5)/400$ always remains finite. 
\paragraph*{Fixed point solutions}
In the small density approximation, the fixed points in terms of the critical energy density can immediately be found (to have a full solution we have to make an ansatz on the form of the energy exchange
parameter). 

The first solution is given by
\bea\label{bulk fix small density}
\sH&=&-\frac{\sB}{w_d}\srho^{1/2}  \non\\
\schi&=&-\frac{w_d}{d}\srho\\
\sT&=&\sB\srho^{3/2}  \non
\ena
where $\sB\equiv-w_d\left(\frac{(d-w_d)2\tcvd V}{d(d-1)M^{10}}\right)^{1/2}$. This represents an inflationary critical point for the cosmological evolution, for $(w_d-d)\tcvd=c_V>0$, i.e. both $\tcvd>0$
and $d>w_d$ or $\tcvd<0$ and $d<w_d$
~\footnote{As examples of negative $\tcvd$ and $(d-w_d)$, both negativeness conditions can be satisfied if $w=1/3$ for $w_\p>8/9$ when $H=F$, but never when $F=0$. If $w=0$ we must have $w_\p>1/2$ in the
equal scale factors limit, while no solution can be found with static compact extra dimensions. We instead get positive $\tcv$ and $(d-w_d)$ if $w=0$ for $-5/6<w_\p<1/2$ with $H=F$ and for $w_\p>1/2$ with
$F=0$. No value of $w_\p$ satisfies the positiveness conditions if $w=1/3$.}
. The acceleration factor at the fixed point is simply given by $q_\star=\sH^2$. Since we assume $w_d$ to be positive, $\sB$ is negative and $\sH$ in \refeq{bulk fix small density} is positive. At this fixed
point the universe is thus expanding.

Another fixed point leaves $\schi$ unchanged, while $\sT$ and $\sH$ have switched signs with respect to \refeq{bulk fix small density}, meaning that $\sH$ is negative and the universe in contracting (there is
a symmetry $T\to-T,H\to-H$)
\bea\label{bulk fix small density 2}
\sH&=&\frac{\sB}{w_d}\srho^{1/2}  \non\\
\schi&=&-\frac{w_d}{d}\srho\\
\sT&=&-\sB\srho^{3/2}  \non
\ena

The trivial critical point is characterized by mirage density equal and opposite to $\srho$, but zero Hubble parameter and energy exchange (in the case we admit for the energy exchange the form $T=A\rho^\n$
all the variables are zero at the trivial fixed point).

For positive critical energy densities $\srho$, we obtain a negative brane--bulk energy exchange parameter at the critical point if $\sH>0$ and, viceversa, we have positive critical energy exchange for a
contracting universe at the critical point. During the evolution, we can expect a change of regime going from negative to positive $T$ as the energy density localized on the brane grows, as for the 5D RS
critical point analysis with energy exchange carried in \cite{Kiritsis:2002zf}. Even though most of the analysis in this section will be performed supposing that the energy exchange parameter has fixed sign
determined by the sign of $A$ (since we will mainly assume $T=A\rho^\n$), we can argue that for small energy density $\rho$ the generic energy exchange is presumably negative, meaning that energy would be
transferred from the bulk onto the brane. In this hypothesis, an equilibrium can be reached, such that the energy density would have a large limiting value for which energy starts to flow back into the bulk
(with positive energy exchange).
\paragraph*{Stability analysis}
The real parts of the eigenvalues of the stability matrix for the $(\d\chi,\d\rho)$ linear perturbations corresponding to the critical point (\ref{bulk fix small density}) can have opposite signs or be
both negative. This depends on the value of $T$ as a function of the energy density $\rho$ at the fixed point \refeq{bulk fix small density} describing an expanding universe with energy influx. The explicit
form of the eigenvalues is
\bea
\l_\pm=\frac{\sB}{2w_d}\srho^{1/2}\left[d+(1-\tn)w_d\pm\sqrt{(d+(1-\tn)w_d)^2-2(3-2\tn)dw_d}\right]  \label{eigenval small}
\ena
where we have defined
\bea\label{def nu tilde}
\tn\equiv\frac{\pa\log|T|}{\pa\log\rho}\bigg|_\star
\ena
Expression \refeq{eigenval small} then shows that the two eigenvalues have negative real part when $\tn<3/2$. There is a second upper bound on $\tn$ derived from requiring negative real part for the
eigenvalues. Nonetheless, for the range of values $-1\le w,w_\p\le1$ and $d=4,6$ in which we're interested, this bound is always equal or greater than $3/2$. If $|T|$ is a decreasing or constant function of
$\rho$ near $\sT$, the non trivial inflationary critical point always is an attractor. Also for growing $|T|$ we can have stable inflationary fixed points, as long as $\tn<3/2$. In particular, the linear
$\tn=1$ case is included in the stable inflationary fixed point window and will be analyzed both solving the Einstein equations numerically, in the next subsection, and deriving an explicit solution, in
subsection \ref{explicit bulk section}. 

Besides, when 
\bea\label{spiral range small}
1-\sqrt{\frac{2d}{w_d}}-\frac{d}{w_d}<\tn<1+\sqrt{\frac{2d}{w_d}}-\frac{d}{w_d}
\ena
the eigenvalues have non zero immaginary part and the critical point will hence be a stable spiral for $\tn<3/2$. For values of $\tn$ out of the range \refeq{spiral range small}, we get a node. As an example,
let's assume the value $\tn=1$ in the equal scale factor background. This gives a stable spiral for $w_\p>1/2$ or $w_\p<1/2$ and $w>-2(1+w_\p)$ (considering $w,w_\p>-1$). This means that in the case
$w\simeq1/3$ and $w_\p=0$ the critical point is a stable spiral, while for both $w$ and $w_\p$ null we instead have a stable node.

For energy outflow $\sT>0$ (which goes along with contraction $\sH<0$), we get a minus sign overall modifying the eigenvalues \refeq{eigenval small} referring to the linearized system around the \refeq{bulk
fix small density 2} critical point (characterized indeed by energy outflow). The eigenvalues cannot be both negative in this case. In fact we should demand $w_d>d/(\tn-1)$ with $\tn>1$ but also $\tn<3/2$ to
get a stable fixed point. Only the trivial point, as we will discuss later, can be attractive for energy outflow dynamics.
\paragraph*{Assumption $T=A\rho^\n$ and numerical solutions}
Assuming the brane--bulk energy exchange parameter to take the form $T=A\rho^\n$ (so that $\tn=\n$ referring to \refeq{def nu tilde}), we can rewrite the system of differential equations in term of
dimensionless quantities $\check\rho=\g^6\rho$, $\check\chi=\g^6\s$, $\check H=\g H$, $\check T=\g^7 T$, where we called $\g^4\equiv\frac{2V}{(d-1)M^{10}}$. The dimensionless variable $\rho/V$ used to perform
the small energy density expansion at the beginnning of this section is related to the dimensionless variable $\crho$ by $\crho=\left(2\tcvd/(d-1)\right)^{3/2} \left(V/M^6\right)^{5/2}(\rho/V)$. So,
considering the small $\rho/V$ approximation is equivalent to considering small $\crho$ approximation if the brane tension $V$ satisfies $V\lesssim M^6$ with respect to the 7D Planck mass and $\tcvd$ is
reasonably of the order $\tcvd\lesssim1$. The complete set of fixed point solutions (discarding the trivial ones) can be calculated in terms of the parameters $A,\n$ characterizing the energy exchange,
$w_d,d$ denoting the background (static extra dimensions or equal scale factors) and the equations of state for both the 3D and internal spaces. 

The Einstein equations become
\bea\label{dimensionless ODEs small}
\cHH^2&=&\tcvd\left(\crho+\cchi\right)  \non\\
\dot\cchi+d\cHH\cchi&=&\cAA\crho^\n\\
\dot\crho+w_d\cHH\crho&=&-\cAA\crho^\n  \non
\ena
where $\cAA=\g^{1+6(1-\n)}A$. 

\begin{figure}[hb]
\begin{tabular}{cc}
\includegraphics[width=0.5\textwidth]{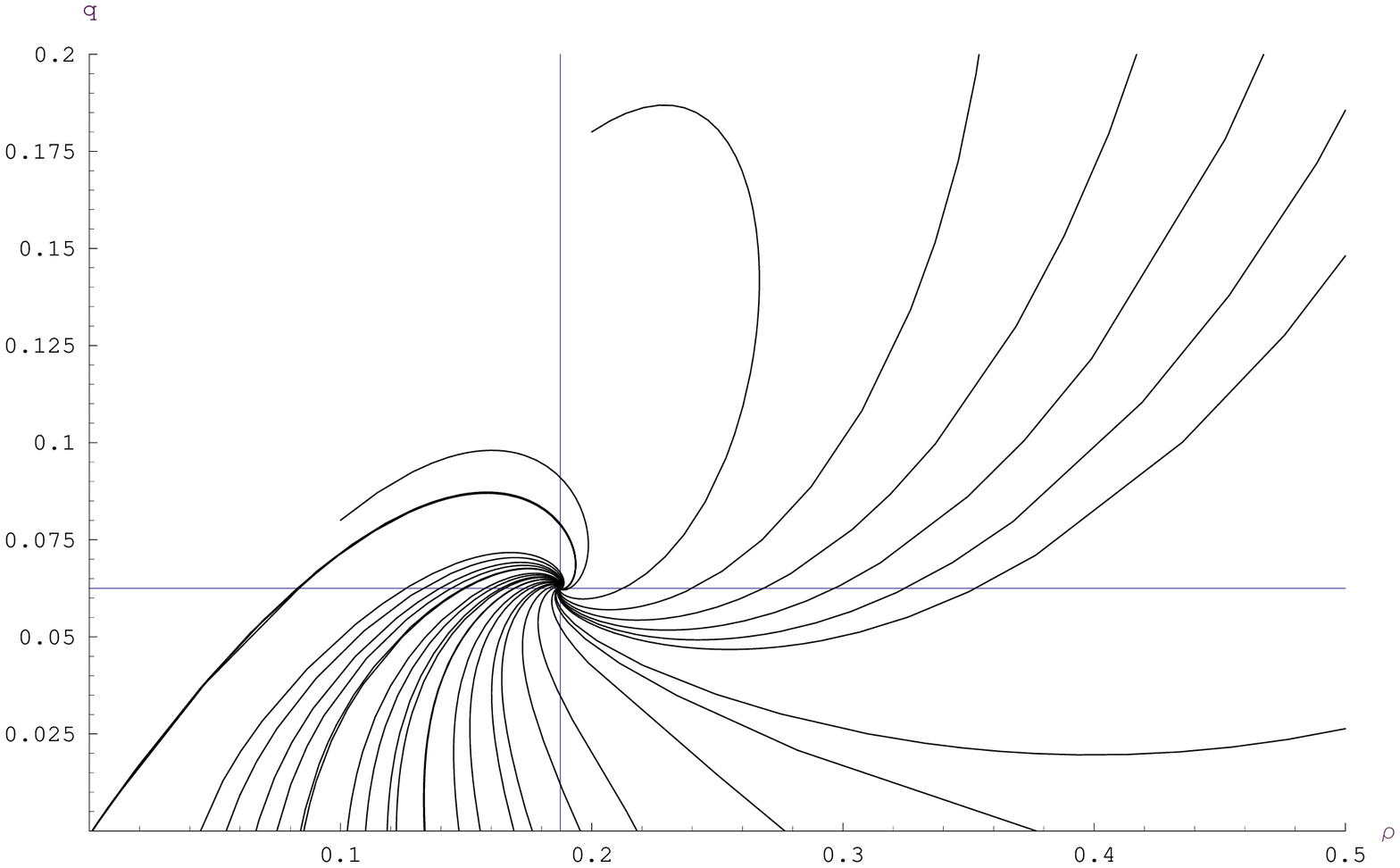}&
\includegraphics[width=0.5\textwidth]{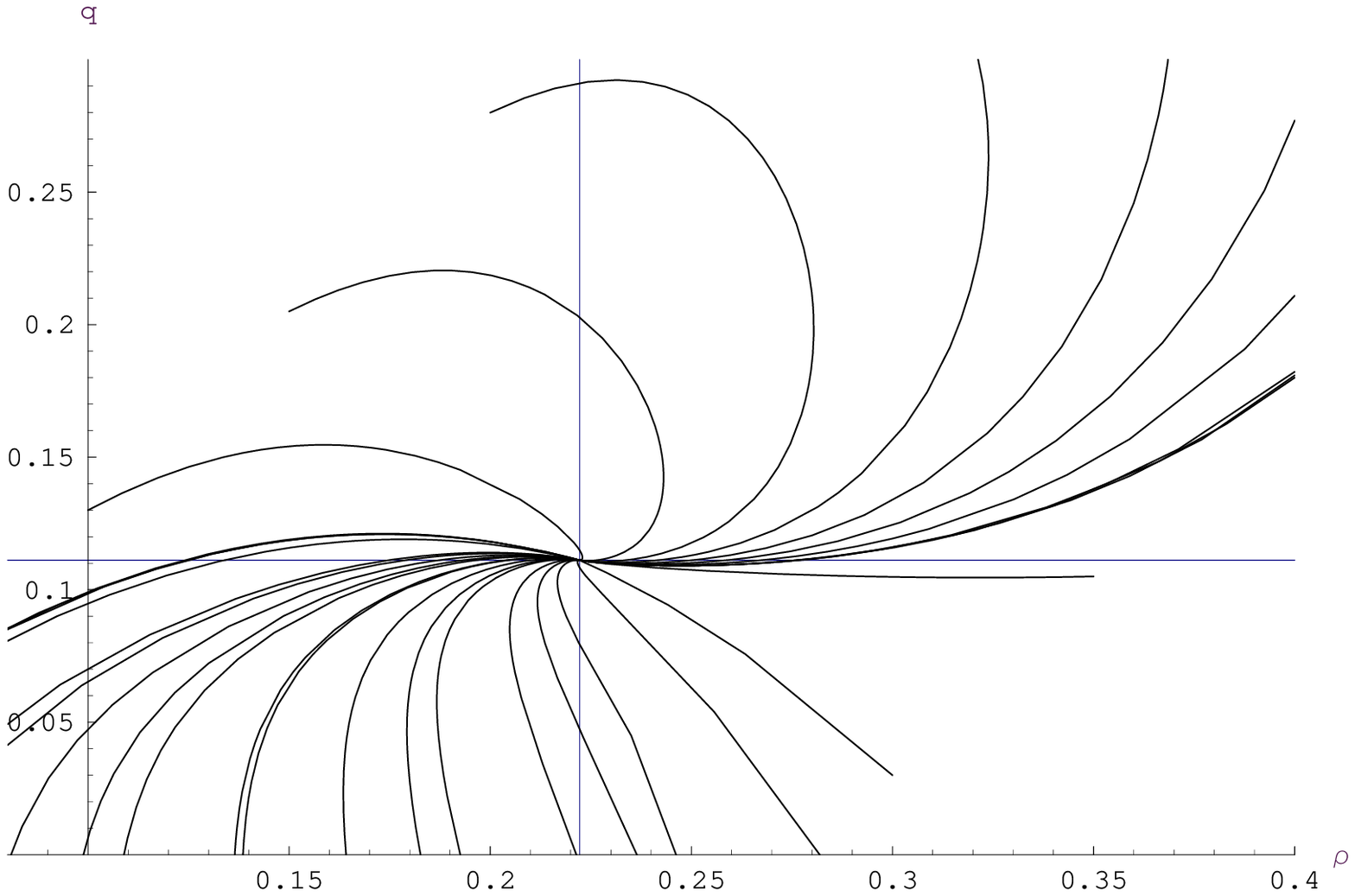}\\
(a)&(b)
\end{tabular}
\caption{\label{phase space small}
Phase spaces $\cqq/\crho$ for different initial conditions $\crho_0$ and $\cqq_0$, with $\cAA=-1$, $\n=1$, $d=6$ (equal scale factors --- analogous pictures come from the static extra dimension
case) and: (a) $w_d=4$ (for istance $w=0,w_\p=-1/2$ or $w=-1/5,w_\p=0$) leading to a spiral like stable critical point, (b) $w_d=3$ (for istance $w=0,w_\p=1/2$) determining a stable node. The extra blue
grid lines intersection represents the fixed point.}
\end{figure}

The acceleration $\cqq$ can be evaluated independently of $\n$ and $\cAA$
\bea
\cqq=\left(1-\frac{w_d}{2}\right)\tcvd\,\crho+\left(1-\frac{d}{2}\right)\tcvd\,\cchi
\ena
as a function of the localized matter density and of the mirage density. Due to the positiveness constraint on $\tcvd\left(\crho+\cchi\right)$ coming from the first equation in \refeq{dimensionless ODEs
small}, the trajectories in the phase space must satisfy
\bea
\cqq\le (d-w_d)\tcvd\,\rho
\ena
as it is indeed showed in the numerical plots of figure \ref{phase space small}. For $w_d>2$ and $\tcvd>0$ we have positive acceleration only if the mirage density $\cchi$ is negative and smaller than
$-\crho(2-w_d)/(2-d)$.  On the other hand, $\cchi$ gets positive (suppose $\tcvd>0$) only if $\cqq<(2-w_d)\crho/2$.

The fixed points are given by
\bea
\check\sH^{3-2\n}&=&(-)^{3-2\n}\left(\frac{\tcvd(d-w_d)}{d}\right)^{1-\n}\frac{\cAA}{w_d}\\
\check\schi^{3-2\n}&=&(-)^{3-2\n}\frac{d^{2(\n-1)}\cAA^2}{\tcvd(d-w_d)w_d^{2\n-1}}\\
\check\srho^{3-2\n}&=&\frac{d\cAA^2}{w_d^2\tcvd(d-w_d)}\label{num fixed rho small}
\ena 
For $\n<3/2$, when the non trivial fixed point is stable, we have two roots of (\ref{num fixed rho small}) with opposite signs if $\n=1/2+m,m\in\mathbb Z$. Only one real root exists for integer $\n$ and it
carries the sign of the r.h.s. in \refeq{num fixed rho small}. Finally, for $\n=(2m+1)/4$, we have a positive root if the r.h.s. in \refeq{num fixed rho small} is negative. The two eigenvalues corresponding
to a negative $\csrho$ always have opposite real part, implying that this fixed point isn't be stable at linear order, it is a saddle. We further note that for $w_d>d$  and positive $\tcvd$ (or alternatively
$d>w_d$ and negative $\tcvd$) we can only have real and positive $\srho$ fixed point if $\n=(2m+1)/4,m\in\mathbb Z$. This implies that for $\n=1$ there is no non trivial fixed point with positive $\srho$,
when $(d-w_d)$ and $\tcvd$ have opposite signs. Moreover, these negative $\srho$ points are always characterized by negative $\sH$, so that they wouldn't be inflationary.

We can check the stability of the critical points by means of a numerical analysis of the differential system of equations (\ref{dimensionless ODEs small}). In the case of energy influx $\cAA<0$, putting
$\n=1$ and different values for the $d,w_d$ parameters, we get the phase spaces in figure \ref{phase space small}, plotting the acceleration factor $\check q(t)\equiv\ddot{\check{a}}(t)/\check a(t)$ as a
function of the energy density $\crho(t)$. We thus check that, solving the system of differential equations for variable initial conditions for $\cchi$ and $\crho$, including both positive and negative
initial $\cqq$, all the different trajectories converge to the non trivial fixed point, designated by the intersection of the two perpendicular lines in the picture. Besides, as we expect, in part \ref{phase space small}(a) they
have a spiral behavior, while in the \ref{phase space small}(b) case they denote a node.

In the limit $w_d\to d$ we numerically recover the analytical solution discussed in subsection \ref{explicit bulk section}, neglecting the very large density behavior.

For $\n>3/2$, i.e. when the non trivial fixed point is no more an attractor, we find that some of the trajectories go to the trivial critical point, while another branch of solutions to the Eintein equations
\refeq{dimensionless ODEs small} are characterized by diverging energy density $\crho$ (they become unreliable when $\crho^2\gtrsim(2\tcvd/(d-1))^{3}(V/M^6)^5$). This happens because, as it is suggested by
the integration of the third equation in \refeq{dimensionless ODEs small} with energy influx hypothesis, for $\crho$ big enough --- precisely for
$\crho^{\n-\frac{3}{2}}\gtrsim\left(w_d/A\right)\tcvd^{\frac{1}{2}}\left(1+\cchi/\crho\right)^{1/2}$ --- the function $\crho(t)$ starts growing, while for small $\crho$, $\crho(t)$ eventually goes to zero.
Depending on the initial conditions we will have solutions ending in the trivial fixed point or diverging. 

The behavior of the system with energy outflow can be analytically deduced. Given the hypothesis $T>0$ it is clear that the trajectories can't be attracted by the fixed point solution
$\sT=\sB\srho^{3/2}$. They may go to the critical point characterized by negative Hubble parameter and $\sT=-\sB\srho^{3/2}$. However, we already determined the non attractive nature of this fixed point.
Another way to describe outflow dynamics is to conclude from the form of the Einstein equations \refeq{bulk limits rho small} that all the trajectories in the phase space $\cqq/\crho$ go toward the trivial
point, since for positive $T$ the density $\rho$ is suppressed at late time. The way in which the trajectories go to the critical point depends as before on the positiveness of the function under the square
root in (\ref{eigenval small}). We numerically checked that for $\n=1$ and $d=6,w_d=4$ the null fixed point is an attractor and, in particular, a stable node. 
\subsection{Critical points with general energy density}\label{bulk critical points gen}
Allowing $\k$ to be different from zero, not restricting the localized energy density to be small and with the assumptions \refeq{small diagonal components} on the bulk matter stress--energy tensor, we have
to solve the following set of equations
\bea
H^2&=&\frac{\tcrd }{(d-1)M^{10}}\rho^2+\frac{2\tcvd V}{(d-1)M^{10}}\left(\rho+\chi\right)-\oneover{2(d-1)}\frac{\k}{b^2}  \label{bulk limits H gen}\\
\dot\chi+d H\chi&=&T\left(1+\frac{\tcrd }{\tcvd }\frac{\rho}{V}\right)  \label{bulk limits chi gen}\\
\dot\rho+w_d H\rho&=&-T  \label{bulk limits rho gen}
\ena 

As in the small energy density regime, we can immediately notice that the analytical behavior for energy outflow is characterized by decreasing $\rho$ in time. The trajectories will thus be attracted to the
trivial fixed point characterized by vanishing $\srho$. 

We will now carry a general critical point analysis. Equation \refeq{bulk limits H gen} exibits divergences if $w_d=d$ or $w_d=d/2$, unless $c_V\propto(w_d-d)$ or $c_\rho\propto(w_d-d/2)$. When the divergence
arises, the right system of equations is \refeq{Friedmann 7D ansatz}. Terms like $(w_d-d)\tcvd=c_V$ and $(w_d-d/2)\tcrd =c_\rho$ are always finite (where we recall $c_V=31w-6w_\p-5$ and
$c_\rho=11w+14w_\p+10-(w-w\p)(3w-7w_\p)$). 
\paragraph*{Fixed point solutions}
If we demand $H$ to be positive, i.e. expanding universe, the solution for $\sH$ and $\sT$ is given by
\bea  \label{fixed line gen}
\sH&=&-\frac{\sB}{w_d}\srho^{1/2} \non\\
\schi&=&-\frac{w_d}{d}\left(1+\frac{\tcrd \srho}{\tcvd V}\right)\srho\\
\sT&=&\sB\srho^{3/2}\non
\ena
where $\sB=\sB\left(\srho\right)$ depends also on $\k$ and $\srho$ and is defined by
\bea\label{B star general}
\sB\left(\srho\right)=-w_d\left[\frac{(d-2w)\tcrd \srho+2(d-w)\tcvd V}{d(d-1)M^{10}}-\frac{\k}{2(d-1)b^2_\star\srho}\right]^{1/2}
\ena
We have a negative energy exchange parameter, as in the small density limit, and a positive Hubble parameter. 

The second fixed point solution is equal to the first except for the $\sH$ and $\sT$ signs reversed (keep in mind the $H\to-H$, $T\to-T$ symmetry), such that $\sH$ would be negative and we would have energy
outflow at the critical point \bea  \label{fixed line gen}
\sH&=&\frac{\sB}{w_d}\srho^{1/2} \non\\
\schi&=&-\frac{w_d}{d}\left(1+\frac{\tcrd \srho}{\tcvd V}\right)\srho\\
\sT&=&\sB\srho^{3/2}\non
\ena

The trivial critical point is characterized by vanishing $\sH$ and $\sT$, while the mirage density becomes 
\bea
\schi=\frac{M^{10}\k}{4\sbb^2\tcvd V}-\left(1+\frac{\tcrd \srho}{2\tcvd V}\right)\srho
\ena
If the energy exchange is supposed to be of the form $T=A\rho^\n$, the trivial fixed point is characterized by zero value for all the variables except for the mirage density $\chi$ which becomes
$\schi=M^{10}\k/4\sbb^2\tcvd V$ and is zero for flat compact spaces.

As in the limit of small energy density considered in the previous section, the constant $\sB$ is negative whenever the argument of the square root in (\ref{B star general}) is positive, i.e. when
\bea
\frac{\k}{2(d-1)\sbb^2}<\frac{(d-2w_d)\tcrd \srho+2(d-w)\tcvd V}{d(d-1)M^{10}}\srho
\ena
If the square root gives an immaginary number, we don't have any real valued fixed point except for the trivial one.
\paragraph*{Stability analysis} 
The positiveness of the eigenvalues of the stability matrix depends now on all the parameters and constants of the theory and not only on $\tn$, as for the small density, flat compact extra dimension simple
case.  We consider the situation in which the variation of $\k/b^2$ vanishes (this happens in the static compact extra dimension limit or for $\k=0$ in the equal scale factor background, as the ratio $\k/b^2$
remains constant). Otherwise we would have a linearized system of two differential equations plus one algebraic equation in the four variables $\d\k,\d\rho,\d\chi,\d H$. 

There are two conditions that must be satisfied, in order to get two negative eigenvalues and hence a stable fixed point. These conditions give two upper bounds for $\tn$ in terms of the constants $w_d,d$ and
$V,\srho,\k/b_\star^2,M^{10}$
\bea\label{bounds gen}
(\tn-1)&<&\frac{w_d}{d}\frac{w_d}{(d-1)\sB^2M^{10}}\left[(d-2w_d)\tcrd \srho+(d-w_d)\tcvd V\right]\\
(\tn-1)&<&\frac{d}{w_d}\non
\ena
The second bound in (\ref{bounds gen}) satisfies $d/w_d+1>3/2$ in the range $-1\le w,w_\p\le1$. Besides, the first bound reduces to $\n<3/2$ when we take the limit $\rho/V\ll1$ and put $\k=0$. So
the results are in agreement with the previous small density analysis. 

The bounds \refeq{bounds gen} depend on the fixed point value of $\rho$, which can't be determined without making any assumption on the form of $T$. However, we can make some remarks on the nature of the
fixed points. For values of $\tn$ in the range
\bea
1-\frac{d}{w_d}-R_\star<\tn<1-\frac{d}{w_d}+R_\star
\ena
where we defined
\bea
R_\star&\equiv&2\sqrt{\frac{w_d}{(d-1)\sB^2M^{10}}\left[(d-2w_d)\tcrd \srho+(d-w_d)\tcvd V\right]}\non
\ena
the stability matrix eigenvalues have non null immaginary part and the trajectories near to the critical point have a spiral like behavior. When $\sR^2<0$ we always have node like fixed points. In agreement
with the small density case (\ref{spiral range small}), when $\rho/V\ll1$ and $\k=0$ we get $\sR\to\sqrt{2d/w_d}$.

As an example, we assign the value $\tn=1$. Since the first bound (\ref{bounds gen}) can be rewritten as
\bea\label{spiral bound gen nu}
(\tn-1)<\frac{w_d}{d}\frac{\sR^2}{4}
\ena
this means that we should have $\sR^2>0$ to get stability. We also find that the fixed points have spiral shape when $\sR>d/w_d$ or $\sR<-d/w_d$, they will be nodes otherwise.
\paragraph*{Assumption $T=A\rho^\n$ and numerical solutions}
To do a more quantitative analysis we have to make an ansatz on the form of the energy exchange parameter $T$. As in the previous section, we suppose a power dependence on the energy density $\rho$ such that
$T=A\rho^\n$. The equations for generic energy densities and internal space curvature can be rewritten introducing dimensionless variables as in (\ref{dimensionless ODEs small})
\bea\label{dif sys gen num}
\cHH^2&=&\tcrd\,\a\crho^2+\tcvd\left(\crho+\cchi\right)-\ck   \non\\
\dot\cchi+d\cHH\cchi&=&\cAA\rho^\n\left(1+2\frac{\tcrd }{\tcvd}\a\crho\right)\\
\dot\crho+w_d\cHH\crho&=&-\cAA\crho^\n   \non
\ena
where $\a$ is a dimensionless constant defined by $\a^2\equiv\frac{(d-1)^3}{64}\left(\frac{M^6}{V}\right)^5$, $\ck$ is the dimensionless variable $\ck=\frac{\g^2\k}{2(d-1)b^2}$ --- we remind that we restrict
to constant $\ck$ approximation.

\begin{figure}[h!]
\begin{tabular}{cc}
\includegraphics[width=0.5\textwidth]{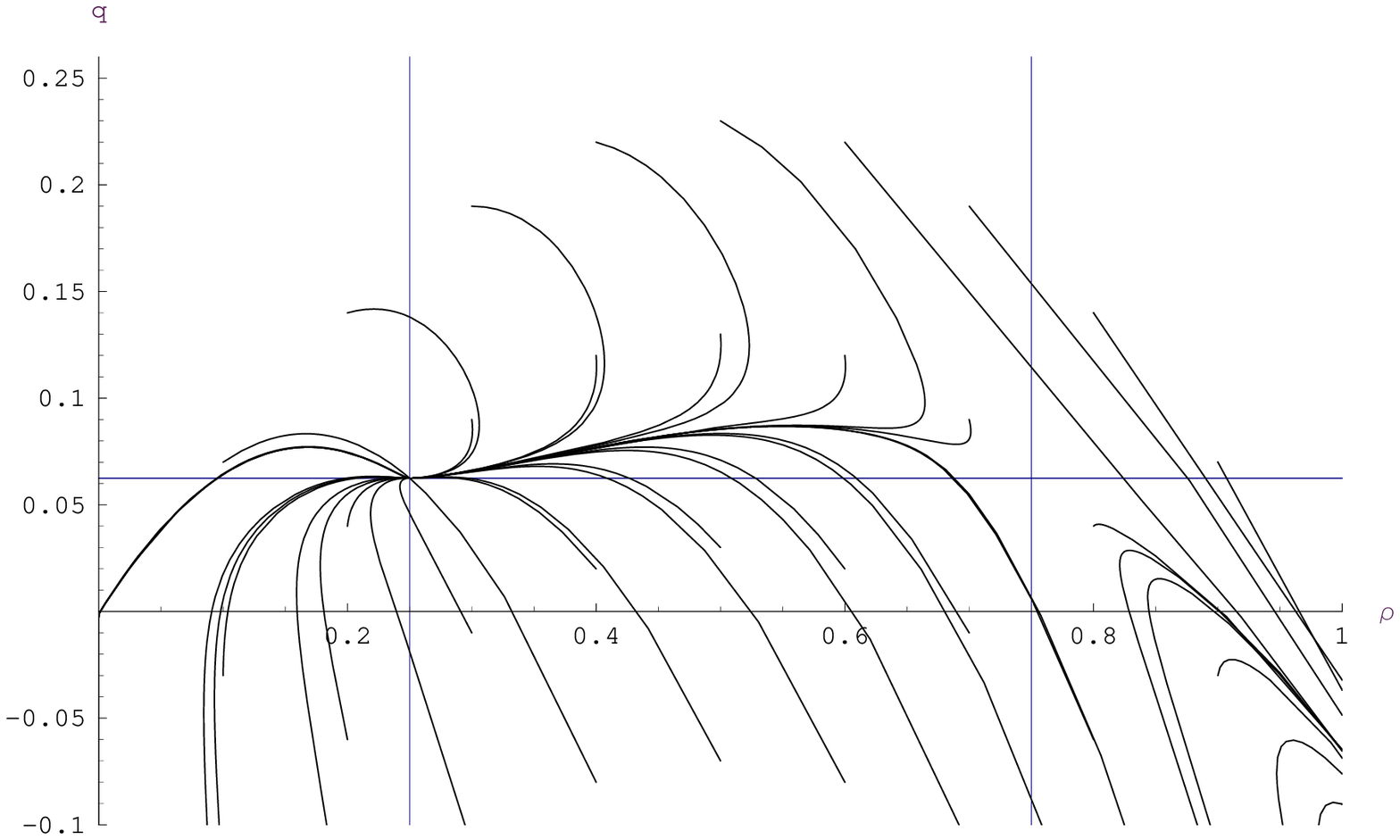}&
\includegraphics[width=0.5\textwidth]{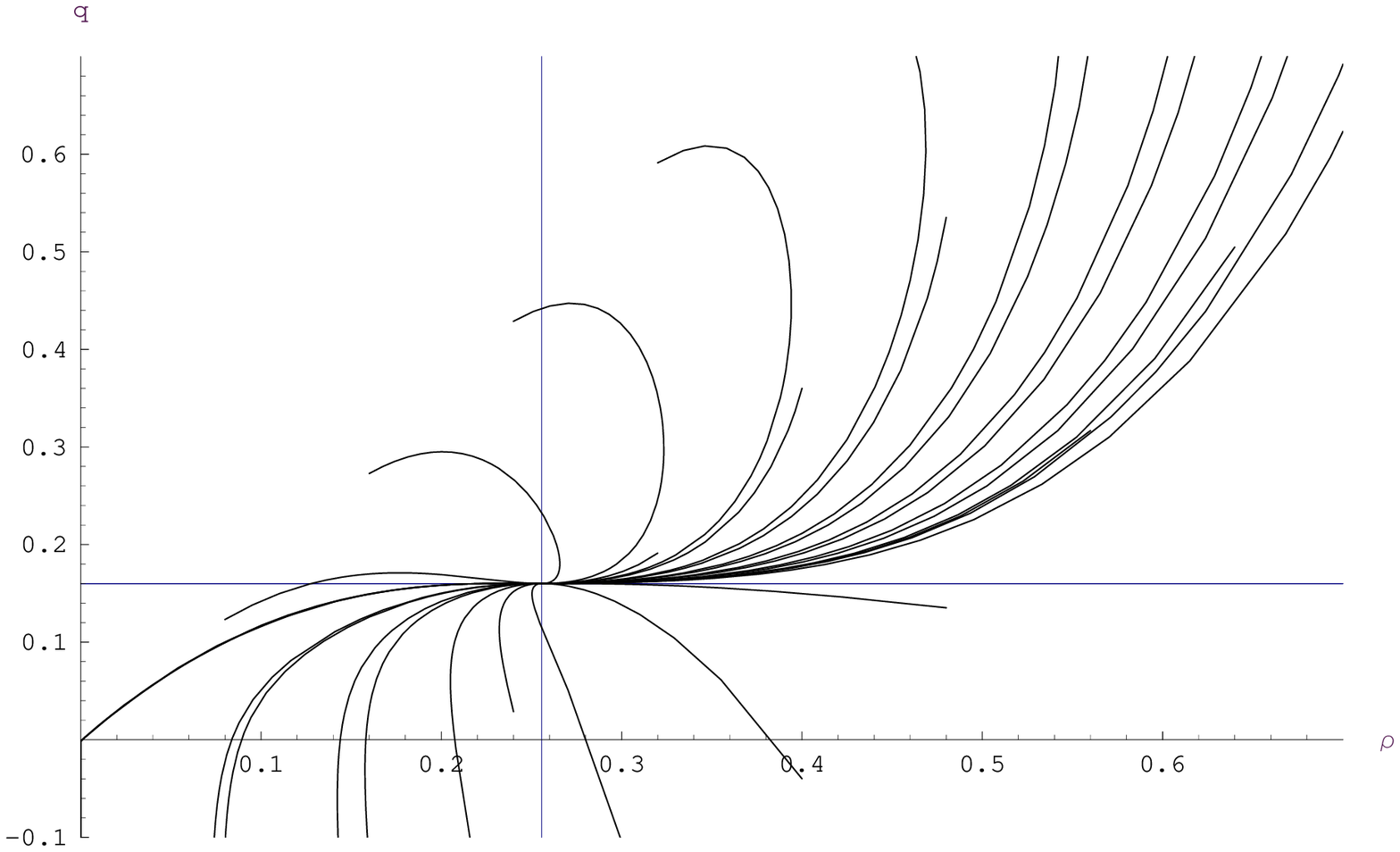}\\
(a)&(b)\\
\includegraphics[width=0.5\textwidth]{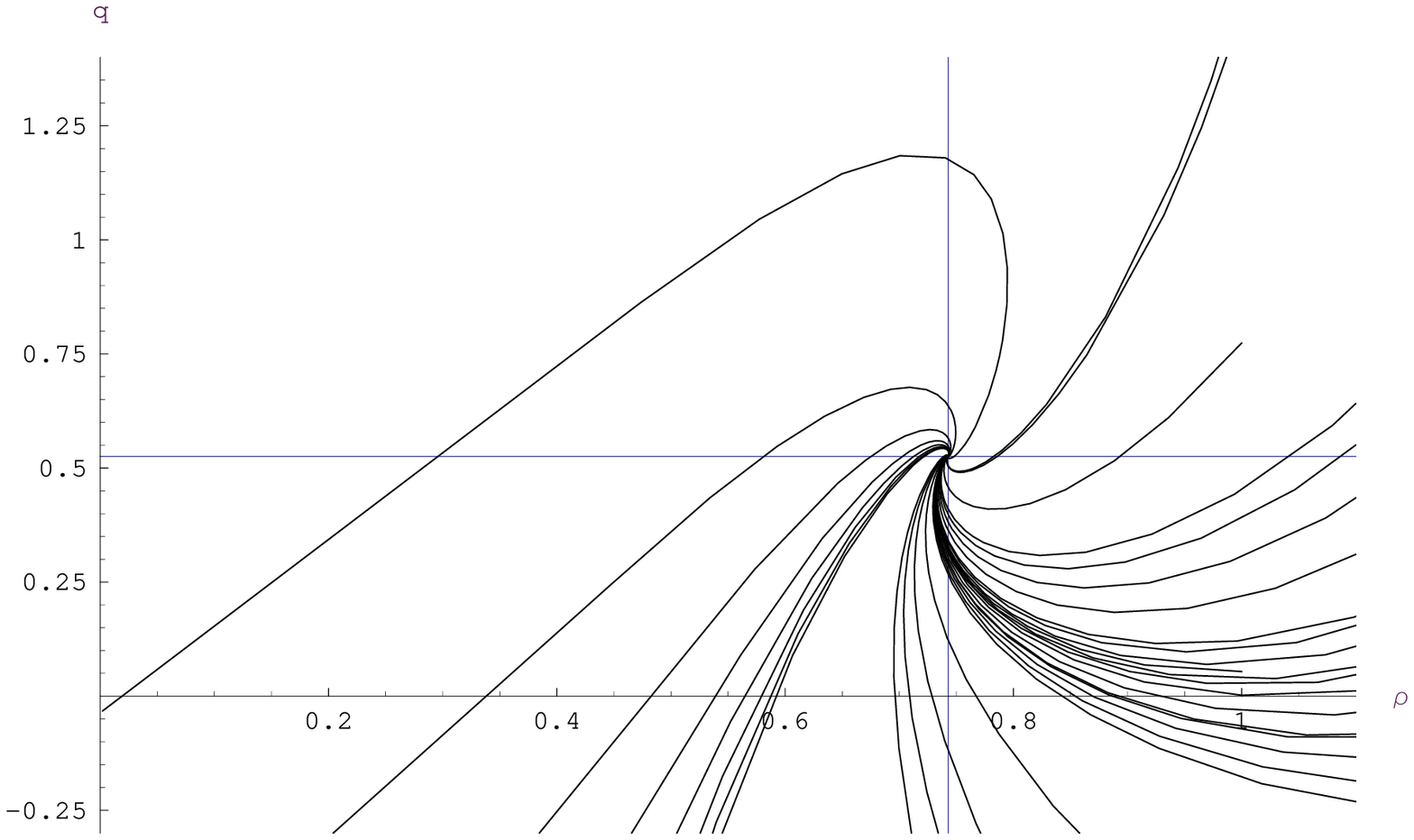}&
\includegraphics[width=0.5\textwidth]{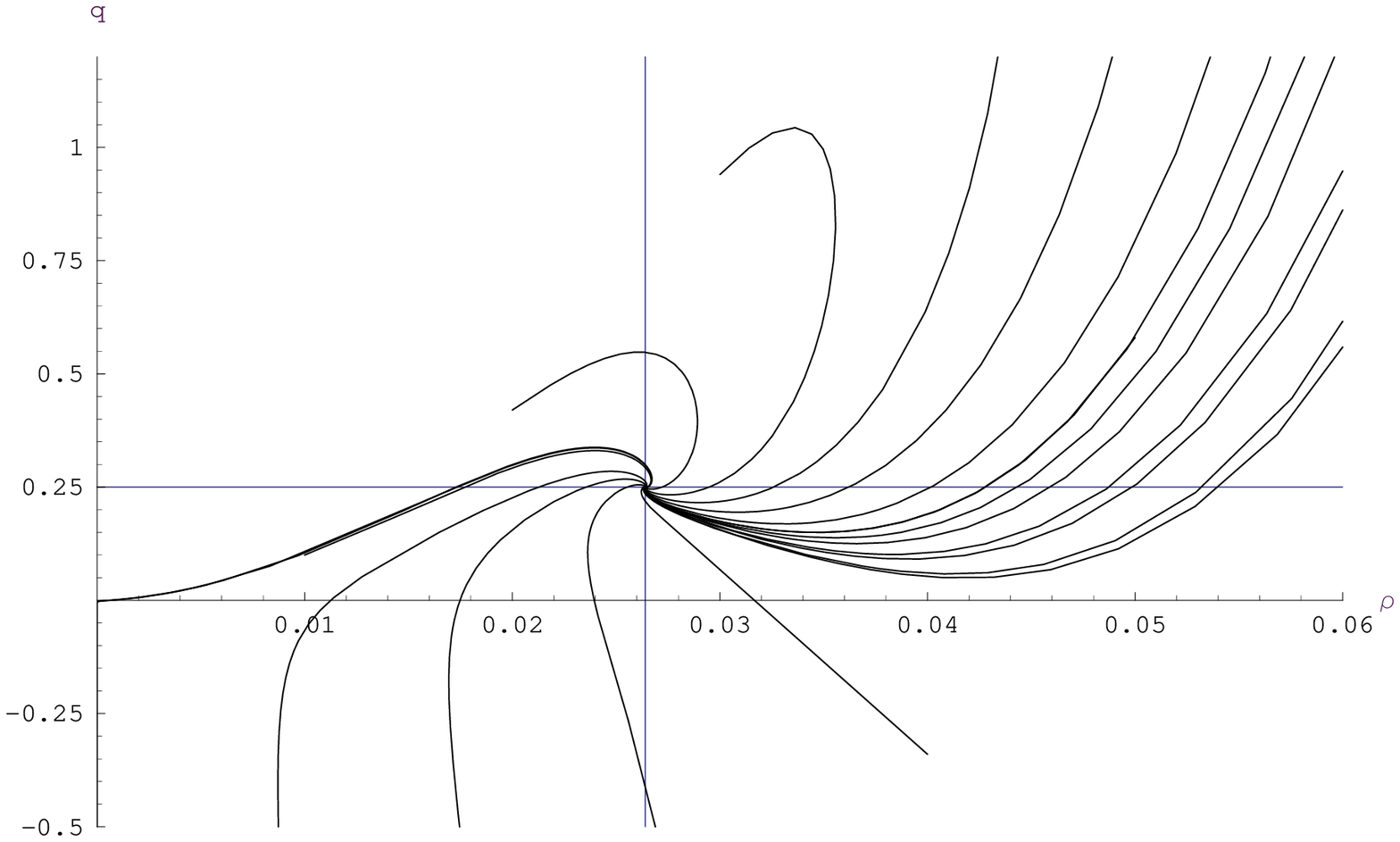}\\
(c)&(d)
\end{tabular}
\caption{\label{phase space gen}Trajectories in the phase spaces $\cqq/\crho$ with varing $\crho_0$ and $\cchi_0$, $d=6,\cAA=-1$ (analogous pictures come from the static extra dimension case) and: (a)
$w_d=4>d/2,\a=1,\n=1$ leading to stable node in $\csrho=1/4$ plus a second repulsive node in $\csrho=3/4$, (b) $w_d=2.5<d/2,\a=1,\n=1$ determining a stable node, (c) $w_d=2.5<d/2,\a=1,\n=-1$ an example of a
stable spiral, (d) $w_d=2<d/2,\a=10^3,\n=1$ a spiral behavior for large $\a$ (large $M^6/V$).}
\end{figure}

To obtain real observables, we have to restrict the possible values for $\crho$ and $\cchi$ such that $\tcrd\,\a\crho^2+\tcvd\left(\crho+\cchi\right)-\ck\ge0$. In fact, the plots show the presence of a
prohibited zone in the phase space --- in particular, in figure \ref{phase space gen}(a) it is clear that the region of the possible trajectories is delimited by a parabola. The relation that must be
satisfied, in terms of the acceleration parameter and the energy density, is
\bea  \label{acceleration parabola gen}
\cqq\le (d-2w_d)\tcrd\,\a\crho^2+(d-w_d)\tcvd\,\crho-d\k
\ena
In fact, the analytical expression for the acceleration $\cqq=\dot\cHH+\cHH^2$ can be written using (\ref{dif sys gen num}) in terms of the visible energy density and the mirage density. For any $\n$
\bea  \label{q dimensionless gen}
\cqq=(1-w_d)\tcrd\,\a\crho^2+\left(1-\frac{w_d}{2}\right)\tcvd\,\crho+\left(1-\frac{d}{2}\right)\tcvd\,\cchi-\ck
\ena
So, taking a specific value for $\crho$, we can have positive acceleration for our universe only if 
\bea
-\left(\tcrd\,\a\crho^2+\tcvd\,\crho-\ck\right)\le\tcvd\,\cchi<-\frac{2(w_d-1)\tcrd\,\a\crho^2+(w_d-2)\tcvd\,\crho+2\ck}{d-2}
\ena
and a necessary condition for this to be possible is a bound on the energy density $(d-2w_d)\tcvd\a\crho^2+(d-w_d)\tcvd\crho>d\ck$, as we can deduce from \refeq{acceleration parabola gen}. The mirage density
$\cchi$ has to be negative to get positive acceleration for $w_d\ge2,\k\ge0$. If instead $w_d\le1,\k\le0$, the mirage density is positive for negative $\cqq$.

Manipulating the set of equations \refeq{dif sys gen num}, we write the following differential equations in terms of the generic energy--exchange parameter $T$
\bea
a\frac{\intd\cchi}{\intd a}&=&-d\cchi+\eta\cTT\left(1+2\frac{\tcrd }{\tcvd}\a\crho\right)\left[\tcrd\, \a\crho^2+\tcvd\left(\crho+\cchi\right)-\ck\right]^{-\frac{1}{2}}\\
a\frac{\intd\crho}{\intd a}&=&-w_d\crho-\eta\cTT\left[\tcrd\, \a\crho^2+\tcvd\left(\crho+\cchi\right)-\ck\right]^{-\frac{1}{2}}
\ena  
We thus come to the differential equation for the acceleration factor
\bea
\left(w_d\crho+\frac{\eta\cTT}{\cHH}\right)\frac{\intd\cqq}{\intd\crho}&=&-\frac{\eta\cTT}{2\cHH}\left(2\a c_\rho\crho+c_V\right)+\\
&&+\left[2\a(1-w_d)c_\rho\crho+\oneover{2}(2-w_d)c_V\right]\crho+d\cqq
\ena
where $\cHH=\sqrt{\left(2\a c_\rho\crho^2+c_V\crho+d\ck\right)/2+\cqq}$ and $\eta=\pm1$ denotes the two possible roots for $\cHH$. Again we note the presence of the symmetry $\cHH\to-\cHH$, $\cTT\to-\cTT$. We
have used the definitions $(w_d-d)\tcvd=c_V$ and $(w_d-d/2)\tcrd =c_\rho$.  From this equation we can infer that positive $\cqq$ implies growing $\cqq$ in an expanding universe ($\eta=+1$) with energy outflow
($\cTT>0$) if $c_V<0,c_\rho>0$ and $\crho<\frac{w_d-2}{w_d-1}\frac{|c_V|}{4\a c_\rho}\equiv\crho_{lim}$. For $\tcvd,\tcrd >0$, $c_V<0,c_\rho>0$ this is realized if $d/2<w_d<d$. In the case of energy influx
($\cTT<0$), we get increasing positive acceleration if $c_V,c_\rho>0$ and $w_d<1$ for all positive energy densities ($\tcvd,\tcrd $ has to be negative). Or else, $\cqq$ grows as $\crho$ grows if
$c_V>0,c_\rho<0$ and $w_d<1$, until the energy density reaches the bound $\crho_{lim}$.

The non trivial fixed points for energy influx are determined by the roots of the equation for $\csrho$
\bea\label{rho fixed gen}
\oneover{d}\left(d-2w_d\right)\tcrd \a\csrho^2+\oneover{d}\tcvd\left(d-w_d\right)\csrho-\frac{\cAA^2}{w_d^2}\csrho^{2(\n-1)}-\ck=0
\ena
while for $\cschi$ and $\csH$ we get the two functions of $\csrho$
\bea
\cschi=-\frac{w_d}{d}\left(1+2\frac{\tcrd }{\tcvd}\a\csrho\right)\csrho, \qquad \csH=-\frac{\cAA}{w_d}\csrho^{\n-1}
\ena
We thus have to fix a particular value for $\n$ in order to establish the precise number of roots and the explicit solution for the critical points. For integer and seminteger $\n$ the number of
roots we can obtain, keeping $d\ne2w_d$, is
\bea
\n\ge2 \qquad &\Longrightarrow& \qquad \mbox{\# roots}=2(\n-1)\ge2  \non\\
1\le\n<2 \qquad &\Longrightarrow& \qquad \mbox{\# roots}=2  \non\\
\n<1 \qquad &\Longrightarrow& \qquad \mbox{\# roots}=2(2-\n)>2  \non
\ena 
These are all the roots of \refeq{rho fixed gen}, including trivial and complex roots. When $d=2w_d$ the critical value for the mirage energy density diverges due to the divergence of $\tcrd $, unless we fix
$w_\p$ to keep it finite. In this case, the number of roots changes to
\bea
\n>1 \qquad &\Longrightarrow& \qquad \mbox{\# roots}=2(\n-1)\ge1  \non\\
\n=1 \qquad &\Longrightarrow& \qquad \mbox{\# roots}=1  \non\\
\n<1 \qquad &\Longrightarrow& \qquad \mbox{\# roots}=3-2\n>1  \non
\ena 
For $\n>1$, one of the roots of equation (\ref{rho fixed gen}) is null if $\k=0$.

There are moreover two trivial fixed point solutions given by $\csrho=\csH=0,\cschi=\tcvd\,\ck$ and $\csrho=\cschi=0,\csH=\sqrt{-\ck}$, that reduce to a unique point with all vanishing variables when the
internal space is flat.

Let's study in more detail the case $\n=1$, since solutions can be written explicitely being the case with the minimum number of roots for \refeq{rho fixed gen}, toghether with the $\n=2$ case. As a result we
get a trivial fixed point solution with $\csH=\csrho=0,\cschi=\tcvd\,\ck$ and the trivial solution, acceptable only for negative and zero curvature, $\csH=\sqrt{-\ck},\csrho=\cschi=0$.  Finally, the two non
trivial solutions (for $d\ne2w_d$) are given by
\bea\label{fix gen nu}
\csH&=&-\frac{\cAA}{w_d} \non\\
\cschi&=&\frac{-\tcvd(d-w_d)-4(d-2w_d)\tcrd \,\a K^2\pm\sqrt{(d-w_d)^2\tcvd^2-4d(d-2w_d)\tcrd \,\a K^2}}{2(d-2w_d)\tcrd \,\a}\\
\csrho&=&\frac{-\tcvd(d-w_d)\pm\sqrt{(d-w_d)^2\tcvd^2-4d(d-2w_d)\tcrd \,\a K^2}}{2(d-2w_d)\tcrd\, \a}  \non
\ena
where $K$ corresponds to a shift and rescaling of $\cAA^2$ due to the nonvanishing value of $\k$ and is defined by $K^2\equiv\cAA^2/w_d^2+\ck$. The two roots are both real only if the argument of the square
root in (\ref{fix gen nu}) is positive, i.e. when $w_d$ lies outside the two roots $-\tilde\a d\left(1\pm\sqrt{(\tilde\a+1)/\tilde\a}\right)$, with $\tilde\a\equiv4\tcrd\,\a K-1$. If $\tilde\a$ is in the
range bounded by $-1$ and $0$ the square root is always real, whatever $d,w_d$ we choose. If two or all among $\tcvd,\tcrd ,(d-w_d),(d-2w_d)$ have equal sign, one of the two solutions \refeq{fix gen nu}
always is characterized by a negative $\csrho$. We note that we can have at list a non trivial fixed point with positive energy density if $(d-2w_d)\tcrd\,\a>0,(d-w_d)\tcvd<0$ --- exactly two positive
$\csrho$ fixed points ---, or for $(d-2w_d)\tcrd\,\a<0$ --- only one critical point with positive energy density. 

The non trivial solution for $d=2w_d$ is 
\bea \label{fixed num gen 2}
\csH=-\frac{\cAA}{w_d}, \qquad  
\csrho=2\tcvd K^2,  \qquad  \cschi=-\tcvd K^2\left(4\tcrd  K^2\a+1\right)
\ena 
For this unique fixed point solution to be characterized by positive $\csrho$ we have to demand a positive $\tcvd K^2$. Moreover, we can derive from (\ref{spiral bound gen nu}) that for $\ck>\tcvd
w_d\csrho/4d$ the fixed point is a spiral, so that for istance, in a flat internal space, we always obtian a node since $\tcvd$ must be positive in order to have a positive $\csrho$ fixed point ($K=A$ in this
case).

The numerical analysis can now show some of the features that we commented for the cosmological evolution with generic density. The differential system of equations (\ref{dif sys gen num}) (substituing some
precise values for $\n$) can be solved numerically in order to check the existence of stable inflationary critical points. In figure \ref{phase space gen} we plot the dimensionless acceleration factor
$\cqq(t)$ as a function of the dimensionless energy density $\crho(t)$, as we did in the previous section for small densities.
In the plots, we consider for simplicity positive $\tcvd,\tcrd $ and flat compact extra dimensions $\k=0$ ($\k\ne0$ results in a shift for $\cAA^2/w_d^2$ in the critical point evaluation and in some scaling
of the suitable value of $\n$ in order to have stability).  

All the critical points we get are characterized by positive $\cqq$, i.e. they represent an inflationary point. In the phase space portrait \ref{phase space gen}(a) trajectories starting with positive acceleration factor and
energy density lower than the critical one pass through an era of larger acceleration and then slow down to the fixed point where we have inflation. There are then solutions starting with negative
acceleration and going to the positive $\cqq$ critical point, eventually passing through a larger acceleration phase or through a smaller density phase. The families of solutions that distinguish the diagram
\ref{phase space gen}(a) from the others are caracterized by both initial and final very high energy density, since they are repelled by the second non attractive fixed point. Some start with high energy density and negative
acceleration at late time, go through an era of larger acceleration (eventually positive) and then, while $\crho$ becomes very large, they go to a region of large and negative $\cqq$. Other go from positive
aceleration to large negative $\cqq$ and large $\crho$. The mirage density $\cchi$ can start from an initial condition smaller or bigger than the critical value, has to be positive for
$\cqq<-\left(3\crho+1\right)\crho$ --- as we deduce from \refeq{q dimensionless gen} plugging in the values for the parameters --- and approaches a negative constant value. 

In diagram \ref{phase space gen}(b), trajectories starting with negative acceleration go to the positive $\cqq$ fixed point, eventually reaching a maximum $\cqq$ before ending into the critical point. Positive acceleration
initial condition lead to growing acceleration at very early times, when usually the energy density grows as well, then both $\cqq$ and $\crho$ decrase to reach the fixed point, passing through a minimum for
acceleration. The mirage density goes to the negative critical value being initially positive for the trajectories that come from negative acceleration conditions, with $\cqq<-\left(4\crho+1\right)\crho/2$.
With the choice of parameters we used, we get $w_d<d/2$. We could also have used the fixed point solution \refeq{fixed num gen 2} if $w_d=d/2=3$ and the diagram for the phase space would have been analogous
to that in plot \ref{phase space gen}(c), keeping the same values of plot \ref{phase space gen}(b) for the other parameters. 

In the phase space \ref{phase space gen}(c) all solutions converge to the fixed point with a spiral behavior. Energy density and acceleration parameter thus oscillate around the critical values. Here $\n=-1$
and $d<2w_d$. The number of roots corresponding to the critical point solutions are six, but four of them are complex roots and one is characterized by negative energy density. Only one real critical point
with positive energy density exists in this case and it has Hubble parameter. Since $\cqq$ can be negative at some time of the evolution, even if starting with a positive value, $\cchi$ can pass through a
positive phase, crossing zero before reaching the negative critical value, if $\cqq<-\left(4\crho+1\right)\crho/2$. A similar plot can be drawn also if $w_d=d/2$. There would be only five roots for $\csrho$,
four of which would be complex conjugated and the last would have positive energy density, representing the stable spiral.
 
Another stable spiral is represented in figure \ref{phase space gen}(d). Here, the dimensionless parameter $\a$, which is proportional to $M^6/V$, is large and $\n=1$. We find only one non trivial fixed point
with positive energy density and spiral behavior, so that trajectories has a shape analogous to the ones in \ref{phase space gen}(c).

For values of $\n$ different from $\n=1$ the number of fixed point roots may vary according to the previous discussion. Nonetheless, (as it is shown as an example in figure \ref{phase space gen}(c) for
$\n=-1$) some of the roots may be complex conjugated and thus not acceptable. Another simple case is $\n=2$, where we get two solutions to \refeq{rho fixed gen}, as with the $\n=1$ assumption. We will not
discuss this situation in detail since the phase spaces we can find are analogous to the $\n=1$ ones. 

The case of energy outflow is analogous to the small energy density analysis. In fact, the fixed point solution \refeq{fixed line gen} can't be characterized by positive energy exchange parameter $T$. We can
nevertheless have a critical point with energy outflowing from the brane into the bulk and negative Hubble parameter (as we can deduce from the expansion~$\to$~contraction, influx~$\to$~outflow symmetry). As
the differential equation for $\rho$ \refeq{bulk limits rho gen} shows, the energy density decreases and go to the trivial fixed point, meaning that the negative $H$ critical point isn't an attractor.

The 4D energy density in the static compact extra dimension case is just given by a constant rescaling of the 6D density. Thus, the phase portraits are given by the plots in figure \ref{phase space small} for
small energy density, and figure \ref{phase space gen} for generic density (up to constant rescaling). However, for equal scale factors $d=6$, the 4D effective energy density $\rhofo$ is dynamically
determined by the energy density localized on the 5--brane $\rho$ and the volume of the compact space $V_{(2)}$: $\rhofo=V_{(2)}\rho$. The 4D mirage density can be similarly defined as $\chifo=V_{(2)}\chi$.
We can single out the volume time dependence defining the dimensionful constant $v$ such that $V_{(2)}\equiv vb^2(t)$, where $b(t)$ is as usual the compact space scale factor --- in this particular case
$b(t)=a(t)$. The generic set of equations for the dimensionless variables $\cHH,\crhofo,\cchifo$ is
\bea
\cHH^2&=&\frac{\tcrsix\,\a}{v^2}\frac{\crhofo^2}{a^4}+\frac{\tcvsix}{v}\frac{\left(\crhofo+\cchifo\right)}{a^2}-\ck   \\
\dot\cchifo+4\cHH\cchifo&=&\cAAfo\,\frac{\rhofo^\n}{a^{2(\n-1)}}\left(1+2\frac{\tcrsix}{\tcvsix}\frac{\a}{v}\crhofo\right) \non\\
\dot\crhofo+(3(1+w)+2w_\p)\cHH\crhofo&=&-\cAAfo\,\frac{\crhofo^\n}{a^{2(\n-1)}}   \label{effective 4D equal gen sys dimensionless}
\ena
where $\Afo\equiv A/v^{\n-1}$. We first note that in the case of zero energy exchange $T=0$ ($\cAAfo=0$) the 4D mirage density satisfies the 4D free radiation equation, as in the static internal space
hypothesis. The 4D energy density $\rhofo$ does not have a definite behavior in the case of energy outflow. While in the static 2D compact space background it is clear that $\rhofo$, just as $\rho$, is
suppressed in time since $w_{\scriptscriptstyle{d=4}}>0$, here we may have a negative coefficient for the linear term in $\crhofo$ in \refeq{effective 4D equal gen sys dimensionless}. If $w_\p<-3(1+w)/2$ ---
which is possible only for $w<-1/3$ if $w_\p>-1$ --- we could have non trivial stable critical points, as in the energy influx context. This scenario would need further investigations.

We can make some considerations regarding the more generic assumption on the relation between the Hubble parameters $H$ and $F$, $F=\x H$, of subsection \ref{proportional hubble}. For positive $\x$ the
qualitative behavior is analogous to what we deduced in the case of static compact extra dimensions and equal scale factors. When $\x$ is negative, meaning that we are using a dynamical compactification
approach \cite{Mohammedi:2002tv}, we could instead have some differences. In particular, it is worth noticing that $w_\x$ (appearing in the conservation equation for $\rho$) can become negative, always
implying a diverging behavior for the localized energy density at late time in the case of energy influx. Thus, there won't be stable critical point in the dynamical compactification scenario with energy
flowing from the bulk onto the brane. 
\subsection{Small density and free radiation equation of state: an explicit solution}\label{explicit bulk section} We can write an explicit solution to the set of equations (\ref{bulk limits H})--(\ref{bulk
limits rho}) in the special limit of small localized energy density $\rho\ll V$ and when $w_d=d$. This last condition is realized if $w=1/3$ with static compact extra dimensions and if $w_\p=(1-3w)/2$ with
equal scale factors. We must be carefull though, because in this limit $\tcvd$ generally diverges. It is thus important to keep it finite by imposing a priori a specific value for $w$ and $w_\p$. With these
assumptions, $H^2$ only depends on the sum $\left(\rho+\chi\right)$ and the equation for this sum can be easily integrated independently of the explicit form of $T$. To be more specific, we get the following
solution \bea\label{explicit bulk} H^2&=&\frac{2\tcvd V}{(d-1)M^{10}}\left(\rho_0+\chi_0\right)\frac{a_0^d}{a^d}-\oneover{2(d-1)}\frac{\k}{b^2}\\ \rho+\chi&=&\left(\rho_0+\chi_0\right)\frac{a_0^d}{a^d} \ena

As in the four dimensional RS model with energy exchange analyzed in \cite{Kiritsis:2002zf}, the evolution is determined by the initial value of the energy density (if we put, for example, $\chi_0=0$)
weighted by the expansion in a (effective six or four dimensional) radiation dominated era.

We deduce from \refeq{explicit bulk} that $|a(t)|$ must have an upper limiting value for $\k>0$. For small positive $a(t)$ the rate $\dot a(t)$ is a positive function and the universe expands until it
reaches the limiting value. If the compactification is over an hyperbolic space, the scale factor grows without bound. In particular, in a universe with extra dimensions evolving according to the same Hubble
parameter as for the observed space--time, the expansion rate goes to a contant positive value as the scale factor grows. In a static extra dimension set--up instead, $a(t)$ exponentially grows at infinity.
When $\k=0$ the explicit solutions for $a(t)$ reduce to $a(t)\sim t^{1/2}$ for static internal space, and $a(t)\sim t^{1/3}$ for equal scale factors. These represent exactly a radiation dominated flat
universe in four or six effective dimensions respectively. The evolution $a(t)\sim t^{1/3}$ can also be traced in 4D to the Friedmann equation for scalar field subject to a null potential.

If we further assume the energy exchange to be linear in the localized energy density $T=A\rho$, also imposing the initial condition $\chi_0=0$ and $w_d=d$, the integration of the $\chi$ and $\rho$ equations
yields 
\bea
\chi=\rho_0\frac{a_0^d}{a^d}\left(1-\ex^{-At}\right),\qquad  \rho=\rho_0\frac{a_0^d}{a^d}\,\ex^{-At}
\ena  
This solution shows that, for energy outflow $A>0$, the initial amount of radiation energy density decays in favor of the mirage energy density. The late time evolution is thus governed by the mirage
density.

For equal scale factors it is interesting to write the explicit solution (remember $w_{\scriptscriptstyle{d=6}}=6$ in this case) in terms of the 4D densities $\rhofo,\chifo$ ($\rhofo=vb^2(t)\rho$,
$\chifo=vb^2(t)\chi$).  The ansatz implying static internal directions only results in a constant rescaling of the 6D quantities. The equal scale factor solution is given by
\bea
H^2&=&\frac{2\tcvsix\, V}{5M^{10}\, v}\left(\rhofz+\chifz\right)\frac{a_0^4}{a^6}-\oneover{10}\frac{\k}{a^2}\\
\rhofo+\chifo&=&\left(\rhofz+\chifz\right)\frac{a_0^4}{a^4}
\ena
The evolution is still weighted by the characteristic effective 6D radiation dominated era $1/a^6$. But we can see, exploring the solutions for $\rhofo$ and $\chifo$ in the case of energy exchange parameter
determined by $T=A\rho$ (with positive $A$), that the 4D localized energy density evolves as 4D radiation, exponentially suppressed in time (as for the static internal space background). In fact, assuming for
example $\chifz=0$, we get
\bea
\chifo=\rhofz\frac{a_0^4}{a^4}\left(1-\ex^{-At}\right),\qquad  \rhofo=\rhofz\frac{a_0^4}{a^4}\,\ex^{-At}
\ena
In this case $\Afo=A$.

The case of energy influx is also analogous to the analysis in \cite{Kiritsis:2002zf}. For both static compact extra dimensions and equal scale factors we rewrite the conservation equation for $\rho$ for flat
internal space as
\bea
\dot\rho+\frac{2}{t}\rho=-T
\ena 
where we have used the equation determining $H$ \refeq{explicit bulk}. This shows that for negative $T$, $\rho$ should increase without bounds at late time. Since we are in the low density approximation, we
can only rely on the generic $\rho$ analysis of the previous section for large $\rho$. However, assuming $T=A\rho^\n$, we can deduce that for $\n>3/2$ the energy density can flow to zero (for certain values
of the parameters). Indeed, $\n>3/2$ conrresponds to non stable crital point in the small density analysis.

We remark that introducing the 4D density $\rhofo$ for a universe with equal scale factors brings to
\bea\label{effective 4D influx equal explicit}
\dot\rhofo+\frac{4}{3t}\rhofo=-T
\ena 
Equation \refeq{effective 4D influx equal explicit} tells us that the 4D localized energy density still grows unlimited at late time, if $T$ is linear in $\rhofo$ --- more general considerations are analogous
to the 6D density case. Again, we would need the full treatment for generic density.

The acceleration parameter $q\equiv\ddot a/a$ in this context is equal to
\bea
q=-\frac{(d-2)\tcvd V}{(d-1)M^{10}}\left(\rho_0+\chi_0\right)\frac{a_0^d}{a^d}
\ena
For non zero $\k$, the value of the acceleration can be either positive or negative. It has to be negative when $\k\ge0$, but may be positive for compactification on hyperbolic spaces ($\k<0$), giving as a
result a loitering universe.
\section{Construction of the holographic dual}\label{dual}
The theory dual to the 7D RS model, via the AdS/CFT correspondence \cite{Maldacena:1997re}, will be derived in complete analogy to the 5D set--up considered in \cite{Kiritsis:2005bm}. The RS model, with a
time independent warped geometry, gives $AdS_7$ metric as a solution to the equations of motion for the gravity action in the bulk. It will be useful to parametrize it according to Fefferman and Graham
\cite{Fefferman} 
\bea\label{FG para}
G_{AB}\intd x^M\intd x^N=\frac{\ell^2}{4}\rho^{-2}\intd \rho^2+\ell^2\rho^{-1} g_{\m\n}\intd x^\m\intd x^\n
\ena
where the indices $M,N.\dots$ run over the 7D bulk space--time, $\m,\n,\dots$ span the 6D space--time on the 5--brane and $\rho$ is a reparametrization of the $z$ coordinate tranverse to the brane. 
The location of the brane, translated to this new set of coordinates, is $\rho=0$ which represents the boundary of the background (\ref{FG para}). Generally, for all seven dimensional asymptotically AdS
space--times the 6D metric $g_{\m\n}$ can be expanded as \cite{Fefferman}
\bea
g=g_{(0)}+\rho g_{(2)}+\rho^2g_{(4)}+\rho^3g_{(6)}+\rho^3\log\rho\, h_{(6)}+\ogr(\rho^4)
\ena
The logarithmic piece appears only for space--times with an odd number of dimensions and is responsible for the cutoff dependent counterterm in the renormalized action.  As a consequence, it is also
responsible for the conformal anomaly of the holographic dual CFT \cite{Henningson:1998gx,deHaro:2000xn} --- in fact, we don't have conformal anomaly in even dimensional backgrounds. The subindices in the
coefficients of the metric expansion stand for the number of derivatives contained in each term.

More precisely, RS background is a slice of $AdS_7$, where the boundary gets replaced by the 5--brane and the IR part is reflected, eliminating the UV slice. To describe seven dimensional gravity we will
take the usual Einstein--Hilbert action in the bulk, adding as usual a Gibbons--Hawking term \cite{Gibbons:1976ue} to take account of the boundary extrinsic curvature. Since the gravitational theory exhibits
divergences in a space--time with boundaries, we also have to regularize the Einstein--Hilbert plus Gibbons--Hawking action, cutting off the boundary of the space--time. We are now going to illustrate the
regularization procedure.
\subsection{Regularization on the gravitational side} The renormalization for a gravitational theory in a background with boundaries has been explained in
\cite{Henningson:1998gx,Skenderis:1999nb,deHaro:2000xn} for a generic number of dimensions. We will apply those computations to the case of a seven dimensional bulk space--time.  

In general, the bulk action  for gravity gets modified with the Gibbons--Hawking boundary term \cite{Gibbons:1976ue} and with some counterterms also localized on the boundary
\bea
S_{gr}=S_{EH}+S_{GH}-S_{count}
\ena
Using the Fefferman and Graham parametrization of the metric \refeq{FG para} and cutting off the boundary at $\rho=\epsilon$, we have
\bea
S_{EH}=M^5\int_{\rho\ge\epsilon}\intd^7x\sqrt{-G}\left(R\left[G\right]+\frac{30}{\ell^2}\right),\quad S_{GH}=2M^5\int_{\rho=\epsilon}\intd^6x\sqrt{-\g}K
\ena
where $R\left[G\right]$ is the bulk Ricci scalar, $K$ is the trace of the extrinsic curvature and $\g_{\m\n}$ is the induced metric on the boundary.  Putting the brane at $\rho=\epsilon$ corresponds to
regularize the gravity action. The counterterm contributions necessary to make it finite in the limit $\epsilon\rightarrow0$ are given by
\bea
S_{count}=S_0+S_1+S_2+S_3
\ena
$S_i$ are terms of order $i$ in the brane curvature $R$ (the curvature of the induced metric $\g_{\m\n}$ on the boundary). In fact they can be written in terms of the induced metric $\g_{\m\n}$ and its
Riemann tensor $R_{\m\n\rho\s}$, using the perturbative expansion relating $\g_{\m\n}$ to $g_{(0)\m\n}$ (see for instance \cite{deHaro:2000xn})
\bea\label{def count 0}
S_0&=&10\,\frac{M^5}{\ell}\int_{\rho=\epsilon}\intd^6x\sqrt{-\g}\\
S_1&=&-\oneover{4}M^5\ell\int_{\rho=\epsilon}\intd^6x\sqrt{-\g}R\\
S_2&=&\oneover{32}M^5\ell^3\int_{\rho=\epsilon}\intd^6x\sqrt{-\g}\left(R_{\m\n}R^{\m\n}-\frac{3}{10}R^2\right)\\
S_3&=&\frac{\log\epsilon}{64}M^5\ell^5\int_{\rho=\epsilon}\intd^6x\sqrt{-\g}\bigg(\oneover{2}RR_{\m\n}R^{\m\n}+\frac{3}{50}R^3+R^{\m\n}R^{\rho\s}R_{\m\rho\n\s}\non\\
&&+\oneover{5}R^{\m\n}\nabla_\m\nabla_\n R-\oneover{2}R^{\m\n}\Box R_{\m\n}\bigg)  \label{def count 3}
\ena
The third order term $S_3$ depends on the cutoff $\epsilon$ and is thus responsible for the breaking of the scale invariance, i.e. it gives rise to the conformal anomaly for the dual 6D CFT in the context of
the AdS/CFT correspondence. We also note that the zeroth order term is related to the brane tension term of the RS model (\ref{7D action}) $S_{tens}$ by $S_{tens}=-2S_0$, if we fine--tune $\l_{RS}=0$. In
fact, in the pure RS set--up, where the effective cosmological constant is null $\l_{RS}=0$, the brane tension is $V=20M^5/\ell$, since the bulk cosmological constant is given by $\L_7=-30M^5/\ell^2$ as a
function of the background length scale $\ell$ and the bulk Planck mass $M$, in our background metric parametrization \refeq{FG para}. We will now use the AdS/CFT correspondence to compute the dual theory.
\subsection{Gauge/gravity duality}
The AdS/CFT duality \cite{Maldacena:1997re} is realized between gravity (string theory or M theory in the decoupling limit) in a background with one or more stacks of some kind of branes and the gauge theory
that lives on the boundary of the near horizon geometry inferred by the branes \cite{Maldacena:1997re,Witten:1998qj}. In our particular case, $AdS_7\times S^4$ is the near horizon geometry of a system of $N$
parallel M5--branes in eleven dimensional M theory. The radius of the AdS space is given in terms of the eleven dimensional Planck lenght $\ell_{Pl}$ and of the number $N$ of M5--branes
\bea
\ell=2(\p N)^{1/3}\ell_{Pl}
\ena
The radius of the four sphere is half the radius of $AdS_7$. The supergravity approximation for M theory is valid if $N\gg1$ and $\ell_{Pl}\sim N^{-1/3}\rightarrow0$, keeping the radius of the AdS large and
finite in units of $\ell_{Pl}$. The six dimensional theory that Maldacena \cite{Maldacena:1997re} conjectured to be dual to M theory in the background described above is a (0,2) SCFT. This theory is realized
as the open string theory in the world--volume of the M5--branes, in the low energy decoupling limit, and it does not contain dimensionless nor dimensionfull parameters.  The $AdS_7\times S^4$ supergravity
backgroud is characterized by a 4--form flux quantized in terms of the number of M5--branes and is not conformally flat, since the radii of the four--sphere and the AdS space are not coincident. 

The AdS/CFT correspondence relates the gravity (M theory) partition function for the bulk fields $\Phi_i$ (which is a function of the value of the fields on the boundary of $AdS_7$, $\phi_i$) to the
generating functional of correlation functions of the dual CFT operators with sources $\phi_i$
\bea
Z_{gr}\left[\phi_i\right]\equiv\int\intD\Phi_i\,\ex^{-S_{gr}}=\ex^{-W_{CFT}\left(\phi_i\right)}
\ena
Knowing that gravity on $AdS_7$ (the $S^4$ geometry can be factored out) corresponds to the specific CFT suggested by Maldacena \cite{Maldacena:1997re}, we can now obtain as a consequence the theory dual to
the 7D RS model, in analogy to \cite{Kiritsis:2005bm}. In fact, the action of the gravitational theory that we want to analyze via holography is 
\bea
S_{RS}=S_{EH}+S_{GH}+S_{tens}+S_m
\ena
We just add the Gibbons--Hawking term to (\ref{7D action}). We expect that the hidden sector of the holographic theory reflects the bulk non trivial contents encoded in the bulk components of the 7D
stress--energy tensor ($T_0^7,T_7^7$) when we go to the non conformal interacting generalization. 

We now have to keep in mind that the duality for gravity on $AdS_7$ can be stated as 
\bea\label{gravity duality}
Z_{gr}\left[\phi_i\right]\equiv\int_{\rho>\epsilon}\intD\Phi_i\,\ex^{-S_{EH}-S_{GH}+S_0+S_1+S_2+S_3}=\ex^{-W_{CFT}\left(\phi_i\right)}
\ena
Secondly, we have to remember that $S_{tens}=-2S_0$. Furthermore, we note that the integration in (\ref{gravity duality}) is over one half of the space--time appearing in the RS model, because of the
$\Zgr_2$ reflection along the $z$ direction. Since the integrals over the two specular regions are independent and equal we can write
\bea
Z_{RS}\left[\phi_i,\chi_i\right]&\equiv&\int_{\mathrm{all }\rho}\intD\Phi_i\intD\chi_i\ex^{-S_{EH}-S_{GH}+2S_0-S_m}\non\\
&=&\int_{\rho>\epsilon}\intD\Phi_i\intD\chi_i\,\ex^{-2S_{EH}-2S_{GH}+2S_0-S_m}
\ena
where $\chi_i$ are the matter fields on the brane.
Finally, putting all toghether, using equation (\ref{gravity duality}), we obtain
\bea
Z_{RS}\left[\phi_i,\chi_i\right]=\int_{\rho>\epsilon}\intD\Phi_i\intD\chi_i\,\ex^{-2W_{CFT}-2S_1-2S_2-2S_3-S_m}
\ena
The RS dual theory is
\bea\label{dual action RS}
S_{\tilde{RS}}=S_{CFT}+S_R+S_{R^2}+S_{R^3}+S_m
\ena
having defined
\bea\label{def RS dual}
S_{CFT}=2W_{CFT},\quad S_R=2S_1,\quad S_{R^2}=2S_2,\quad S_{R^3}=2S_3
\ena
The 6D Planck mass is thus given by $\Mpl=\frac{M^5\ell}{2}$.

We are now ready to calculate the equations of motion for the holographic 6D RS cosmology.
\section{Holographic evolution equations}\label{6D equations}
As we know, the RS classical solution in a 7D bulk with a warped geometry is $AdS_7$. Since our purpose is to study the cosmology associated to the 7D RS set--up, we have generalized the ansatz for the metric
to be time dependent in section \ref{setup} and \ref{bulk cosmology}. We have successively reviewed the notion of holographic dual theory in the previous section. What we want to do now is to describe the
cosmology of the seven dimensional RS model from the six dimensional holographic point of view, using the correspondence relation obtained in the previous section and generalizing the ansatz for the 6D
induced metric on the 5--brane to a time dependent geometry, as we had done for the 7D bulk analysis.

We consider a 6D space--time, compactified on a 2D internal space, with a FRW metric for the four large dimensions. The induced metric tensor can be expressed as
\bea\label{metric 4+2}
\g_{\m\n}\intd x^\m\intd x^\n&=&
-\intd t^2+\frac{a^2(t)}{1-k r^2}\intd r^2+a^2(t)\,r^2\intd\th^2+a^2(t)\,r^2\sin^2\th\intd\phi^2+\non\\
&&+\frac{b^2(t)}{1-\k\rho^2}\intd\rho^2+b^2(t)\,\rho^2\intd\psi^2
\ena
where $k$ and $\k$, $a(t)$ and $b(t)$, $H(t)$ and $F(t)$ are, respectively, the curvatures, the scale factors, the Hubble parameters for the 3D and 2D spaces.

The action we are considering is 
\bea\label{action}
S_{\tilde{RS}}=S_{CFT}+S_{m}+S_{\l}+S_{R}+S_{R^2}+S_{R^3}
\ena
$S_{R}$, $S_{R^2}$, $S_{R^3}$ being respectively twice the first, second and third order terms in the curvature contributing to the counterterm action, as defined in (\ref{def count 0})--(\ref{def count 3}).
$S_{\l}$ is an effective cosmological term on the brane --- that represents a generalization to the case of a non exact RS fine--tuning with respect to the action (\ref{dual action RS}). $S_m$ and $S_{CFT}$
are the matter and (twice the) CFT actions of the 6D description.

We want to solve the Friedmann equations, imposing the conservation and anomaly equations, defining 
\bea
T_{\m\n}&=&\oneover{\sqrt{-\g}}\frac{\d S_{m}}{\d \g^{\m\n}} \qquad\qquad
W_{\m\n}=\oneover{\sqrt{-\g}}\frac{\d S_{CFT}}{\d \g^{\m\n}} \\
Y_{\m\n}&=&\oneover{\sqrt{-\g}}\frac{\d S_{R^2}}{\d \g^{\m\n}} \qquad\qquad
Z_{\m\n}=\oneover{\sqrt{-\g}}\frac{\d S_{R^3}}{\d \g^{\m\n}}
\ena
and $V_{\m\n}=W_{\m\n}+Y_{\m\n}+Z_{\m\n}$. The equations of motion take the form
\bea\label{Fried}
M_{Pl}^4G_{\m\n}+\l \g_{\m\n}&=&T_{\m\n}+V_{\m\n}\non\\
\nabla^\n T_{\m\n}&=&0\non\\
\nabla^\n V_{\m\n}&=&0\\
V^\m_\m&=&\ano+Y\non
\ena
Here $\ano$ is the general anomaly for a 6D conformal theory \cite{Deser:1993yx,Bastianelli:2000rs} that comes uniquely from the $S_{3}$ contribution to the renormalized action
\cite{Henningson:1998gx}--\cite{deHaro:2000xn}
~\footnote{The scheme dependent contribution to the anomaly --- the type D anomaly --- gets cancelled by the equal and opposite scheme dependent contribution to $S^3$ --- which are local covariant counterterms
and can be derived from appendix C of \cite{Bastianelli:2000rs}. For an explicit example of how this happens, see \cite{Kiritsis:2005bm} in the four dimensional case.}
, while $Y$ is the trace $Y^\m_\m$ of the variation of (twice) the second order counterterm action $S_{R^2}$. The trace of $Z_{\m\n}$ is null
~\footnote{It is indeed proportional to the traceless tensor $h_{(6)}$ that appears in the Fefferman and Graham metric parametrization \refeq{metric 4+2} \cite{deHaro:2000xn}. We thank K. Skenderis for helpful
discussion on this subject.}
. The trace of $Y_{\m\n}$ is quadratic in the curvature of the metric \refeq{metric 4+2} 
\bea
Y=\oneover{32}M^5\ell\left(R^{\m\n}R_{\m\n}-\frac{3}{10}R^2\right)
\ena
The explicit form for the anomaly is a complicated expression of dimensions 6, cubic in the curvature, and is discussed in appendix \ref{anomaly}. The effective cosmological constant on the brane is $\l$.
The stress--energy tensors are parametrized as
\bea\label{stress-en}\begin{array}{rclrclrcl}
T_{00}&=&\rho(t),\quad T_{ij}&=&p(t)\,\g_{ij}, \quad T_{ab}&=&\p(t)\,\g_{ab}\\
V_{00}&=&\s(t),\quad V_{ij}&=&\s_p(t)\,\g_{ij}, \quad V_{ab}&=&\s_\p(t)\,\g_{ab}\\
\end{array}\ena
where the indices $ij\dots$ parametrize the space part of the 4D FRW space--time and run from 1 to 3, while $ab\dots$ belong to the 2D internal space and take values in $(4,5)$.~
\footnote{The energy density $\rho(t)$ should not be confused with the 2D coordinate $\rho$, since the radius of the extra dimensions does not appear in the calculations.}

Equations (\ref{Fried}) take the following form when we choose the metric (\ref{metric 4+2}) and the stress--energy tensors written in (\ref{stress-en}).  The Friedmann equations become
\bea\label{Friedmann eqs}
M_{Pl}^4\left(3H^2+6H\,F+F^2+3\frac{k}{a^2}+\frac{\k}{b^2}\right)&=&\rho+\s+\l\non\\
M_{Pl}^4\left(2\dot H+3H^2+4H\,F+2\dot F+3F^2+\frac{k}{a^2}+\frac{\k}{b^2}\right)&=&-p-\s_p+\l\\
M_{Pl}^4\left(3\dot H+6H^2+3H\,F+\dot F+F^2+\frac{k}{a^2}\right)&=&-\p-\s_\p+\l\non
\ena
then the conservation equations
\bea\label{conservation eqs}
\dot\s+3(\s+\s_p) H+2(\s+\s_\p)F&=&0\non\\
\dot\rho+3(\rho+p) H+2(\rho+\p)F&=&0
\ena
and finally the anomaly equation
\bea\label{anomaly eq}
&&\s-3\s_p-2\s_\p=\ano+Y
\ena
As we said, the anomaly comes from the cubic counterterm, so that it is cubic in the curvature. We make some more precise statement about its form in appendix \ref{anomaly}, where we also explicitely give
$Y$.
\subsection{Simplifications and ansatz}\label{simplifications}
The set of equations (\ref{Friedmann eqs})--(\ref{anomaly eq}) does not contains six independent equations. Plugging the conservation in the first Friedmann equation differentiated w.r.t. time, we get a
linear combination of the other two Friedmann equations. So we will discard the last of \refeq{Friedmann eqs} from now on. We further note that the system contains only one algebraic equation: the Friedmann
equation. We will start by solving the anomaly equation in terms of one of the pressures coming from the hidden theory. 

Plugging the expression for $\s_p$ obtained evaluating (\ref{anomaly eq}) into the first of the conservation equations (\ref{conservation eqs}), we get a differential equation for $\s$ depending on $\s_\p$
\bea\label{conservation + anomaly}
\dot\s+2\left(2H+F\right)\s+2\left(H-F\right)\s_\p=\ano+Y
\ena
To obtain a solvable decoupled equation for $\s$ we can consider the limit in which the internal space has the same CFT pressure $\s_\p=\s_p$ of the three large dimensions~
\footnote{This limit comes from classical evaluation of stress--energy tensor derived from the action $S\propto\int\intd^6x\sqrt{-\g}\,H_{\m\n\rho}H^{\m\n\rho}$}
. Else, we can also consider the limit of zero pressure --- for the CFT --- in the internal space.  Putting these two limits toghether, we can try to solve the Friedmann equations imposing a more general
ansatz
\bea
\s_\p=\O\s_p
\ena
So that
\bea\label{manipulate anomaly eq}
\s-\oneover{\o}\s_p=\ano+Y
\ena
where $\o\equiv1/(3+2\O)$ ($\o$ is equal to $1/5,1/3$ in the two limits considered above) and the differential equation for $\s$ becomes
\bea\label{conservation eq}
\dot\s+3\left[(1+\o)H+(1-\o)F\right]\s-\left[3\o\,H+(1-3\o)F\right]\left(\ano+Y\right)=0
\ena
We could now evaluate $\s$ solving the following integral
\bea\label{sigma anal sol}
\s&=&\chi+\oneover{a^{3(1+\o)}b^{3(1-\o)}}\int\intd t\;a^{3(1+\o)}b^{3(1-\o)}\left[3\o\frac{\dot a}{a}+(1-3\o)\frac{\dot b}{b}\right]\cdot\non\\
&&\cd \left[c_A\,E_{(6)}+c_B\,I_{(6)}+Y\right]
\ena
where $\chi$ is a solution for the homogeneous equation
\bea\label{homo eq}
\dot\chi+3\left[(1+\o)H+(1-\o)F\right]\chi=0\quad\Rightarrow\quad\chi=\frac{\chi_0}{a^{3(1+\o)}b^{3(1-\o)}}
\ena
We observe that generally (\ref{sigma anal sol}) is not explicitely integrable. In \refeq{sigma anal sol} we have written the anomaly in terms of its contributions that are the Euler density in six dimensions
$E_{(6)}$ and the local covariants included in $I_{(6)}$ (see appendix \ref{anomaly} for further details).

The set of independent equations we finally have to solve, once we use (\ref{manipulate anomaly eq}) to eliminate $\s_p$ by means of
\bea\label{eliminate sigma_p}
\s_p=\o\s-\o\left(\ano+Y\right) 
\ena
is then
\bea\label{final system}
M_{Pl}^4\left(3H^2+6HF+F^2+3\frac{k}{a^2}+\frac{\k}{b^2}\right)&=&\rho+\s+\l\\
M_{Pl}^4\left(2\dot H+2\dot F+3H^2+4HF+3F^2+\frac{k}{a^2}+\frac{\k}{b^2}\right)&=&-w\rho-\o\sigma+\o\left(\ano+Y\right)+\l\non\\
\dot\s+3\left[(1+\o)H+(1-\o)F\right]\s&=&\left[3\o H+(1-3\o)F\right]\left(\ano+Y\right)\non\\
\dot\rho+\left[3(1+w)H+2(1+w_\p)F\right]\rho&=&0 \label{final system rho}
\ena
(the anomaly $\ano$ will be written explicitely --- in terms of $H$ and $F$ --- in the particular cases that we will take under examination in the following). We also use the three following ansatz relating
the pressures and the energy densities 
\bea\label{ansatz}
p&=&w\rho\non\\
\p&=&W\,p=w_\p\rho\\
\s_\p&=&\O\,\s_p \quad (3+2\O=1/\o)\non
\ena
Now we are left with a system of four equations (\ref{final system})--\refeq{final system rho} in four variables ($H,F,\rho,\s$). The other variables ($\s_p,\s_\p,p,\p$) are determined by the ansatz
(\ref{ansatz}) and by the equation (\ref{eliminate sigma_p}). In the next section, this system of differential equations will be studied restricting to some special limits, such as flat or static internal
space, or equal scale factors. We will find the critical point solutions and analyze the associated stability matrix. 
\section{Holographic critical point analysis}\label{critical brane}
The fixed points of the cosmological evolution of the universe we are considering may represent its inflationary eras --- for instance the early time or the late time acceleration ---, since the Hubble
parameters, just as the energy densities, are constant. If the constant value for the Hubble parameter is positive we have inflation. In this section, we are going to look for the existence of such
inflationary points for our specific holographic model and to find what kind of dependence they have on the parameters of the theory.

We will describe the fixed point solutions in the special limits of flat extra dimensions, curved static extra dimensions and equal scale factors for the internal and extended spaces. We will then study the
stability matrix associated with the critical points. Since the fixed points represents inflationary eras in the universe evolution, they could offer an explanation to the early inflation or to the late time
acceleration. In the first case they will have to be unstable or saddle points to allow the trajectory of the cosmological evolution to flow away from inflation and exit from this phase. In the second case,
on the other hand, the fixed points must be stable and act as attractors for the nearby trajectories. 

In what follows we will always suppose that the effective cosmological constant on the brane $\l$ will be zero, unless we specify it differently.
\subsection{Flat compact extra dimensions}\label{brane critical flat}
In the limit of zero spatial curvature for the extra dimensions and for the extended space, the Friedmann plus conservation equations (\ref{final system}) take the form
\bea\label{final system flat}
M_{Pl}^4\left(3H^2+6HF+F^2\right)&=&\rho+\s+\l\\
M_{Pl}^4\left(2\dot H+2\dot F+3H^2+4HF+3F^2\right)&=&-w\rho-\o\sigma+\o\left(\ano+Y\right)+\l\\
\dot\s+3\left[(1+\o)H+(1-\o)F\right]\s&=&\left[3\o H+(1-3\o)F\right]\left(\ano+Y\right)\\
\dot\rho+\left[3(1+w)H+2(1+w_\p)F\right]\rho&=&0  \label{final system flat}
\ena
We find the fixed points of this system of differential equations and study stability with some further restrictions (see appendix \ref{fixed points} in the flat extra dimension subsections for the
explicit calculations). As we point out in appendix \ref{anomaly}, the anomaly $\ano$ generally depends on the Hubble parameters of the model, on their time derivatives up to the third order and on the
spatial curvatures. This remains true also for the flat extra dimension limit that we are examining, ignoring the curvatures.
\paragraph*{Fixed point solutions}
With the assumption of flat internal space and zero curvature for the 3D space as well, we can find different fixed points depending on the value of the extra dimensions Hubble parameter. They can be
summarized as follows.
\begin{enumerate}
\item
As we illustrate in appendix \ref{fixed points}, there are two non trivial time independent solution with $F$ non vanishing at the fixed point, $\sF\ne0$, one for $\o\ne1/5$ and one for $\o=1/5$.  The values
of the 3D Hubble parameter (the measurable Hubble parameter) are given in terms of the constants $\o,c_A,c_B,c_Y$ and the mass scale $M_{Pl}$. We note that the anomaly parameters $c_A,c_B$ are given by the
CFT, while $\o$ relates the hidden sector pressure of the internal space to the hidden pressure of the 3D space (\ref{ansatz}). In particular, the two Hubble parameters are related by the equality
$H_\star=(\cep +1)F_\star$ (where $\cep$ is a rather complicated function of $\o,c_A,c_B,c_Y\mpl$) when $\o\ne1/5$, while for $\o=1/5$ we have $H_\star=F_\star$. For $\o=1/5$, $F_\star\neq0$ and we can choose
it to be positive or negative, implying that the extra dimensions scale exponentially at those points, with respectively either positive or negative velocity. Consequently $H$ would also describe either a
contracting or expanding universe.
\item
We also found a fixed point solution for which the extra dimensions are static, i.e. $F_\star=0$ ($\o\ne0,-1$ and $c_B\ne0$), meaning that, while our visible universe is exponentially growing (or, in
principle, decreasing) the internal space isn't expanding nor collapsing. The corresponding solution is given by 
\bea\label{fixed flat holo}
H_\star^2&=&-\frac{20}{3c_B}\frac{\o}{\o+1}M_{Pl}^2\left[48c_Y\pm\sqrt{6\left(384c_Y^2-c_B\frac{\o}{\o+1}\right)}\right] \\
\ssig&=&-\ssigpp=\frac{2\o}{3\o-1}\ssigpp=-\frac{20}{c_B}\frac{\o}{\o+1}\MPl\left[48c_Y\pm\sqrt{6\left(384c_Y^2-c_B\frac{\o}{\o+1}\right)}\right]\non\\
\srho&=&0\non
\ena
The roots are real if $384c_Y^2-c_B\o/(\o+1)>0$ and cannot be both positive. We never have a couple of fixed points in the phase space diagram.
\item
A third fixed point is characterized by zero extra dimension Hubble parameter and $\o=-1$. In this case the critical point exists only if the conformal field theory is characterized by a positive coefficient
for the type B anomaly. The solution is 
\bea\label{fixed flat omega holo}
H_\star^2=\frac{640c_Y}{c_B}\mpl, \quad \ssig=-\ssigp=\oneover{2}\ssigpp=\frac{640c_Y}{3c_B}\MPl,\quad \srho=0 
\ena
If $c_B$ is zero (i.e. the anomaly vanishes at the fixed point) we are left only with the trivial fixed point. 
\item
For vanishing $\o$, $\l$ should be non zero to get the inflationary fixed point $H_\star^2=\frac{\l}{3\Mpl}$.
\item
There also exists a trivial fixed point with $\sH=\sF=0$, where also the anomaly and the trace $Y^\m_\m$ become zero, and $\srho=\ssig=0$ if $\o\neq w$ or $\srho=-\ssig$ if $\o=w$.
\end{enumerate}

In any case --- i.e. for every $\o$ and $\l$ --- the solution does not depend on $w,w_\p$ (in fact, the system of equations (\ref{Friedmann rho})--(\ref{rho}) doesn't contain $w,w_\p$). So, if a solution
exists for some $\o$ and $c_A,c_B,c_Y$, that solution always exists whatever values the two parameters relating matter pressures to energy density take. This marks a difference with the bulk analysis of
section \ref{critical bulk}, since here we don't have any bulk dynamics perturbing the conservation equations, being the hidden sector theory conformal and non interacting.

All the critical points have zero localized energy density $\srho$ (except for the trivial point with $\o=w$). Also, when the Hubble parameter is non vanishing, we don't get a positive value for $\srho$.
\paragraph*{Stability analysis}
In appendix \ref{fixed points} we analyze the stability of the $\sF=0$ critical points linearizing the system of differential equations around the fixed point. We conclude that the studied fixed point
(\ref{fixed flat holo})--(\ref{fixed flat omega holo}), characterized by vanishing $\sF$ and $\o=-1$ or $\o\ne-1$ can both be saddles or attractors, depending on the value of the anomaly parameter $c_B$, of
$c_Y$ and of $\o$ (relating the two CFT pressures). The trivial fixed point cannot be analyzed at linear order, since its stability matrix is null. It is obvious from \refeq{final system flat} that for
positive Hubble parameters the energy density goes to zero at late time.

We can thus observe that, starting the cosmological evolution with a hidden energy density $\s$ different from zero and with a suitable value of the anomaly coefficient $c_B$, choosing $\o$ to be such that
the $\sF=0$ stability matrix has negative eigenvalues, the corresponding $\sF=0$ fixed point could be a global attractor for the flat extra dimension universe. This critical point could eventually represent
the present accelerated era. However, it's a zero energy density critical point.
\subsection{Static compact extra dimensions}\label{brane critical static}
The Einstein equations of motion (\ref{final system}) get simplified when we take the $b=\mathrm{const}$ ansatz for the internal space scale factor 
\bea\label{final system static}
M_{Pl}^4\left(3H^2+\frac{\k}{b_0^2}\right)&=&\rho+\s+\l\\
M_{Pl}^4\left(2\dot H+\frac{\k}{b_0^2}\right)&=&-w\rho-\o\sigma+\o\left(\ano+Y\right)+\l\\
\dot\s+3(1+\o)H\s&=&3\o H\left(\ano+Y\right)\\
\dot\rho+3(1+w)H\rho&=&0
\ena
From these equations we can deduce the corresponding fixed points and their criticality, following the calculations in appendix \ref{fixed points}.
\paragraph*{Fixed point solutions}
We have found the inflationary fixed points for a universe with non evolving internal dimensions, i.e. with constant scale factor $b(t)\equiv b_0$. Besides the trivial fixed point $\sH=0$ there are other solutions.
\begin{enumerate}
\item 
The existence of non trivial fixed points is determined by the values of the parameter of the specific conformal theory $c_A,c_B,c_Y$ and $\o$, but also by the mass scales $M_{Pl}$ and $\k/b_0$. The form of
the fixed point solutions are derived in appendix \ref{fixed points}. An easy critical point solution can be derived in the case of zero type B contribution to the anomaly and $\o=-1$
\bea\label{fixed static cA=cB}
H^2_\star&=&\frac{\k}{b_0^2}\left[9\pm\sqrt{90+\frac{5c_A}{192c_Y}\oneover{\mpl}\frac{\k}{b_0^2}}\right]^{-1}\\
\ssig&=&-\ssigp=2\ssigpp=\Mpl\left(3\left[9\pm\sqrt{90+\frac{5c_A}{192c_Y}\oneover{\mpl}\frac{\k}{b_0^2}}\right]^{-1}+1\right)\frac{\k}{b_0^2}\non\\
\srho&=&0\non
\ena
For $\k=0$ it reduces to the trivial fixed point. We have real roots for $\sH$ if $192c_Y^2\mpl>-c_A\k/45b_0^2$ and they will be both positive when $\k>0$ and $-192c_Y^2\mpl<c_A\k/45b_0^2<-192c_Y^2\mpl/252$, so
that $c_A$ must be negative.
\item
We also have a trivial critical point with $\sH=0$. In particular both $\srho$ and $\ssig$ can be different from zero, they are functions of $\k/b_0^2$ as we can see from the equation
$\srho+\ssig=\Mpl\k/b_0^2$, and are also related by 
\bea
(1+w)\srho+(1+\o)\ssig=\o\left(-\frac{c_B}{800}\frac{\k}{b_0^2}+\frac{4c_Y}{5}\mpl\right)\frac{\k^2}{b_0^4}
\ena
\item 
Other fixed points can be found, for example for $\o=0$. The internal space curvature $\k$ would have to be negative in those cases, since $H^2_\star=-\k/3b_0^2$ and we would have to compactify on
an hyperbolic space.
\end{enumerate}
\paragraph*{Stability analysis}
For the static extra dimension fixed points we found that, depending on the values of $c_A,c_B,cY\mpl$ and $\o$, we can get an attractor or a saddle. 

We can thus choose the hidden sector parameters such that we can get a stable fixed point. There exists however another critical point, i.e. the trivial one characterized by $\sH=0$ but generally non zero
$\srho$ and $\ssig$, which always is a saddle. So, trajectories may either be attracted by the non trivial critical point or flow away from the saddle point.
\subsection{Equal scale factors}\label{brane critical equal}
Another limit that simplifies some of the calculations is the equal scale factor assumption. In this case the Hubble parameters of the internal space and the 3D space are equal, $F=H$, and we remain with the
following set of equations for the variables $H,\rho,\s$
\bea\label{final system equal}
M_{Pl}^4\left(10H^2+\frac{\k}{b^2}\right)&=&\rho+\s+\l\\
M_{Pl}^4\left(4\dot H+10H^2+\frac{\k}{b^2}\right)&=&-w\rho-\o\sigma+\o\left(\ano+Y\right)+\l\\
\dot\s+6H\s&=&H\left(\ano+Y\right)\\
\dot\rho+(5+3w+2w_\p)H\rho&=&0
\ena
\paragraph*{Fixed point solutions}
We find an inflationary fixed point where the internal space is also staying in an inflationary era, since the two Hubble parameters are equal (see appendix \ref{fixed points} for calculations). The existence
of such fixed point depends on the values of $c_A,c_B,c_Y$, $\k$ and $\o$.

Is is necessary to impose $\o=1/5$, implying that the two pressures characterising the CFT stress--energy tensor must be equal $\s_\p=\s_p$, in order to obtain the following fixed point solutions.
\begin{enumerate}
\item
We write the explicit solution for a flat internal space $\k=0$, where the critical value for the Hubble parameter is given in terms of the anomaly coefficients and of the only dimensionful parameter which is
the 6D Planck mass. This solution restricts the possible values of $c_A$ and $c_B$, coming from the conformal field theory anomaly
\bea
\sH^2&=&-\frac{24}{c_A+2c_B}\mpl\left[24c_Y\pm\sqrt{576c_Y^2+(c_A+2c_B)}\right]\\  
\ssig&=&-\ssigp=-\ssigpp=-\frac{240}{c_A+2c_B}\MPl\left[24c_Y\pm\sqrt{576c_Y^2+(c_A+2c_B)}\right],  \qquad  \srho=0  \non
\ena
For $(24c_Y)^2>-(c_A+2c_B)$ the two roots are real. We must satisfy the condition $(c_A+2c_B)<0$ on the anomaly parameters in order to have two positive $\sH$ critical points (both denoted by zero energy
density $\srho$).
\item
For $\k\ne0$ there exists no fixed points, since $\left(\ano+Y\right)$ and $\left(10\sH^2+\frac{\k}{a^2}\right)$ cannot be both constant unless we enforce the staticity condition on the Hubble parameter
$H_\star=0$ and $\k=0$.
\end{enumerate} 
\paragraph*{Stability analysis}
In this last case of equal scale factors we have again carried the stability analysis in appendix \ref{fixed points}. As a result, we found that the equal scale factor critical point with flat internal
space is an attractor. It could thus represent the eternal acceleration of the universe.
\subsection{Comments}
We want to summarize the interesting features of the critical analysis of sections \ref{brane critical flat}--\ref{brane critical equal}, in the perspective of the comparison with the 7D bulk gravitational
dual description that we illustrated basically in sections \ref{bulk cosmology} and \ref{critical bulk}.
\begin{itemize}
\item[-]
All the critical points we can find in the brane description are characterized by exactly zero value for the localized matter energy density $\rho$ (except for some $\sH=0$ trivial points). To have a non
vanishing energy density it is necessary to introduce an interaction term between the matter fields and the hidden sector fields. The reason for this is that it modifies the conservation equation for $\rho$,
allowing for a non zero time independent solution. This intuitively corresponds to turning on the brane--bulk energy exchange on the bulk gravity side.
\item[-]
In most of the simple critical point solutions, the hidden sector pressure $\ssigp$ is related to the energy density $\ssig$ by $\ssigp=-\ssig$. This indicates a vacuum behavior for the equation of state of
the hidden sector of the holographic dual theory at the inflationary fixed points.
\item[-]
There also appears to exist more than one critical point solution in some of the explicitely examined limits. Since the stability matrix analysis reveals that we can have either stable or saddle points
(depending on the CFT and counterterm parameters $c_A,c_B,c_Y$ and on the Plack mass), we can expect two kinds of behavior (if the number of critical point solution is two). Whenever one of the fixed points
is attractive and the other one is a saddle, we can generally depict a phase portrait such that some of the trajectories are attracted by the stable point, while others can be repulsed by the saddle and go
toward the large density region. If, on the other hand, we get two saddle points, trajectories bend near to the saddles and flow away. We note that the trivial critical point has undefined stability at linear
order in the perturbations, so that it may either attract or repel trajectories in its neighborhood. However, if $H>0$ and $w,w_\p>-1$, late time evolution always is described by $\rho\to0$.
\item[-]
Comparing these results with the bulk cosmology we note that we get both trivial and non trivial accelerating critical points in the CFT description. The non trivial fixed points are associated to the so
called Starobinski branch of the solution to the Einstein, conservation and anomaly equations. Acceleration is produced due to conformal anomaly in higher derivative terms. These accelerating points can be
interpreted to be in correspondence with the accelerating fixed points found in the gravity description with inflow. The trivial fixed points come from the smooth branch, instead, and they are smoothly
connected to the gravity description. An analogous map can be found in the 5D-4D model \cite{Kiritsis:2005bm}.
\item[-]
In the bulk description we only found positive acceleration critical points since $q=H^2$. With the holographic approach, we could in principle also get non constant $H$ critical point solutions. In fact we
could solve the system of Einstein equations on the brane asking time independence for $q=\dot H+H^2,\rho$ and $\s$. We would get a new system of first order (non linear) differential equations in $H$, which
in principle could have non trivial solutions.
\end{itemize}
\section{Brane/bulk correspondence}\label{examples}
We derive some solutions that can illustrate interesting aspects of the cosmological model we considered and in particular of the duality that relates the two descriptions. We will be able to
find explicit expressions by making special simplifying assumptions on the parameters of the holographic theory and on the space--time background. These examples allow us to make a comparison between the
results we will find in the holographic set--up and the expressions we derived in the bulk gravity theory. 

Since we are interested in comparing the two dual approaches, in the sense of AdS/CFT correspondence, we have to derive some expressions for $H$ in terms of the localized matter energy density $\rho$ and
of a mirage density $\chi$. In section \ref{bulk cosmology}, we performed the cosmological analysis in the 7D bulk description, expressing the 3D Hubble parameter $H$ in terms of the localised energy density
$\rho$, reducing the first order ODE in $H$ to an algebraic equation for $H^2$ plus a first order ODE for the mirage density $\chi$. In the holographic dual theory, the mirage density is identified with the
solution to the homogeneous equation associated to the conservation equation for the hidden sector density $\s$. It will thus have the property of obeying to the free radiation conservation equation in $d$
effective dimensions (where $d$ is the effective number of dimensions equal to $4$ in a static compact space background and to $6$ when the $a(t)$ and $b(t)$ scale factors are equal).

To obtain the explicit result for $H^2$ in the brane dual description, however, it is necessary to integrate the differential equation \refeq{sigma anal sol} for the energy density $\s$. It is established
that the anomaly and the trace of the quadratic contribution to the variation of the dual theory action are highly non trivial functions of the Hubble parameters and the spatial curvatures and contain
derivatives of $H,F$ up to order three. So, an analytical integration of the $\s$ conservation equation is in general apparently unachievable. However, it is possible to neglect the $\ano$ and $Y$
contributions if we are in the slowly scaling approximation, which corresponds to a small curvature approximation. This is what we are going to discuss in the following.
\subsection{Slowly scaling approximation}
We give a rough idea on how the correspondence between the brane and the bullk dual theories works. In fact, we will neglect all the higher order terms in the holographic description, which is equivalent to
ask that the Hubble parameter is negligeable with respect to the Planck mass $H^2\ll\mpl$ (i.e. $\dot a\ll M_{Pl}\,a$). In this approximation, all the higher order curvature terms --- including the anomaly
and the trace $Y^\m_\m$ --- can be neglected in favor of contributions proportional to the Einstein tensor. The integration of the $\s$ equation would give terms of the order of $\mpl H^4$ (from $Y$) and
$H^6$ (from $\ano$). Once we plug the result for $\s$ in the Einstein equation \refeq{final system} these terms are suppressed, since the l.h.s. of \refeq{final system} is of order $\Mpl H^2$. 

We anticipate that, as a consequence of the small Hubble parameter approximation on the brane, we get a linear dependence of $H^2$ on the mirage plus visible matter energy densities (the hidden sector density
$\s$ is identified with the mirage density $\chi$). Higher order contributions due to the anomaly and to $Y$ would give rise to higher power dependences in a small density expansion for $H^2$. Since in the
holographic description we truncate to linear order in density, we also keep only linear terms in $\rho$ for the bulk gravity results. The bulk equations for $H$ and $\chi$ \refeq{bulk limits H}--\refeq{bulk
limits rho}, can be formulated independently of $w$ if we ignore higher (quadratic) order terms in $\rho$ and assign a specific value to $w_\p$. Neglecting the second order term in $\rho$ we will only have
one condition to determine $M$ and $V$ in terms of the brane parameter $M_{Pl}$, so that only the ratio $M^{10}/V$ will be identified.

Since in this approximation the quadratic and higher order dependence of $H^2$ on $\rho$ are absent, we don't capture the eventual Starobinsky \cite{Starobinsky:1980te} behavior of the solutions to the
Einstein equations \cite{Vilenkin:1985md}. The higher derivative terms are necessary in that case to calculate the exit from inflation to a matter dominated universe and the subsequent thermalization to
radiation dominated era. In Starobinsky model the higher derivative terms are represented by the type D conformal anomaly contribution to the trace of the stress--energy tensor. Nonetheless, in the 5D RS
holographic dual analysis of \cite{Kiritsis:2005bm}, where these terms are cancelled, stringy corrections like Gauss--Bonnet terms can play the same role. We note that in our set--up, the 6D conformal anomaly
contains suitable higher derivative terms, not only in the total derivative contributions but also in type B anomaly.

We now explain the results to which the slowly scaling approximation leads in some particular limits.
\subsubsection{Equal scale factors}\label{equal comparison}
We start looking at the Einstein, conservation and anomaly equations in the equal scale factor limit. The system of equations \refeq{final system}--\refeq{final system rho} for the theory on the brane takes
the form
\bea\label{final system equal compare}
\Mpl\left(10H^2+3\frac{k}{a^2}+\frac{\k}{a^2}\right)&=&\rho+\s+\l\\
\Mpl\left(4\dot H+10H^2+\frac{k}{a^2}+\frac{\k}{a^2}\right)&=&-w\rho-\o\sigma+\o\left(\ano+Y\right)+\l\\
\dot\s+6H\s&=&H\left(\ano+Y\right)  \label{final system sigma equal compare}\\
\dot\rho+(3(1+w)+2(1+w_\p))H\rho&=&0 \label{final system rho equal compare}
\ena
We note that the set of equations is independent of the hidden sector parameter $\o$ and it is furthermore interesting that the homogeneous equation associated to \refeq{final system sigma equal compare} is
indeed precisely the 6D free radiation equation, independently of the value for $\o$.  In particular, the two conservation equations can be written in the integral form
\bea
\s&=&\chi+a^{-6}\int\intd t\,a^6\,H\left(\ano+Y\right), \qquad \chi=\chi_0\,\left(\frac{a_0}{a}\right)^6\\
\rho&=&\rho_0\,\left(\frac{a_0}{a}\right)^{3(1+w)+2(1+w_\p)}\non
\ena
Plugging the result for $\s$ in the first equation of the system \refeq{final system equal compare} and neglecting the curvature higher order terms that come from the integration of $\left(\ano+Y\right)$,
we obtain the following expression for $H^2$, toghether with the $\rho$ and $\chi$ equations in their differential form
\bea
H^2+\oneover{10}\left(3\frac{k}{a^2}+\frac{\k}{a^2}\right)&=&\oneover{10\Mpl}\left(\rho+\chi\right)+\oneover{10\Mpl}\l\label{H 6D equal linear}\\
\dot\chi+6H\chi&=&0\\
\dot\rho+(3(1+w)+2(1+w_\p))H\rho&=&0\label{rho 6D equal linear}
\ena
It is now easy to compare \refeq{H 6D equal linear}--\refeq{rho 6D equal linear} with the corresponding system of equations in the bulk description of the equal scale factor universe, with zero brane--bulk
energy exchange $T^0_7$ and bulk ``self interaction'' $T^7_7$. The expression for $H^2$ can also be written in a $w$--independent way fixing $w_\p$ (in this particular case we could also include the quadratic
term in $\rho$ in the $w$--independent formulation, i.e. when $w_\p=w$). Neglecting the second order term in the energy densities we obtain
\bea\label{H 7D equal linear}
H^2+\oneover{10}\left(3\frac{k}{a^2}+\frac{\k}{a^2}\right)&=&\frac{2\tcveq V}{5M^{10}}\left(\rho+\chi\right)+\l_{RS}\\
\dot\chi+6H\chi&=&0\\
\dot\rho+(3(1+w)+2(1+w_\p))H\rho&=&0\label{rho 7D equal linear}
\ena

The two systems of equations \refeq{H 6D equal linear}--\refeq{rho 6D equal linear} and \refeq{H 7D equal linear}--\refeq{rho 7D equal linear} perfectly agree at this order in the approximation. The matching
between the scales on the two sides of the duality is then
\bea\label{relation small density equal}
\frac{M^{10}}{V}=4\tcveq\Mpl  \qquad\stackrel{w_\p=w}{\longrightarrow}\qquad  \frac{M^{10}}{V}=\frac{\Mpl}{20}
\ena
As we announced, only the ratio $M^{10}/V$ can be determined, sice we only have one condition to match the two descriptions. When higher order corrections are included in the brane description we would
generally find a matching for both $M$ and $V$, which in principle would depend on the particular CFT parameters ($c_A,c_B,c_Y$). We can guess that the ratio $M^{10}/V$ would not depend on them (indeed, for
the pure RS set--up we get $\Mpl V\propto M^{10}$). When $w_\p=w$ (i.e. when the pressures of the matter perfect fluid relative to the 2D internal space and the 3D space are equal $\p=p$) the coefficient
$\tcv$ becomes $\tcveq=1/80$. It is interesting to note that for $w=w_\p$ the matching exactly reduces to the RS condition for zero effective cosmological constant on the brane $\l_{RS}=0$. Since we are in
the limit $F=H$, it seems natural to have $\p=p$ too.
\subsubsection{Static compact extra dimensions}\label{static comparison}
We consider the static extra dimension limit $F=0$. The Einstein equations plus conservation and anomaly equations in this limit read
\bea\label{final system static compare}
\Mpl\left(3H^2+3\frac{k}{a^2}+\frac{\k}{b_0^2}\right)&=&\rho+\s+\l\\
\Mpl\left(2\dot H+3H^2+\frac{k}{a^2}+\frac{\k}{b_0^2}\right)&=&-w\rho-\o\sigma+\o\left(\ano+Y\right)+\l\\\non\\
\s+3(1+\o)H\s&=&3\o H\left(\ano+Y\right)  \label{final system sigma static compare}\\
\dot\rho+3(1+w)H\rho&=&0 \label{final system rho static compare}
\ena
The homogeneous equation associated to \refeq{final system sigma static compare} is the 4D free radiation equation only if $\o=1/3$, implying that the hidden sector pressure of the internal space $\s_\p$ must
be zero. With this assumption, the two conservation equations become
\bea\label{sigma integrate static compare}
\s&=&\chi+a^{-4}\int\intd t\,a^4\,H\left(\ano+Y\right), \qquad \chi=\chi_0\,\left(\frac{a_0}{a}\right)^4\\
\rho&=&\rho_0\,\left(\frac{a_0}{a}\right)^{3(1+w)}\non
\ena
The results for $\chi$ and $\rho$ agree with the bulk formulation for zero energy exchange. Plugging \refeq{sigma integrate static compare} into \refeq{final system static compare} and neglecting the
curvature higher order term, as we are in the slowly scaling approximation, we find
\bea
H^2+\oneover{3}\frac{\k}{b_0^2}&=&\oneover{3\Mpl}\left(\rho+\chi\right)\label{H 6D static linear}\\
\dot\chi+4H\chi&=&0\\
\dot\rho+3(1+w)H\rho&=&0
\ena
Although conservation equations agree, the Friedmann like equation doesn't give the expected $1/6$ coefficient in front of the $\k/b_0^2$ term in \refeq{H 6D static linear} that instead follows from the
equations on the bulk gravity side
\bea\label{H 7D static linear}
H^2+\oneover{6}\frac{\k}{b_0^2}&=&\frac{2\tcvst V}{3M^{10}}\left(\rho+\chi\right)\\
\dot\chi+4H\chi&=&0\\
\dot\rho+3(w+1)H\rho&=&0
\ena
The bulk equations are derived in the density linear approximation and for vanishing energy exchange $T^0_7$ and $T^7_7$. The coefficient $\tcvst$ can be written in a $w$--independent way if we fix $w_\p$.

As a consequence, the ratio of the two bulk parameters can be identified with
\bea\label{relation small density static}
\frac{M^{10}}{V}=2\tcvst\Mpl  \qquad\stackrel{w_\p=\frac{w+5}{6}}{\longrightarrow}\qquad  \frac{M^{10}}{V}=\frac{\Mpl}{20}
\ena
The $\k/b_0^2$ terms differ in the two dual descriptions (in the static extra dimension background). The matching \refeq{relation small density static} gives a result that depends on the valus of $w,w_\p$ in
a different way if compared to the equal scale factor limit \refeq{relation small density equal}. It is thus interesting to further examine how the matching varies according to the value of the internal space
Hubble parameter. We are indeed going to consider the proportionality ansatz $F=\x H$ to better understand this behavior. We note that in the limit $w_\p=(w+5)/6$ we recover in \refeq{relation small density
static} the RS fine--tuning determining zero effective cosmological constant on the brane $\l_{RS}=0$, since $\tcvst=1/40$.
\subsubsection{Proportional Hubble parameters}\label{proportional hubble comparison}
Following the computations in the last two sections and generalizing them, we derive the set of equations for $H^2,\chi$ and $\rho$ for proportional and small Hubble parameters. Since, as before, $\s=\chi$ if
we neglect higher order terms in the small curvature approximation, equations \refeq{final system}--\refeq{final system rho} lead to the following equations
\bea
H^2+\oneover{(\x_b^2+6\x_b+3)}\frac{\k_b}{b^2}&=&\oneover{(\x_b^2+6\x_b+3)\Mpl}\left(\rho+\chi\right)\label{H 6D prop linear}\\
\dot\chi+d_{\x_b} H\chi&=&0\\
\dot\rho+w_{\x_b} H\rho&=&0
\ena
where $d_{\x_b}\equiv3(1+\o)+3\x_b(1-\o)$, $w_{\x_b}\equiv3(1+w)+2\x_b(1+w_\p)$ and $\x_b$ is the proportionality factor $F=\x_b H$ (or $b=a^{\x_b}$). The bulk equations where derived in \refeq{bulk limits H
prop}--\refeq{bulk limits rho prop}. For the moment we will keep two different proportionality factors: in the bulk $F=\x_B H$ as in section \ref{proportional hubble} (putting zero $T^7_7$ and $T^0_7$) 
\bea\label{H 7D prop linear}
H^2+\oneover{(\x_B^2+3\x_B+6)}\frac{\k_B}{b^2}&=&\frac{2\tcvx V}{(2\x_B+3)M^{10}}\left(\rho+\chi\right)\\
\dot\chi+d_{\x_B} H\chi&=&0\\
\dot\rho+w_{\x_B} H\rho&=&0
\ena
We have used the following definitions: $d_{\x_B}\equiv6(\x_B^2+2\x_B+2)/(2\x_B+3)$, $w_{\x_B}\equiv3(1+w)+2\x_B(1+w_\p)$ and we recall that $\tcvx$ is $\tcvx\equiv c_V/(w_{\x B}-d_{\x B})$, where
$c_V=(31w-6w_\p-5)/400$.

In order for the two descriptions to be equivalent w.r.t. the $\chi$ and $\rho$ differential equations (which don't get any correction from higher order contributions), we have to put $\x_b=\x_B=\x$, as it
was expected. Besides, the parameter relating $\s_\p$ to $\s_p$ must be $\o=1/(2\x+3)$. However, assuming an equal proportionality relation on the two sides of the duality, the coefficient of the $\k$ term in
\refeq{H 6D prop linear} and \refeq{H 7D prop linear} differ if the two curvatures in the bulk and brane descriptions are equal, unless $\x=1$. So, the only set--up that predicts the same effective
spatial curvature for the internal space in the brane and bulk descriptions is the equal scale factor background (neglecting higher order corrections). However, we can determine an effective spatial curvature
for the internal space in the brane description, given by $\k_b=(\x^2+6\x+3)\k_B/(\x^2+3\x+6)$.

The matching for the scales of the two dual theories is given by
\bea\label{relation small density prop}
\frac{M^{10}}{V}=\frac{\x^2+6\x+3}{2\x+3}2\tcvx\Mpl  
\ena
It is always possible to choose a $w_\p$ such that the matching relation \refeq{relation small density prop} gives the RS fine--tuning condition $\Mpl=20M^{10}/V$. Otherwise, missing the fine--tuning would
amount to introducing a non vanishing effective cosmological constant on the RS brane.
\section{Non--conformal and interacting generalization}\label{general}
To examine the general cosmological evolution that reflects the presence of energy exchange between the brane and the bulk in the seven dimensional bulk picture and the non zero value of the bulk component
of the stress tensor $T^7_7$, we will drop the assumption of having a conformal non interacting field theory living on the brane. Intuitively, a non vanishing $T^0_7$ in the bulk description corresponds to
interactions between the gauge theory and the visible matter. The diagonal $T^7_7$ component appears in the brane description as dual to a new trace term spoiling the conformal invariance. The generalization
of the 6D RS dual action will be modified adding the new interaction term and substituing the hidden sector CFT with a strongly coupled gauge theory (SCGT). The following analysis will be done in analogy with
the 4D holographic dual generalization of the 5D RS cosmology exposed in \cite{Kiritsis:2005bm}.

Using the notations of section \ref{dual} we write the generalized action as
\bea\label{action 7D gen}
S_{gen}=S_{SCGT}+S_R+S_{R^2}+S_{R^3}+S_{m}+S_{int}
\ena
where the new entry is the interaction term $S_{int}$ and $S_{CFT}$ has been changed into $S_{SCGT}$.
The strongly coupled fields can be integrated out, transforming the sum of the strongly coupled theory action plus the interaction term into an effective functional of the visible fields (and of the metric)
$W_{SCGT}$. As a result, the action \refeq{action 7D gen} becomes
\bea
S_{gen}=W_{SCGT}+S_R+S_{R^2}+S_{R^3}+S_{m}
\ena

As in the conformal non interacting case, we are now ready to calculate the general 6D equations of motion for the holographic generalized RS cosmology. 
\subsection{Generalized evolution equations}
The stress--energy tensors are defined in an analogous way
\bea
T_{\m\n}&=&\oneover{\sqrt{-\g}}\frac{\d S_m}{\d\g^{\m\n}},\quad W_{\m\n}=\oneover{\sqrt{-\g}}\frac{\d W_{SCGT}}{\d\g^{\m\n}}\non\\
Y_{\m\n}&=&\oneover{\sqrt{-\g}}\frac{\d S_{R^2}}{\d\g^{\m\n}},\quad Z_{\m\n}=\oneover{\sqrt{-\g}}\frac{\d S_{R^3}}{\d\g^{\m\n}}
\ena
\bea
V_{\m\n}&=&W_{\m\n}+Y_{\m\n}+Z_{\m\n}, \qquad Y^\m_\m=Y\non
\ena
and the Einstein equation, the (non) conservation conditions, the anomaly equation read
\bea\label{Fried gen}
M_{Pl}^4G_{\m\n}&=&T_{\m\n}+W_{\m\n}+Y_{\m\n}+V_{\m\n}\non\\
\nabla^\n T_{\m\n}&=&T\non\\
\nabla^\n V_{\m\n}&=&-T\non\\
V^\m_\m&=&\ano+X+Y
\ena
The total stress--energy tensor is still conserved. Taking account of the interactions between the hidden theory and the matter generally amounts to have non separately conserved $V_{\m\n}$ and $T_{\m\n}$.
This is reflected by the introduction of a non homogenous term in the conservation equations. The anomaly equation contains the general expression for the conformal anomaly in six dimensions $\ano$ and
the trace term $Y$.  Furthermore, it gets modified including an extra term $X$ that accounts for classical and quantum breaking of the conformal symmetry in a FRW plus compact space background. The
stress--energy tensors are parametrized as before
\bea\label{gen stress-en}\begin{array}{rclrclrcl}
T_{00}&=&\rho(t),\quad T_{ij}&=&p(t)\,\g_{ij}, \quad T_{ab}&=&\p(t)\,\g_{ab}\\
V_{00}&=&\s(t),\quad V_{ij}&=&\s_p(t)\,\g_{ij}, \quad V_{ab}&=&\s_\p(t)\,\g_{ab}\\
\end{array}\ena

The consequent changes in the equations written in terms of the Hubble parameters, of the energy densities and pressures are the following. The Friedmann equations remain the same (as for the non interacting
conformal case)
\bea\label{gen Friedmann eqs}
M_{Pl}^4\left(3H^2+6H\,F+F^2+3\frac{k}{a^2}+\frac{\k}{b^2}\right)&=&\rho+\s+\l\non\\
M_{Pl}^4\left(2\dot H+3H^2+4H\,F+2\dot F+3F^2+\frac{k}{a^2}+\frac{\k}{b^2}\right)&=&-p-\s_p+\l
\ena
the conservation equations now involve the quantity $T$
\bea\label{gen conservation eqs}
\dot\s+3(\s+\s_p) H+2(\s+\s_\p)F&=&T\non\\
\dot\rho+3(\rho+p) H+2(\rho+\p)F&=&-T
\ena
and the anomaly equation includes the conformal breaking term, as a consequence of the masses and $\b$--functions of the strongly coupled gauge theory
\bea\label{gen anomaly eq}
&&\s-3\s_p-2\s_\p=\ano+Y+X
\ena
$X$ has to be written in terms of the $\b$-functions and operators of the SCGT and matter theory.
Taking the same ansatz for the pressures as for the non interacting conformal theory 
\bea
\s_\p&=&\O\s_p \quad (\o^{-1}\equiv3+2\O)\non\\
p&=&w\rho,\qquad \p=w_\p\rho
\ena
and using the anomaly equation to eliminate $\s_p=\o\left(\s-\ano-Y-X\right)$ from the set of remainig equations, we get
\bea
M_{Pl}^4\left(3H^2+6H\,F+F^2+3\frac{k}{a^2}+\frac{\k}{b^2}\right)&=&\rho+\s+\l  \label{general H ODE 1}\\
M_{Pl}^4\left(2\dot H+3H^2+4H\,F+2\dot F+3F^2+\frac{k}{a^2}+\frac{\k}{b^2}\right)&=&-w\rho-\o\s+\l+\o\left(\ano+Y+X\right) \non\\
\ena\bea
\dot\s+\left[3(1+\o)H+3(1-\o)F\right]\s&=&\left[3\o H+(1-3\o)F\right]\left(\ano+Y+X\right)+T  \label{general sigma ODE}\\
\dot\rho+\left[3(1+w)H+2(1-w_\p)F\right]\rho&=&-T  \label{general rho ODE}
\ena

The cosmological evolution described by these differential equations which include non conformality (represented by the $X$ term) and matter/hidden sector interactions (related to the $T$ term) could be now
investigated. In the spirit of AdS/CFT correspondence, the CFT generalization amounts to introducing non trivial dynamics in the bulk and brane--bulk energy exchange (and bulk self--interaction) in the 7D
picture. The bulk cosmology with $T^0_7$ parameter turned on has been analyzed in section \ref{critical bulk}.
\subsection{Critical points and stability}
The fixed points can be derived as we have done in appendix \ref{fixed points} for the conformal non interacting theory and their stability can then be studied for specific theories. We won't discuss this
topic here. Since the new  deformation parameters $X,T$ depend on the 6D space--time curvature, they contain functions of the Hubble parameters and spatial curvatures and of the intrinsic energy scale of
the background, the $AdS_7$ radius (or $M_{Pl}$). They will thus in general modify the equations for the fixed points and their stability in a sensible way, depending on the specific generalization one wants
to consider. 

We will instead try to understand how the comparison with the bulk description gets changed when we go to the generalized scenario. This will be the subject of the next subsection.
\subsection{Comparison to 7D cosmology with energy exchange in slowly scaling regime}
As we have done for $T=X=0$, we want to illustrate some explicit examples with the aim of understanding the peculiar features of this cosmological model and its two dual descriptions. We will therefore make
some assumptions simplifying the set of equations including Einstein, conservation and anomaly equations. First of all, we are going to neglect terms containing higher orders in the background curvature ---
namely the anomaly coming from $S_{R^3}$ and the trace contribution $Y$ coming from the second order action $S_{R^2}$. 
\paragraph*{Equal scale factors}
The correspondence works as in the conformal non interacting analysis of subsection \ref{equal comparison}. On the brane side (referring to eqs \refeq{general H ODE 1}--\refeq{general rho ODE}), the slowly
scaling approximation leads to the equations
\bea
H^2+\oneover{10}\left(3\frac{k}{a^2}+\frac{\k}{a^2}\right)&=&\oneover{10\Mpl}\left(\rho+\chi\right)+\oneover{10\Mpl}\l\label{H 6D equal linear gen}\\
\dot\chi+6H\chi&=&HX+T\\
\dot\rho+(3(1+w)+2(1+w_\p))H\rho&=&-T  \label{rho 6D equal linear gen}
\ena
To get the bulk description expressions for $H,\rho,\chi$ we truncate equations \refeq{bulk limits H}--\refeq{bulk limits rho} to first order in the density, neglecting $\rho/V$ w.r.t. order 1 terms 
\bea\label{H 7D equal linear gen}
H^2+\oneover{10}\left(3\frac{k}{a^2}+\frac{\k}{a^2}\right)&=&\frac{2\tcveq V}{5M^{10}}\left(\rho+\chi\right)+\l\\
\dot\chi+6H\chi&=&2T^0_7-\frac{40M^5}{V}HT^7_7\\
\dot\rho+(3(1+w)+2(1+w_\p))H\rho&=&-2T^0_7
\ena

The matching with the system of equations on the brane \refeq{H 6D equal linear gen}--\refeq{rho 6D equal linear gen} is exact if we have the following relations among the brane and bulk parameters
\bea
\Mpl&=&\frac{M^{10}}{2\tcveq V}  \quad\stackrel{w_\p=w}{\longrightarrow}\quad  \Mpl=20\frac{M^{10}}{V}\\
T&=&2T^0_7\\
X&=&-\frac{M^5}{2\tcveq V}T^7_7  \quad\Longrightarrow\quad  \frac{X}{\Mpl}=-2\frac{T^7_7}{M^{10}} 
\ena
In the previous equations we have also explicitely evaluated the matching for equal matter pressures $w_\p=w$, that gives the RS fine--tuning $\l_{RS}=0$, as in the non interacting conformal theory.
\paragraph*{Static compact extra dimensions}
The condition of static internal space $F=0$, toghether with the small Hubble parameter approximation, brings equations \refeq{general H ODE 1}--\refeq{general rho ODE}, relative to the brane description, and
equations \refeq{bulk limits H}--\refeq{bulk limits rho}, relative to the bulk description, in the form
\bea
H^2+\oneover{3}\frac{\k}{b_0^2}&=&\oneover{3\Mpl}\left(\rho+\chi\right)\label{H 6D static linear gen}\\
\dot\chi+4H\chi&=&HX+T\\
\dot\rho+3(1+w)H\rho&=&-T
\ena
(where we assumed $\o=1/3$) and
\bea\label{H 7D static linear gen}
H^2+\oneover{6}\frac{\k}{b_0^2}&=&\frac{2\tcvst V}{3M^{10}}\left(\rho+\chi\right)\\
\dot\chi+4H\chi&=&2T^0_7-\frac{40M^5}{V}HT^7_7\\
\dot\rho+3(w+1)H\rho&=&-2T^0_7
\ena

The parameters in the gauge and gravity descriptions are thus related by the following expressions
\bea
\Mpl&=&\frac{M^{10}}{4\tcvst V}  \quad\stackrel{w_\p=\frac{w+5}{6}}{\longrightarrow}\quad  \Mpl=20\frac{M^{10}}{V}  \label{matching M equal general linear}\\
T&=&2T^0_7\\
X&=&-\frac{M^5}{2\tcvst V}T^7_7  \quad\Longrightarrow\quad  \frac{X}{\Mpl}=-\frac{T^7_7}{M^{10}}  \label{matching X equal general linear}
\ena
For $w_\p=(w+5)/6$ we get the zero effective cosmological constant on the RS brane, as before.
\paragraph*{Proportional Hubble parameters}
In the limit of proportional Hubble parameters or equivalently scale factors related by $b=a^{\x_b}$, we use the set of equations \refeq{general H ODE 1}--\refeq{general rho ODE}, substituing $F=\x_bH$,
$\k\to\k_b$, and expanding in the slowly scaling approximation
\bea
H^2+\oneover{(\x_b^2+6\x_b+3)}\frac{\k_b}{b^2}&=&\oneover{(\x_b^2+6\x_b+3)\Mpl}\left(\rho+\chi\right)\label{H 6D prop linear gen}\\
\dot\chi+d_{\x_b} H\chi&=&\left(3\o+\x_b(1-3\o)\right)HX+T\\
\dot\rho+w_{\x_b} H\rho&=&-T
\ena
As before $d_{\x_b}\equiv3(1+\o)+3\x_b(1-\o)$, $w_{\x_b}\equiv3(1+w)+2\x_b(1+w_\p)$. The bulk dynamics is described by \refeq{bulk limits H prop}--\refeq{bulk limits rho prop} with $F=\x_BH$, $\k\to\k_B$
\bea\label{H 7D prop linear gen}
H^2+\oneover{(\x_B^2+3\x_B+6)}\frac{\k_B}{b^2}&=&\frac{2\tcvx V}{(2\x_B+3)M^{10}}\left(\rho+\chi\right)\\
\dot\chi+d_{\x_B} H\chi&=&2T^0_7-\frac{40M^5}{V}HT^7_7\\
\dot\rho+w_{\x_B} H\rho&=&-2T^0_7
\ena
with $d_{\x_B}\equiv6(\x_B^2+2\x_B+2)/(2\x_B+3)$, $w_{\x_B}\equiv3(1+w)+2\x_B(1+w_\p)$.

If $w$ and $w_\p$ are the same on both sides of the duality, then we must make the identification $\x_b=\x_B=\x$ to match $\rho$ equations. As a consequence, $\o=1/(2\x+3)$ in order to have agreement for the
mirage density (non) conservation equations and $\k_b=(\x^2+6\x+3)\k_B/(\x^2+3\x+6)$. With these conditions, the comparison between the two sets of equations thus gives the following matching relations
\bea
\Mpl&=&\frac{2\x+3}{\x^2+6\x+3}\frac{M^{10}}{2\tcvx V}  \\
T&=&2T^0_7\\
X&=&-\frac{2\x+3}{2\x^2+3}\frac{M^5}{2\tcvx V}T^7_7  \quad\Longrightarrow\quad  \frac{X}{\Mpl}=-\frac{\x^2+6\x+3}{2\x^2+3}\frac{T^7_7}{M^{10}}  \label{matching X equal general linear}
\ena
\section{Summary and conclusions}\label{summary}
In the context of holographic cosmology, we have investigated the specific background of 7D RS gravity, including an energy exchange interaction between brane and bulk. Some novel features arise both on
the bulk side of the duality and in the conformal holographic theory on the brane. In particular, we found distinctive results in the comparison between the two descriptions that need a better understanding.
The originality with respect to the 5D/4D holographic cosmology \cite{Kiritsis:2002zf,Kiritsis:2005bm} is due to the compactification over a 2D internal space, around which we wrap the 5--brane. The 6D
space--time filled by the brane acquires a dishomogeneity that distinguishes the 3D visible space from the 2D internal directions. Evolution can generally be different in the two spaces and pressures are
individually related to the energy density by the usual ansatz $p=w\rho$ and $\p=w_\p\rho$ ($p$ and $\p$ are respectively the 3D and 2D pressures), with $w_\p\neq w$ in general.

Concerning the gravity theory in the bulk, we studied the Friedmann like equation that comes along with the introduction of a mirage energy density satisfying to the non homogeneous radiation equation in some
effective number of dimensions (which is six for equal scale factors in both 3D and 2D spaces and is four when the internal space is static). The bulk cosmological evolution is then determined by the
Friedmann equation and by the (non) conservation equations for the mirage density and the localized matter density on the 5--brane. Making use of some simple ansatz for the evolution of the 2D
compactification space (such as putting the corresponding Hubble parameter $F$ equal to the Hubble parameter of the visible space $H$ or to make it vanish) we found a wide spectrum of possible cosmologies
that reduce to the RS vacuum in the absence of matter (i.e. we imposed the RS fine--tuning $\l_{RS}=0$).

Assuming small density approximation, we have described the explicit analytical solution in case of radiation dominated universe. The Hubble parameter evolves as in an effective 6D (4D) radiation dominated
era for equal scale factors (static compact extra dimensions), independently of the form of the brane--bulk energy exchange. The effective 4D mirage and matter energy densities obey to the 4D free radiation
equation in absence of energy exchange. If energy flows from the brane into the bulk, the 4D localized energy density is suppressed in time, in favor of the mirage density, even with zero mirage initial
condition. For influx to the bulk, the 4D matter energy density apparently grows unbounded (if $T$ is linear in $\rho$, otherwise energy density may go to zero for suitable values of the theory parameters),
eventually diverging at a finite time. The small density approximation must break down and the full analysis is needed. On the other hand, still for small densities but generic perfect fluid equation of state
(non necessarily pure vacuum energy) and energy influx, we found inflationary fixed point solutions that are stable for a large class of energy exchange parametrizations $T=A\rho^\n$. These thus represent
stable de Sitter solutions for (the four visible space--time directions of) our universe. We moreover argued that, differently than in the 5D RS approach \cite{Kiritsis:2002zf}, we may have a stable de Sitter
critical point solution even for energy outflowing to the bulk, in the case of equal scale factors with $w<-1/3$ (and $\n=1$). For dynamical compactification (i.e. $F=\x H$, $\x<0$) \cite{Mohammedi:2002tv} we
could in principle also get an outflow stable inflationary fixed point. We note that the 4D mirage energy density evolution without energy exchange is governed by the effective 4D free radiation equation only
in the two limits of equal scale factors $F=H$ and static compact extra dimensions $F=0$. 

Having dropped the small density approximation, more elaborate models of cosmologies were developed. The number of possible inflationary critical point solutions can be larger than one, depending on the
parametrization for the brane--bulk energy exchange. For energy influx we showed the 6D picture of a scenario with two fixed points, where trajectories in the phase space can either always be characterized by
positive acceleration, either remain at all time with negative acceleration, or alternate acceleration and deceleration phases. The portraits are rigorously valid for the effective 4D energy density in the
case of static internal space (up to a constant rescaling). If we have equal scale factors, the evolution equations become much more complicated functions of the 4D densities and the computation is beyond the
scope of the paper. For $\n=1$ there seems to exist only the trivial critical point characterized by vanishing Hubble parameter, so that the energy density should grow without bounds as predicted by the
small density approximation, until the full analysis is needed. For energy outflow, the 6D energy density localized on the 5--brane decreases and the trajectories in the phase space go toward the trivial
fixed point, eventually passing through an accelerated era. The effective 4D picture may differ from this description in the equal scale factor case, since it would be possible in principle not to have
decreasing density at all times.

The rather detailed study of the various cosmologies emerging from the 7D RS model with energy exchange has been embodied in the context of the AdS/CFT correspondence. We have examined the role played in the
holographic critical point analysis by the 6D anomalous CFT coupled to 6D gravity --- with the addition of the higher order counterterms to the dual action. The dual theory on the brane is conformal
(classically) and non interacting (with the matter theory on the brane). This CFT would then correspond to the RS set--up with no energy exchange. Despite this fact, we may find inflationary critical point
solutions, depending on the anomaly parameters and on the coefficient of the second order counterterm (in the curvature). All this fixed points are characterized by zero matter energy density. Clearly,
neglecting all the higher order contributions (including the anomaly $\ano$ and the trace term $Y$) we recover the trivial fixed points of the pure RS gravity background.

The comparison between the two dual descriptions has been achieved in the approximation where all the higher order terms can be neglected, i.e. for small Hubble parameters. Since higher order terms are
truncated, we cannot access to the typical non conventional $\rho^2$ dependence in the expression for $H^2$ --- only linear terms are present in this approximation. Comparing the bulk Friedmann equation with
the corresponding equation derived in the holographic description, we have to match the ratio $M^{10}/V$ ($M$ is the 7D Planck Mass and $V$ is the tension of the RS brane) in terms of the 6D Planck mass
$\Mpl$ (the 4D Planck mass $M_{(4)}$ is related to $M_{Pl}$ by $M_{(4)}^2=V_{(2)}\Mpl$). The RS fine--tuning $\Mpl=20M^{10}/V$ is restored when we recover homogeneity in the background, imposing $F=H$ and
$w_\p=w$. With these assumptions indeed, the matching is exact also with respect to the spatial curvature terms. As we move away from homogeneity, we have to define an effective spatial curvature for the
compact extra dimensions in the holographic description. The matching between the scales of the theories reflects the RS fine--tuning for a specific value of the matter pressure in the internal space
(determined by $w_\p$), depending on the proportionality factor $\x$ relating the two Hubble parameters $F(t)=\x H(t)$, or $b(t)=a^\x(t)$.  We finally matched the evolution equations in the generalized
holographic dual theory with the general bulk description. The interactions between hidden and visible sectors encode the dynamics of the brane--bulk energy exchange, $T^0_7\ne0$, on the bulk gravity side,
while the breaking of conformal invariance (via non zero $\b$--functions or masses) amounts to turning on the bulk ``self--interaction'', $T^7_7\ne0$. 

The 7D RS background has been quite accurately studied on the bulk side, though many profound cosmological aspects have not been explored. There could be space, however, to fit the cosmological evolution of
the universe in this model, since one of the stable de Sitter critical point solutions we found could represent the actual accelerated era. Besides, trajectories can end into the stable point first passing
through a decelerated phase representing the matter or radiation dominated universe. Another accelerated era may be present at early times, eventually corresponding primordial inflation. Still, there is no
rigorous construction of such a precise evolution.  The holographic dual theory could also give an interesting cosmological description of the brane--world. It would be interesting to exploit Starobinky
argument of graceful exit from primordial inflation via higher derivatives term in this context. In conclusion, we showed the basics of the brane--bulk duality in 7D RS background with energy exchange and of
its cosmological features, but many further questions can be adressed within the framework of the 7D RS holographic cosmology. 
\begin{acknowledgments}
I would like to thank A. Butti, R. Casero, U. Gursoy, A. Mariotti, M. Musso, A. Paredes, M. Pirrone, A. Romagnoni, G. Tartaglino--Mazzucchelli, and in particular F. Nitti and K. Skenderis for useful
suggestions and discussions. I am especially greatful to E. Kiritsis for his fundamental contributions and to S. Penati for carefully reading the paper and giving helpful suggestions.

This work has been partly supported by INFN, PRIN prot. 2005024045-002 and the European Commission RTN program MRTN-CT-2004-005104 and by ANR grant NT05-1-41861, INTAS grant, 03-51-6346, RTN contracts
MRTN-CT-2004-005104 and MRTN-CT-2004-503369, CNRS PICS 2530 and 3059 and by a European Excellence Grant, MEXT-CT-2003-509661.
\end{acknowledgments}
\appendix
\section{Conformal anomaly and traces in six dimensions}\label{anomaly}
\paragraph*{Conformal anomaly} The conformal anomaly for 6D theories has been studied in \cite{Deser:1993yx}. It can be derived using AdS/CFT and the gravitational renormalization procedure as in
\cite{Henningson:1998gx,Skenderis:1999nb,deHaro:2000xn}.

In our notations, the general expression for the trace anomaly in a six dimensional CFT is 
\bea\label{gen anomaly} 
\ano=-\left(c_A E_{(6)}+c_B I_{(6)}+ \nabla_\m J^\m_{(5)}\right)
\ena 
$E_{(6)}$ is the Euler density in six dimensions (type A anomaly), $I_{(6)}$ is a fixed linear combination of three independent Weyl invariants of dimension six (type B anomaly) and $\nabla_iJ^\m_{(5)}$ is an
linear combinations of the Weyl variation of six independent local functionals (type D anomaly), so that at the end we have eight free coefficients in the general form of the anomaly, depending on the
specific CFT. The type D anomaly is a trivial (it is a total derivative, indeed) scheme dependent term that can be cancelled by adding local covariant counteterms to the action \cite{Bastianelli:2000rs}. 

For our metric we obtain as a result that $E_{(6)}$ depends on the Hubble parameters and on their time derivatives up to order one (that is, up to the second time derivative of the scale factors). $I_{(6)}$
instead depends on $H$ and $F$ time derivatives up to order three, and so does the divergence term. To be more specific, $I_{(6)}$ is made up by three contributions $I_1,I_2,I_3$, with fixed coefficients; the
first two are two different contractions of three Weyl tensors (and contain only derivatives of the Hubble parameters up to order one), while in $I_3$ there are second order derivatives of the Weyl tensor
(i.e. third order time derivatives of the Hubble parameters). 

For a 6D FRW background (i.e. requiring homogeneity in all six dimensions) the sum of type A plus type B anomalies depends on the Hubble parameters only up to the first time derivative, while the type D
anomaly contains time derivatives up to order three.

In terms of the Riemann tensor, Ricci tensor and scalar curvature, the A, B and D contributions to the anomaly read \cite{Henningson:1998gx}
\bea
E_{(6)}&=&\frac{1}{6912}E_0\non\\
I_{(6)}&=&\frac{1}{1152}\left(-\frac{10}{3}I_1-\oneover{6}I_2+\oneover{10}I_3\right)\non\\
J^\m_{(5)}&=&-\frac{1}{1152}\left[-R^{\m\n\rho\s}\nabla^\t R_{\t\n\rho\s}+2\left(R_{\n\rho}\nabla^\m R^{\n\rho}-R_{\n\rho}\nabla^\n R^{\m\rho}\right)\right]+\non\\
&&-\oneover{2880}R^{\m\n}\nabla_\n R+\oneover{5760}R\nabla^\m R
\ena
where
\bea
E_0&=&K_1-12K_2+3K_3+16K_4-24K_5-24K_6+4K_7+8K_8\non\\
I_1&=&\frac{19}{800}K_1-\frac{57}{160}K_2+\frac{3}{40}K_3+\frac{7}{16}K_4-\frac{9}{8}K_5-\frac{3}{4}K_6+K_8\non\\
I_2&=&\frac{9}{200}K_1-\frac{27}{40}K_2+\frac{3}{10}K_3+\frac{5}{4}K_4-\frac{3}{2}K_5-3K_6+K_7\non\\
I_3&=&K_1-8K_2-2K_3+10K_4-10K_5-\frac{1}{2}K_9+5K_{10}-5K_{11}\non
\ena
and
\bea
(K_1,\dots,K_{11})&=&\left(R^3,RR_{\m\n}R^{\m\n},RR_{\m\n\rho\s}R^{\m\n\rho\s},R_\m{}^\n R_\n{}^\rho R_\rho{}^\m,R^{\m\n}R^{\rho\s}R_{\m\rho\s\n},\right.\cr
&&R_{\m\n}R^{\m\rho\s\t}R^\n{}_{\rho\s\t},R_{\m\n\rho\s}R^{\m\n\t\l}R^{\rho\s}{}_{\t n},R_{\m\n\rho\s}R^{\m\t\l\s}R^\n{}_{\t\l}{}^\rho,\cr
&&\left.R\Box R,R_{\m\n}\Box R^{\m\n},R_{\m\n\rho\s}\Box R^{\m\n\rho\s}\right)\non
\ena
In the analysis of the solutions to the Friedmann equations we plug in the specific expression for the Riemann tensor obtained considering our ansatz (\ref{metric 4+2}) for the metric. But before doing
this, we use the anomaly equation and some standard assumptions on the pressures that parametrize the stress--energy tensors to manipulate our system of differential equations.

To give an explicit result for the conformal anomaly in the specific case of 6D CFT on curved space--time, with the ansatz \refeq{metric 4+2} for a 4D FRW plus a 2D compact internal space background, we write
the type A contribution, in terms of the 3D and 2D spaces Hubble parameters $H,F$ and spatial curvatures $k,\k$:
\bea
E_{(6)}&=&-\oneover{48}\bigg\{\kb\left(\dot H+H^2\right)\left(H^2+\ka\right)+F^2\left(\dot H+H^2\right)\left(3H^2+\ka\right)+\non\\
&&+2\left(\dot F+F^2\right)\left(H^2+\ka\right)\bigg\}
\ena
The type B and D contributions have a more complicated form and we write them when it is necessary, in the specific limits we consider throughout the paper. 

For the (0,2) SCFT dual to the $N$ M5 background, the anomaly coefficients are given by $c_A=c_B=4N^3/\p^3$ \cite{Henningson:1998gx}.
\paragraph*{Counteterm traces}
The dual RS theory action contains the three counterterms $S_1$, $S_2$, $S_3$ written in \refeq{def count 0}. Varying these contributions w.r.t. the six dimensional induced metric $\g_{\m\n}$ on the brane
yields
~\footnote{Just for this formula we set $\ell=1$ to simplify the notation}
 \cite{deHaro:2000xn}
\bea \label{counterT}
T^{ct}_{\m\n}&=&-2M^5\left(5\g_{\m\n}+\frac{1}{4}\left(R_{\m\n}-\oneover{2}R\g_{\m\n}\right)\right.\non\\ 
&&\left.-\frac{1}{32}\left[-\Box R_{\m\n}+2R_{\m\s\n\rho}R^{\rho\s}+\frac{2}{5}\nabla_\m \nabla_\n R-\frac{3}{5}RR_{\m\n}\right.\right.\non\\
&&\left.\left.-\oneover2\g_{\m\n}\left(R_{\rho\s}R^{\rho\s}-\frac{3}{10}R^2-\frac{1}{5}\Box R\right)\right]-T^a_{\m\n}\log\e\right)
\ena
Where $T^a_{\m\n}$ is a traceless tensor of cubic order in the curvature. The trace of the conformal variation of $S_1$ (corresponding to the linear part of \refeq{counterT} in the curvature) gives a term
proportional to the Einstein tensor. The variation of the $S_3\propto \int\sqrt{-\hat g_{(0)}}a_{(6)}$ action (related to $T^a_{\m\n}$, where we introduced $a_{(3)}$ as in the standard notation of
\cite{Skenderis:1999nb}) is traceless because it is proportional \cite{deHaro:2000xn} to the traceless tensor $h_{(6)\m\n}$ that enters into the parametrization of the metric \refeq{FG para} due to Fefferman
and Graham. However, the variation under conformal transformations of the cutoff dependent counterterm is non trivial and is the only contribution to the conformal anomaly, so that $\ano\propto a_{(6)}$ (see
for instance \cite{deHaro:2000xn,Mazzanti:2007yb} for a more detailed derivation). Finally, the trace of $S_2$ (equal to the trace of the quadratic contributions in \refeq{counterT}) is 
\bea
Y=\oneover{32}M^5\ell\left(R^{\m\n}R_{\m\n}-\frac{3}{10}R^2\right)
\ena
which can be expressed in terms of the Hubble parameters $H,F$ and of the spatial curvatures $k,\k$ of the (4D FRW + 2D compact space) background \refeq{metric 4+2} as
\bea
Y&=&-\frac{2M^5\ell}{160}\bigg\{-3\frac{k^2}{a^4}+6\frac{k}{a^2}\left(3\frac{\k}{b^2}+F^2+8FH+3H^2+6\dot F+4\dot H\right)+\non\\
&&+2\frac{\k}{b^4}+2\frac{\k}{b^2}\left[-\left(F-6H\right)\left(F+3H\right)+\dot F+9\dot H\right]+\non\\
&&-3F^4+48F^3H+6FH\left(21H^2+7\dot F+13\dot H\right)+F^2\left(111H^2-4\dot F+24\dot H\right)+\non\\
&&+3\left[6H^2-\left(\dot F-\dot H\right)^2+2H^2\left(7\dot F+3\dot H\right)\right]\bigg\}
\ena

We define $c_Y\equiv M^5\ell/32\mpl$ wich is given as a function of the number $N$ of M5--branes in the gravity background by $c_Y=\sqrt{2N^3/\p^3}$.
\section{Fixed points in the holographic description}\label{fixed points}
In this appendix, we are going to look for the existence of inflationary points for our specific holographic model and to find what kind of dependence they have on the parameters of the theory.  We will also
study the stability matrix determining --- in some special limits --- whether the critical points are stable or saddles.

In the calculations, we suppose that the effective cosmological constant on the brane $\l$ is null.
\subsection{Flat compact extra dimensions}
We start by considering the case of (locally) flat internal space, which could be for example a two--torus. 
The three spatial dimensions of the 4D FRW are already supposed to be flat, so that the system of equations of motion simplify, having dropped the terms proportional to both spatial curvatures.

The general flat extra dimension fixed points ($F_\star\neq0$) are not easy to characterize. We choose to analyze the case in which the extra dimensions Hubble parameter is zero at the fixed point, meaning
that the fixed point represents a universe with static flat extra dimensions.   
\subsubsection{$F_\star\ne0,\;\o\ne\frac{1}{5}$}
\paragraph*{Fixed point solutions}
Looking for the solution to the Friedmann plus conservation set of equations with constant Hubble parameters ($H\equiv H_\star,F\equiv F_\star$) and zero curvatures ($k=\k=0$), we have to
consider the simplified system of equations (where we have already solved the equation for $\s$)
\bea
&&M_{Pl}^4\left(3H_\star^2+6H_\star\,F_\star+F_\star^2\right)-\left[3\o H_\star+\left(1-3\o\right)F_\star\right]\left(\stano+\stY\right)
=\l  \label{Friedmann rho}\\
&&M_{Pl}^4\left(3H_\star^2+4H_\star\,F_\star+3F_\star^2\right)-\o\left(3H_\star+2F_\star\right)\left(\stano+\stY\right)
=\l  \label{Friedmann p}\\
&&\s_\star=\left[3\o H_\star+\left(1-3\o\right)F_\star\right]\stano=\Mpl\left(3\sH^2+6\sH\sF+\sF^2\right)-\l  \label{sigma}\\
&&\s_{p\star}=-\o\left(3H_\star+2F_\star\right)\left(\stano+\stY\right)=-\Mpl\left(3\sH^2+4\sH\sF+3\sF^2\right)+\l  \label{sigma_p}\\
&&\rho_\star=0, \quad \chi_\star=0 \label{rho}
\ena
where the relation between $\O$ and $\o$ was defined to be $1/\o=2\O+3$ and the trace contributions $\sano,\sY$ take the form
\bea
\sano&=&\frac{c_A}{48}\left[2F_\star^3H_\star^3+3F_\star^2H_\star^4\right]+\frac{c_B}{4800}\left[12F_\star^6-128F_\star^5H_\star\right.+\non\\
&&\left.+291F_\star^4H_\star^2+184F_\star^3H_\star^3+557F_\star^2H_\star^4+138FH_\star^5-54H_\star^6\right]\\
\sY&=&\frac{6}{5}c_Y\mpl\left(2\sH^2+2\sH\sF+\sF^2\right)\left(3\sH^2+18\sH\sF-\sF^2\right)
\ena
and $\stano,\stY$ have been defined as 
\bea
\stano&\equiv&\sano/\left[3(1+\o)H_\star+3(1-\o)F_\star\right] \non\\
\stY&\equiv&\sY/\left[3(1+\o)H_\star+3(1-\o)F_\star\right]\non 
\ena
for $(1+\o)H_\star+(1-\o)F_\star\ne0$.

For $F_\star\ne0$ and $\o\ne1/5$, we can reformulate eqs (\ref{Friedmann rho}) and (\ref{Friedmann p}) in order to get 
\bea
&&(3-21\o)H_\star^2+(4-30)F_\star H_\star-(3-11\o)F_\star^2=(1-5\o)\,\frac{\l}{\Mpl}\\
&&\left(\stano+\stY\right)=2\Mpl\,\frac{H_\star-F_\star}{1-5\o}\label{eq ii+ii ano}
\ena
Imposing $\l=0$ (no effective constant on the brane) greatly simplifies the solution since $H_\star\propto F_\star$. Under that assumption, defining $\cep$ and $\dep$ --- as functions of $\o$ and
$\epsilon=\pm1$ ($\dep$ is a function of the anomaly parameters $c_A$ and $c_B$, of $c_Y$ and the Planck mass as well) --- such that $H_\star-F_\star=\cep F$ and $\left(\stano+\stY\right)=\dep F_\star^5$, the
solution takes the form
\bea
H_\star^2=M^2_{Pl}\left(\cep+1\right)^2\left[\frac{2\cep}{(1-5\o)\dep}\right]^{\frac{1}{2}}, \quad\label{sol H F}
F_\star^2=M^2_{Pl}\left[\frac{2\cep}{(1-5\o)\dep}\right]^\oneover{2} 
\ena
This solution exists for the values of $\o$ such that $\cep/\dep>0$ (for $\o<1/5$) or $\cep/\dep<0$ (for $\o>1/5$). 

The CFT energy density and pressures are then given by
\bea
&&\s_\star=M_{Pl}^6\left(1+3\o\cep\right)\dep\left[\frac{2\cep}{(1-5\o)\dep}\right]^\frac{3}{2}\label{sol sigma}\\
&&\s_{p\star}=\frac{2\o}{1-3\o}\s_\p=-M_{Pl}^6\,\o\left(5+3\cep\right)\dep\left[\frac{2\cep}{(1-5\o)\dep}\right]^\frac{3}{2},\quad \srho=0  \label{sol sigma_p}
\ena
(for $\o=1/3$ we have $\s_\p=0$).
\subsubsection{$F_\star\ne0,\;\o=\frac{1}{5}$}
\paragraph*{Fixed point solutions}
To analyze the case $\o=1/5$, it's better to reformulate equations (\ref{Friedmann rho}) and (\ref{Friedmann p}) in the following way:
\bea
&&2\Mpl \left(H_\star-F_\star\right)F_\star-(1-5\o)F_\star\left(\stano+\stY\right)=0\label{Fried no l}\\
&&M_{Pl}^4\left(3H_\star^2+6H_\star\,F_\star+F_\star^2\right)-\left[3\o H_\star+\left(1-3\o\right)F_\star\right]\left(\stano+\stY\right)=\l \label{Fried l}
\ena 
From the first equation we get $H_\star=F_\star$ and substituing in the second we find the equation for $H_\star$
\bea
-\frac{5}{288}(c_A+2c_B)H_\star^6-120c_Y\mpl \sH^4+10\Mpl H_\star^2=\l
\ena
For $\l=0$ it has a non trivial solution only if $(24c_Y)^2>(c_A+2c_B)$ 
\bea
H_\star^2=F_\star^2=-\frac{24}{c_A+2c_B}\mpl\left[24c_Y\pm\sqrt{576c_Y^2+(c_A+2c_B)}\right]
\ena
The energy density and pressures are equal to
\bea
\s_\star&=&-\ssigp=-\ssigpp=-\frac{240}{c_A+2c_B}\MPl\left[24c_Y\pm\sqrt{576c_Y^2+(c_A+2c_B)}\right] \\
\srho&=&0\non
\ena
\subsubsection{$F_\star=0$}
\paragraph*{Fixed point solutions}
Supposing instead $F_\star=0$, the fixed point solution is 
\bea\label{F zero fix}
H_\star^2&=&-\frac{20}{3c_B}\frac{\o}{\o+1}M_{Pl}^2\left[48c_Y\pm\sqrt{6\left(384c_Y^2-c_B\frac{\o}{\o+1}\right)}\right] \\
\ssig&=&-\ssigpp=\frac{2\o}{3\o-1}\ssigpp=-\frac{20}{c_B}\frac{\o}{\o+1}\MPl\left[48c_Y\pm\sqrt{6\left(384c_Y^2-c_B\frac{\o}{\o+1}\right)}\right]\\
\srho=0\non
\ena
for $\o\ne-1$. This gives real Hubble parameter for $384c_Y^2-c_B\o/(\o+1)>0$. If $\o=-1$, we find 
\bea\label{F zero omega -1 fix}
H_\star^2=\frac{640c_Y}{c_B}\mpl, \quad \ssig=-\ssigp=\oneover{2}\ssigpp=\frac{640c_Y}{3c_B}\MPl,\quad \srho=0 
\ena
If the CFT is characterized by a positive $c_B$, there is no non trivial critical point. For vanishing $c_B$ the only fixed point with $\l\ne0$ is the trivial one.

When $F_\star=-H(1+\o)/(1-\o)$, there is only one possible solution, for which the parameters must have the values: $\o=-1$ (i.e. $\s_\p=-2\s_p$), $c_B=0$, $F_\star=0$ and the fixed point is thus the one in
(\ref{F zero omega -1 fix}).

\paragraph*{Stability analysis}
For both fixed points characterized by the zero value of the extra dimension Hubble parameter, i.e. both for $\o\ne-1$ or $\o=-1$, we must find the eigenvalues of a $4\times4$ matrix. In fact we have a third
order linearized differential equation for the perturbation $\delta H(t)$ and a first order ODE for the energy density $\delta\rho$, while $\delta F(t)$ is found to be proportional to $\delta H(t)$ solving an
algebraic equation
\bea
\delta H^{(3)}&=&-a_2\,\delta\ddot H-a_1\,\delta\dot H-a_0\,\delta H+c_0\,\delta\rho\\
\delta\dot\rho&=&-3(1+w)H_\star\delta\rho\\
\delta F&=&\a\,\delta H
\ena
The coefficients in the differential equations are functions of the anomaly parameters $c_A,c_B$, of the trace parameter $c_Y$, of $\o$ and of $M_{Pl}$.

The eigenvalues $\l_1,\l_2,\l_3$ are then given by the roots of the third degree polynomial
\bea
\l^3+a_2\l^2+a_1\l+a_0=0
\ena
while $\l_4=-3(1+w)H_\star<0$. The coefficients $a_0,a_1,a_2$ are given by
\bea
a_0&=&\frac{12}{25}\frac{8000\Mpl(3+2\a+3(1+\a)\o)}{c_BH_\star(1-\a)\o}-\left(c_BH_\star^4(23\a-54)+144c_Y\mpl\sH^2\right)\o\non\\
a_1&=&\frac{1}{25}\frac{960000\Mpl(1+\a)}{c_BH_\star^2(1-\a)\o}-\left(c_BH_\star^4(137\a-222)+36c_Y\mpl\sH^2\right)\o\non\\
a_2&=&\frac{7-6\a}{1-\a}H_\star
\ena
where
\bea
\a=-3\,\frac{800\Mpl(1+\o)-\left(-3c_BH_\star^4+480c_Y\mpl\sH^2\right)\o}{800\Mpl(5-3\o)-\left(-3c_BH_\star^4+480c_Y\mpl\sH^2\right)(1-3\o)}
\ena

The sign of the eigenvalues $\l_1,\l_2,\l_3$ ($\l_3=\overline\l_2$ iff $27a_0^2+4a_1^3-18a_0a_1a_2-a_1^2a_2^2+4a_0a_2^3>0$, otherwise we get three real roots) determines the nature of the fixed point. Since
$\l_4<0$, we find that we can only have a completely stable fixed point or a saddle. In the case of one real and two complex conjugated roots the critical point can be attractive only if $a_2>0$ and
\bea
-\frac{2a_2}{3}<A+B<\frac{a_2}{3}
\ena
where
\bea
&&A=\sgn(R) \left(|R|+\sqrt{R^2-Q^3}\right)^{\frac{1}{3}},\quad B=\frac{Q}{A}\\
&&R\equiv\oneover{54}(2a_2^3-9a_1a_2+27a_0),\quad Q\equiv\oneover{9}(a_2^2-3a_1)
\ena
When the three roots are real, they are negative (corresponding to an attractive fixed point) iff $a_0,a_1,a_2>0$. For the other values of $a_0,a_1,a_2$ the critical point is a saddle.

The coefficients of the linearized differential equation don't depend on the anomaly parameter $c_A$ corresponding to the type A anomaly, so that only type B anomaly influences the characteristics of this
fixed point.
\subsection{Static compact extra dimensions}
We analyze the set of differential equations when the extra dimensions are (locally) compactified on a sphere, i.e. $\k>0$, supposing that the corresponding acceleration factor $b(t)$ remains constant, so
that $F(t)\equiv0$.

Beside the $H_\star=0$ fixed points, we only have two acceptable time independent solutions to the Friedmann equations. 
The $H_\star=0,\k\ne0$ fixed points are always saddle points as we can conclude from the linear order analysis, since the eigenvalues of the stability matrix (or their real parts) are one opposite to the
other.
\subsubsection{$\o\ne0$}
\paragraph*{Fixed point solutions}
The energy density $\rho_\star$ is zero, $H_\star$ and $\s_\star$ are then determined by
\bea
\left(\sano+\sY\right)&\equiv&\frac{c_A}{48}\frac{\k}{b_0^2} H^4_\star-\frac{c_B}{4800}\left(54H^6_\star-98\frac{\k}{b_0^2} H^4_\star+42\frac{\k^2}{b_0^4} H^2_\star-6\frac{\k^3}{b_0^6}\right)+ \non\\
&&+\frac{4c_Y}{5}\mpl\left(9\sH^4+18\frac{\k}{b_0^2}\frac{\k^2}{b_0^4}\right)= \non\\
&=&\frac{1+\o}{\o}\Mpl\left(3H^2_\star+\frac{\k}{b_0^2}\right)\label{ano=some}\\
\s_\star&=&\Mpl\left(3H^2_\star+\frac{\k}{b_0^2}\right)\\
\rho_\star&=&0
\ena
Restricting the possible values of $\o$, we can obtain at least one positive root $H_\star^2$ of eq (\ref{ano=some}), without having limitations on the anomaly coefficients $c_A,c_B$.

We can illustrate an example, choosing the simple case $c_B=0$ (i.e. there is no contribution from the conformal invariants in the anomaly) and also $\o=-1$, which simplifies the equation \refeq{ano=some}.
The fixed point is thus determined by
\bea
H^2_\star&=&\frac{\k}{b_0^2}\left[9\pm\sqrt{90+\frac{5c_A}{192c_Y}\oneover{\mpl}\frac{\k}{b_0^2}}\right]^{-1}\\
\ssig&=&-\ssigp=2\ssigpp=\Mpl\left(3\left[9\pm\sqrt{90-\frac{5c_A}{192c_Y}\oneover{\mpl}\frac{\k}{b_0^2}}\right]^{-1}+1\right)\frac{\k}{b_0^2}\\
\srho&=&0\non
\ena
which is real for $192c_Y^2\mpl>-c_A\k/45b_0^2$. We can moreover have two distinct positive $\sH$ fixed points if $5c_A\k/9b_0^2<-192c_Y^2\mpl$.

\paragraph*{Stability analysis}
We now analyze the $H_\star\neq0$ fixed points behavior.

Regarding the fixed point determined by (\ref{ano=some}), we get a negative eigenvalue $\l_4=-3(1+w)H_\star$ (given that $w>-1,H_\star>0$) and the other three are the roots of the third degree polynomial
\bea\label{eigenval eq}
\l^3+a_2\l^2+a_1\l+a_0=0
\ena
where $a_{i}=\tilde a_{i}/\tilde a_3, i=0,1,2$ and
\bea
\tilde a_0&=&-\frac{c_A\o}{12}\frac{\k}{b_0^2} H^3_\star+\frac{c_B\o}{1200} H_\star\left(81H^4_\star-98\frac{\k}{b_0^2}H^2_\star+21\frac{\k^2}{b_0^4}\right)\non\\
&&+\frac{144c_Y\o}{5}\mpl\left(\sH^2+\frac{\k}{b_0^2}\right)+6\Mpl(1+\o)H_\star,\non\\
\tilde a_1&=&-\frac{c_A\o}{48}\frac{\k}{b_0^2} H^2_\star+\frac{c_B\o}{4800} \left(111H^4_\star-68\frac{\k}{b_0^2} H^2_\star+21\frac{\k^2}{b_0^4}\right)
+\frac{36c_Y\o}{5}\mpl\left(\sH^2+\frac{\k}{b_0^2}\right)+2\Mpl,\non\\
\tilde a_2&=&\frac{7c_B\o}{1920}H_\star\left(H^2_\star+\frac{\k}{b_0^2}\right),\quad \tilde a_3=\frac{c_B\o}{1920}\left(H^2_\star+\frac{\k}{b_0^2}\right)
\ena
We get a $4\times4$ stability matrix --- despite the fact that we should have only 2 variables ($H,\rho$) --- because the differential equations are of third order: the $\rho$ eigenvalue is $\l_4$, but $H$ is
a superposition of the four modes corresponding to the four eigevalues of the matrix.

As in the previous analysis for the flat extra dimensions, the solutions $\l_{1,2,3}$ of the equation (\ref{eigenval eq}) are such that $\l_1\in \mathbb{R}$, $\l_3=\overline \l_2$ or
$\l_1,\l_2,\l_3\in\mathbb{R}$. Besides, when we have the complex conjugated pair, there are only three possibilities:
\begin{enumerate}
\item $\l_1,\Re(\l_2)=\Re(\l_3),\l_4\le0$ $\Rightarrow$ the solution is stable (even if one of the eigenvalues are null, because that mode won't then contribute to the expression for $H$)
\item $\l_1,\l_4<0,\;\Re(\l_2)=\Re(\l_3)>0$ $\Rightarrow$ we get a saddle point
\item $\l_1>0,\;\Re(\l_2)=\Re(\l_3),\l_4<0$ $\Rightarrow$ in this case too, the fixed point is a saddle
\end{enumerate}
The equalities $\l_1=0$ and $\Re(\l_2)=\Re(\l_3)$ in the case (i) are possible, but not simultaneousely.
When the roots are all real we can get a stable point iff $a_0,a_1,a_2>0$, which implies $\l_1,\l_2,\l_3<0$, and a saddle otherwise, with one negative two positive, or two negative one positive roots. 
\subsection{Equal scale factors}
Another limit that simplifies some of the calculations is the equal scale factor assumption. In this case the Hubble parameters of the internal space and the 3D space are equal, $F=H$, and we remain with a
set of equations for the variables $H,\rho,\s$, as in the static extra dimension limit.
\subsubsection{$\o\ne0$}
\paragraph*{Fixed point solutions}
We observe that when $\o=0$ there is no acceptable solutions to the time independent Einstein equations. So, we want to find the fixed points in the $H=F$, $\o\ne0$ limit. The time independent Friedmann
plus conservation equations lead to
\bea
\left(\sano+\sY\right)&\equiv&\frac{5}{48}(c_A+2c_B)\sH^4\left(\sH^2+\frac{\k}{a^2}\right)-\oneover{192}c_B\frac{\k^2}{a^4}\sH^2+\oneover{800}c_B\frac{\k^3}{a^6}+  \non\\
&&+\frac{4c_Y}{5}\mpl\left(150\sH^4+20\frac{\k}{b_0^2}\sH^2-\frac{\k^2}{b_0^4}\right)  \non\\
&=&6\Mpl\left(10\sH^2+\frac{\k}{a^2}\right)\\\label{H roots equal}
\ssig&=&\Mpl\left(10\sH^2+\frac{\k}{a^2}\right)\\
\srho&=&0
\ena
provided that $5+3w+2w_\p\neq0$
~\footnote{If the reasonable range of values for the pressures is given by $w,w_\p\ge-1$, this condition is satisfied whenever $w$ (or $w_\p$) is strictly greater than $-1$ and, since we are interested in
fixed points with a general $w\ne-1$, we will assume that this is the case.}
. As a consequence of solving the system of equations we also obtain a constraint on the CFT pressure of the hidden sector $\s_\p$, as we must impose $\o=1/5$, i.e. $\s_\p=\s_p$. The equation (\ref{H roots
equal}) yields the value of the Hubble parameter at the fixed point as a function of $c_A,c_B,c_Y,M_{Pl},\k$. 

In the case of flat extra dimensions $\k=0$ we immidiately solve the system of equations finding (discarding the trivial $\sH=0$ solution)
\bea
\sH^2&=&-\frac{24}{c_A+2c_B}\mpl\left[24c_Y\pm\sqrt{576c_Y^2+(c_A+2c_B)}\right]\\  
\ssig&=&-\ssigp=-\ssigpp=-\frac{240}{c_A+2c_B}\MPl\left[24c_Y\pm\sqrt{576c_Y^2+(c_A+2c_B)}\right],  \qquad  \srho=0  \non
\ena
The solution is acceptable if $-(c_A-2c_B)<(24c_Y)^2$ and the two roots are both positive when $(c_A-c_B)<0$.
\paragraph*{Stability analysis}
The last situation that we are considering is the case of equal scale factors, as in the fixed point analysis. We only found one fixed point with $F=H$, that entails a relation
between the two CFT pressures $\s_\p=\s_p$ ($\o=1/5$). We could thus calculate the stability matrix eigenvalues corresponding to this particular limit.

When the extra dimensions spatial curvature is zero $\k=0$, in addition to the vanishing 3D curvature ($k=0$), the stability matrix can be studied straightforward. All the eigenvalues are coincident since
$\d H\propto\d\s\propto\d\rho$. They are given by
\bea
\l=-(5+3w+2w_\p)\sH<0
\ena
For $\sH>0$ the fixed point is hence stable.

\end{document}